\documentclass[a4paper,11pt]{article}

\usepackage[utf8x]{inputenc}
\usepackage{graphicx}
\usepackage{float}
\usepackage[T1]{fontenc}
\usepackage{epsfig}
\usepackage{color}
\usepackage{amsmath,jheppub_0,mathtools}
\usepackage{subfloat}
\usepackage{amsfonts}
\usepackage{braket}
\usepackage{cleveref}
\usepackage{epstopdf}
\usepackage{caption}
\usepackage{subcaption}
\usepackage{enumitem}
\usepackage[titletoc,toc,title]{appendix}
\usepackage{hyperref}
\hypersetup{
	colorlinks=true,
	linkcolor=blue,
	filecolor=red,      
	urlcolor=blue,
	citecolor=blue
} 

	\title{Reflected Entropy and Entanglement Negativity for Holographic Moving Mirrors}

\author[a]{Jaydeep Kumar Basak}
\author[b]{Debarshi Basu}
\author[c]{Vinay Malvimat}
\author[d]{Himanshu Parihar}
\author[e]{and Gautam Sengupta}

\affiliation[a,b,d,e]{
	Department of Physics,\\Indian Institute of Technology Kanpur,\\208016, India}
\affiliation[c]{Theory Division, Saha Institute of Nuclear Physics,
	Homi Bhaba National Institute (HBNI),
	1/AF, Bidhannagar, Kolkata 700064, India.}
\emailAdd{jaydeep@iitk.ac.in}
\emailAdd{debarshi@iitk.ac.in}
\emailAdd{vinay.malvimat@saha.ac.in}
\emailAdd{himansp@iitk.ac.in}
\emailAdd{sengupta@iitk.ac.in}
	\date{}

\abstract{
	\noindent We investigate the time evolution of reflected entropy and entanglement negativity for mixed state configurations involving two adjacent and disjoint intervals in the radiation flux of moving mirrors by utilizing the $AdS/BCFT$ duality. These measures are computed for the required mixed state configurations by using the respective replica techniques in the large central charge limit of the $BCFT_{1+1}$. We demonstrate that the results obtained exactly agree with the corresponding holographic computations in the dual bulk $AdS_3$ geometry with an end of the world brane. In this context, the analogues of the Page curves for these measures are obtained for the required configurations in the radiation flux of kink and escaping mirrors which mimic the Hawking radiation from evaporating and  eternal black holes respectively.
}
	
\begin{document}
	
	\maketitle
	
		\section{Introduction}
	\label{sec:intro}
	The study of black hole information loss paradox has played a pivotal role in deepening our understanding of the quantum structure of spacetime. The paradox occurs post Page time when the fine grained entropy of the radiation emitted from a black hole becomes larger than the coarse grained entropy of the black hole leading to a violation of unitarity.
	Recent progress in this regard was initiated by the proposal of a novel \textit{island} formula for the fine grained entropy of subsystems in quantum field theories coupled to semi-classical theories of gravity \cite{Penington:2019npb,Almheiri:2019psf,Almheiri:2019hni,Almheiri:2019yqk,Almheiri:2020cfm}. The crucial insight of the island formula involves certain regions in the black hole geometry termed islands, whose contribution to the fine grained entropy of a bath subsystem coupled to a black hole leads to the reproduction of the Page curve for the Hawking radiation. This in turn led to a wide range of exciting developments through the application of the island formula to a variety of black hole/bath systems and their corresponding dual $CFT$s in the context of holography \cite{Almheiri:2019psy,Anderson:2020vwi,Chen:2019iro,Balasubramanian:2020hfs,Chen:2020wiq,Gautason:2020tmk,Bhattacharya:2020ymw,Anegawa:2020ezn,Hashimoto:2020cas,Hartman:2020swn,Krishnan:2020oun,Alishahiha:2020qza,Geng:2020qvw,Li:2020ceg,Chandrasekaran:2020qtn,Bak:2020enw,Krishnan:2020fer,Karlsson:2020uga,Hartman:2020khs,Balasubramanian:2020coy,Balasubramanian:2020xqf,Sybesma:2020fxg,Chen:2020hmv,Ling:2020laa,Hernandez:2020nem,Marolf:2020rpm,Matsuo:2020ypv,Akal:2020twv,Caceres:2020jcn,Raju:2020smc,Deng:2020ent,Anous:2022wqh,Bousso:2022gth,Hu:2022ymx,Grimaldi:2022suv,Akers:2022max,Yu:2021rfg,Geng:2021mic,Chou:2021boq,Hollowood:2021lsw,He:2021mst,Arefeva:2021kfx,Ling:2021vxe,Bhattacharya:2021dnd,Azarnia:2021uch,Saha:2021ohr,Hollowood:2021wkw,Sun:2021dfl,Li:2021dmf,Aguilar-Gutierrez:2021bns,Ahn:2021chg,Yu:2021cgi,Lu:2021gmv,Caceres:2021fuw,Akal:2021foz,Arefeva:2022cam,Arefeva:2022guf,Bousso:2022ntt,Krishnan:2021ffb,Zeng:2021kyb,Teresi:2021qff,Okuyama:2021bqg,Chen:2021jzx,Pedraza:2021ssc,Guo:2021blh,Kibe:2021gtw,Renner:2021qbe,Dong:2021oad,Raju:2021lwh,Nam:2021bml,Kames-King:2021etp,Chen:2021lnq,Sato:2021ftf,Kudler-Flam:2021alo,Wang:2021afl,Ageev:2021ipd,Buoninfante:2021ijy,Cadoni:2021ypx,Marolf:2021ghr,Chu:2021gdb,Urbach:2021zil,Li:2021lfo,Neuenfeld:2021bsb,Aalsma:2021bit,Ghosh:2021axl,Bhattacharya:2021jrn,Geng:2021wcq,Krishnan:2021faa,Verheijden:2021yrb,Bousso:2021sji,Karananas:2020fwx,Goto:2020wnk,Bhattacharya:2020uun,Chen:2020jvn,Agon:2020fqs,Laddha:2020kvp,Akers:2019nfi,Chen:2019uhq,Uhlemann:2021nhu,Uhlemann:2021itz,Geng:2021iyq,Geng:2021hlu,Geng:2020fxl,Espindola:2022fqb}. Substantial evidence for the island formula for the entanglement entropy was provided by taking certain replica wormhole contributions to the corresponding gravitational path integral in \cite{Penington:2019kki, Almheiri:2019qdq, Almheiri:2020cfm, Kawabata:2021vyo}.
	
	Interestingly, it was recently demonstrated that the island formula naturally arises for an  $AdS$ geometry dual to
	a conformal field theory on a manifold with a boundary (BCFT)\cite{Suzuki:2022xwv,Anous:2022wqh}. The bulk
	geometry dual to the $d$ dimensional BCFT on a manifold $\Sigma$ in this scenario, is described by a bulk 
	$AdS_{d+1}$ spacetime with a boundary $Q \cup \Sigma$ where $Q$ is a codimension one surface termed as
	the end-of-the-world (EOW) brane \cite{Takayanagi:2011zk,Fujita:2011fp}. The application of the RT/HRT prescription for the holographic entanglement entropy in the above $AdS/BCFT$ scenario is then expected to lead to results consistent with the Island formulation. Consequently this implies that it should be possible to obtain the Page curve 
	for subsystems in the dual $BCFT$ through the utilization of the usual RT/HRT formula for the holographic entanglement entropy. Recently an intriguing model involving a holographic moving mirror configuration, which mimics the Hawking radiation from an evaporating black hole was explored in \cite{Akal:2020twv,Akal:2021foz}
	in the context of the $AdS/BCFT$ correspondence \footnote{Note that historically, the radiation from a moving mirror was investigated in \cite{Davies_Fulling} ( see also \cite{birrell_davies_1982} and \cite{Good:2019tnf, Good:2020nmz} for more recent discussions.).}. The authors in 
	\cite{Akal:2020twv,Akal:2021foz} considered various moving mirror profiles which simulate the Hawking radiation from eternal and evaporating black holes and demonstrated that the entanglement entropy of a subsystem in
	the radiation flux of such moving mirrors described by $BCFT$s leads to unitary Page curves. Following this advancement several interesting investigations have been explored in \cite{Reyes:2021npy,Sato:2021ftf,Kawabata:2021hac,Ageev:2021ipd}.
	
	In the light of the developments described above, the significant issue of the mixed state entanglement structure between parts of the radiation/bath system arises naturally. Since the entanglement entropy is not a valid measure for mixed state entanglement it is required to investigate other appropriate quantum information theoretic measures in this context. This is an active area of investigation in quantum information theory and several such mixed state entanglement and correlation measures have been proposed to address this critical issue \cite{Vidal:2002zz,2002,plenio2006introduction,2009,Dutta:2019gen}. Some of these measures such as  the entanglement negativity, the reflected entropy and the entanglement of purification have also been explored in the context of conformal field theories \cite{Calabrese:2012ew,Calabrese:2012nk,Calabrese:2014yza,Hirai:2018jwy,Caputa:2018xuf,Dutta:2019gen} and holography \cite{Rangamani:2014ywa,Chaturvedi:2016rcn,Chaturvedi:2016rft,Jain:2017aqk,Takayanagi:2017knl,Malvimat:2018txq,Malvimat:2018ood,Kudler-Flam:2018qjo,Kusuki:2019zsp,Dutta:2019gen,KumarBasak:2020eia,KumarBasak:2021lwm,Dong:2021clv,Basu:2021axf,Basu:2021awn,Basu:2022nds}. Specifically the reflected entropy and the entanglement negativity which are of interest in the present article have been shown to possess rich entanglement phase structures and are expected to lead to corresponding analogues of the Page curves \cite{Shapourian:2020mkc,KumarBasak:2021rrx,Akers:2022max,Vardhan:2021mdy} for black hole/bath systems. Furthermore, recently the island formulae for the reflected entropy and the entanglement negativity have been proposed and explored in \cite{Chandrasekaran:2020qtn,KumarBasak:2020ams,Dong:2021oad,Li:2020ceg,Ling:2021vxe}. 	
	
	The holographic moving mirror configuration provides us with an interesting model to investigate the structure of mixed state entanglement for the Hawking radiation without any direct reference to the island formula as the holographic computations in the $AdS/BCFT$ scenario naturally encode the island contributions. Consequently the investigation of the time evolution of measures such as the entanglement negativity and the reflected entropy to probe the mixed state entanglement structure of the radiation flux from a moving mirror is expected to lead to interesting insights into the corresponding entanglement structure of the Hawking radiation. In this context, we compute the reflected entropy and the entanglement negativity of mixed state configurations involving  disjoint and adjacent intervals in the radiation flux of  moving mirrors by utilizing the replica technique for the corresponding $BCFT_{1+1}$s. Subsequently these quantities
	are computed through appropriate holographic techniques involving the bulk $AdS_{3}$ spacetime with a EOW brane utilizing a Banados map \cite {Banados:1998gg,Roberts:2012aq,Shimaji:2018czt,Akal:2021foz}. We demonstrate that the holographic computations involving the bulk geometry precisely match with the corresponding replica technique results in the  $BCFT_{1+1}$ for various phases in the moving mirror configurations.  Finally we determine the Page curves for the reflected entropy and the entanglement negativity of the above mixed state configurations in the radiation flux of the kink and the escaping mirrors which mimic the Hawking radiation from evaporating and eternal black holes respectively. 
	
	A brief outline of our article is as follows. In section \ref{sec:review} we review the computation of entanglement entropy in the holographic moving mirror configuration. Subsequently, we recapitulate the replica technique and the holographic proposals for the mixed state correlation and entanglement measures involving the reflected entropy and the entanglement negativity. Following this in section \ref{sec:Reflected-MM} we employ  the replica technique in a $BCFT_{1+1}$ to determine the reflected entropy for the mixed state configurations of adjacent and disjoint intervals in the radiation flux of a moving mirror. Furthermore, we determine the corresponding holographic dual and demonstrate that the results from the $AdS_3$ match exactly with that obtained from the replica technique. Subsequently we determine the analogues of the Page curves for the reflected entropy of the mixed state configurations in question for various mirror profiles. In section \ref{secENMM} we determine the entanglement negativity of adjacent and disjoint intervals through the corresponding replica technique in a $BCFT_{1+1}$. Following this, we compute the holographic entanglement negativity for several phases of the required configurations and demonstrate that the results  precisely match with those obtained through the replica technique in the $BCFT_{1+1}$. We then obtain the corresponding Page curves for the entanglement negativity for different mirror profiles. Finally, in section \ref{sec:sum} we summarize the results of our article and present our conclusions.
	%-------------------------------------------------------
	%-------------------------------------------------------
	\section{Review of holographic moving mirror, reflected entropy and entanglement negativity}\label{sec:review}
	In this section we review the salient features of the moving mirror setup and subsequently the $BCFT$ computations of the entanglement entropy of a subsystem which is placed in the radiation flux from such a moving mirror. Subsequently, we provide a brief overview of  the  replica techniques for reflected entropy and entanglement negativity and describe the corresponding holographic proposals .
	\subsection{Moving mirror from a $BCFT_{1+1}$}	
	We begin with a brief review of the computation of the entanglement entropy in a moving mirror model described in \cite{Akal:2020twv,Akal:2021foz}. The authors in \cite{Akal:2020twv,Akal:2021foz} considered a moving mirror in two dimensions with the mirror trajectory given by the profile $x=Z(t)$. In this model, the region  $x\geq Z(t)$ which is to the right of the mirror trajectory (see fig.\ref{fig:disjoint}) is described by a boundary conformal field theory. The authors employed light cone coordinates $u=t-x$, $v=t+x$ and subsequently performed the following conformal transformations
	\begin{equation}\label{c-map}
	\tilde{u}=p(u), \quad \tilde{v}=v,
	\end{equation}
	where  $\tilde{u}=\tilde{t}-\tilde{x}$ and $\tilde{v}=\tilde{t}+\tilde{x}$ are the light cone variables in the tilde coordinates. The function $p(u)$ is chosen such that the mirror trajectory is given by $v=p(u)$ and may be written in the original coordinates as
	\begin{equation}
	t+Z(t)=p(t-Z(t)).
	\end{equation}
	Under the conformal transformation in \cref{c-map}, the moving mirror trajectory in the original coordinates $u,v$ is mapped to that of a static mirror in the tilde coordinates, given as $\tilde{u}-\tilde{v}=0$. Note that, as mentioned earlier, the right half plane (RHP) $\tilde{x}\geq 0$ is described by a BCFT. Furthermore, in these tilde coordinates the BCFT is in its vacuum state, and hence the stress energy tensor in the original coordinates may be obtained in terms of the conformal anomaly described by the Schwarzian as given below
	\begin{equation}
	T_{uu}=\frac{c}{24 \pi}\left[\frac{3}{2}\bigg(\frac{p''(u)}{p'(u)}\bigg)^2-\frac{p'''(u)}{p'(u)}\right].
	\end{equation}

	The authors in \cite{Akal:2020twv,Akal:2021foz} considered two distinct moving mirror profiles which mimic Hawking radiation from black holes. These profiles are discussed below:
	
	\smallskip
	\noindent\textbf{Escaping mirror} 
	
	\noindent The trajectory of an escaping mirror is described by the  profile 
	\begin{equation}\label{escprof}
	p(u)= -\beta \log(1+e^{-\frac{u}{\beta}}).
	\end{equation}
	Note that this profile is such that in the far past $t\to -\infty$, the mirror is static $Z(t)\simeq 0$, whereas at late times $t\to \infty$ the trajectory is given by $Z(t)\simeq -t-\beta e^{-2t/\beta}$. The energy flux from this mirror mimics the Hawking radiation from an eternal black hole.
	
	\smallskip
	
	\noindent\textbf{Kink mirror} 
	The authors in \cite{Akal:2020twv,Akal:2021foz} also investigated the radiation from a kink mirror which is described by the following profile
	\begin{equation}\label{kinkprof}
	p(u)=-\beta \log(1+e^{-\frac{u}{\beta}})+\beta\log(1+e^{\frac{(u-u_0)}{\beta}}),
	\end{equation}
	where $\beta\geq 0$ and $u_0\geq 0$. The mirror trajectory in the limit $\beta\to 0$ is given by
	\begin{equation}
	Z(t)\simeq 
	\begin{cases}
	0,\quad (t< 0)\\
	-t \quad (0\leq t \leq \frac{u_0}{2})\\
	-\frac{u_0}{2} \quad (t >\frac{u_0}{2}).
	\end{cases}
	\end{equation}
	The energy flux from the kink mirror mimics the Hawking  radiation emitted during the evaporation of a single sided black hole.
	
	\subsection{Entanglement entropy for moving mirrors from a $BCFT_{1+1}$}
	Having reviewed the moving mirror setup we now briefly discuss the computation of the entanglement entropy of a single interval in the presence of radiation from a moving mirror described by a $BCFT_{1+1}$ with a large central charge. To this end we consider a single interval $A\equiv [x_0,x_1]$ at a time slice $t$. It is well known that the entanglement entropy for an interval $A$ in a CFT may be expressed in terms of twist field correlators using the replica technique described in \cite{Calabrese:2004eu,Calabrese:2009qy} as
	\begin{equation}\label{entropy-def}
	S_A=\lim_{n\to 1}\frac{1}{1-n}\log\left<\sigma_n(t,x_0)\bar{\sigma}_n(t,x_1)\right>,
	\end{equation}
	where the conformal weights of the twist fields $\sigma_n$ and $\bar{\sigma}_n$ are given by
	\begin{equation}
	h_n=\bar{h}_n=\frac{c}{24}\left(n-\frac{1}{n}\right).
	\end{equation}
	As described earlier, the moving mirror setup may be mapped to that of a static mirror via the conformal transformation given in \cref{c-map} and hence the above two point twist correlator is related to the two point function on the right half plane (RHP) as follows
	\begin{equation}\label{2pt}
	\left<\sigma_n(t_0,x_0)\bar{\sigma}_n(t_1,x_1)\right>=(p'(u_0) p'(u_1))^{h_n} 
	\left<\tilde{\sigma}_n(\tilde{t}_0,\tilde{x}_0)\tilde{\bar{\sigma}}_n(\tilde{t}_1,\tilde{x}_1)\right>_{\mathrm{ RHP}}.
	\end{equation}
	For holographic CFTs, the two point twist correlator on the RHP may be written in terms of two different OPE channels of the four point function of a chiral CFT on the full complex plane which can be re-expressed as the product of 2 two point correlators on $\mathbb{R}^{1,1}$ as follows \cite{Sully:2020pza}
	\begin{align}\label{2pt-RHP}
	\langle\tilde{\sigma}_{n}\left(\tilde{t}_0,\tilde{x}_0\right)\tilde{\bar{\sigma}}_{n}\left(\tilde{t}_1,\tilde{x}_1\right)\rangle_{\rm RHP} 
	= \max
	\begin{cases}
	\langle\tilde{\sigma}_{n}\left(\tilde{t}_0,\tilde{x}_0\right)\tilde{\bar{\sigma}}_{n}\left(\tilde{t}_1,\tilde{x}_1\right)\rangle_{\mathbb{R}^{1,1}}\langle\tilde{\sigma}_{n}\left(\tilde{t}_0,-\tilde{x}_0\right)\tilde{\bar{\sigma}}_{n}\left(\tilde{t}_1,-\tilde{x}_1\right)\rangle_{\mathbb{R}^{1,1}} \\
	e^{2(1-n)S_{\mathrm{bdy}}}
	\Pi_{i\in\{0,1\}}\,
	\langle\tilde{\sigma}_{n}\left(\tilde{t}_i,\tilde{x}_i\right)\tilde{\bar{\sigma}}_{n}\left(\tilde{t}_i,-\tilde{x}_i\right)\rangle_{\mathbb{R}^{1,1}}.
	\end{cases}
	\end{align}
	The Renyi entropy of the interval $A$ was then obtained by using eq. (\ref{2pt-RHP}),  (\ref{2pt}) in eq. (\ref{entropy-def}) as follows \cite{Akal:2020twv}
	\begin{equation}
	S^{(n)}_A=\mathrm{Min}[S_A^{(n)\mathrm{dis}},S_A^{(n)\mathrm{con}}],
	\end{equation}
	where 
	
	\begin{align}\label{SAMM}
	S_A^{(n)\mathrm{dis}}&=\frac{c}{12}\bigg(1+\frac{1}{n}\bigg)\log\left[\frac{(t+x_0-p(t-x_0))(t+x_1-p(t-x_1))}{\epsilon^2\sqrt{p'(t-x_0)p'(t-x_1)}}\right]+2 S_{\text{bdy}}, \\
	S_A^{(n)\mathrm{con}}&=\frac{c}{12}\bigg(1+\frac{1}{n}\bigg)\log\left[\frac{(x_1-x_0)\left(p(t-x_0)-p(t-x_1)\right)}{\epsilon^2\sqrt{p'(t-x_1)p'(t-x_0)}}\right].
	\end{align}	
	where the superscripts $dis$ and $con$ will be explained shortly in the next subsection on holographic description. The entanglement entropy may be obtained from the above equation by taking the replica limit $n\to1$.
	Following this, the authors in \cite{Akal:2020twv} demonstrated the time evolution for entanglement entropy  of a single interval for the escaping and the kink mirror, follow the Page curves for eternal and evaporating black holes respectively.
	
	\subsection{Entanglement entropy for  holographic moving mirrors}
	
	We now discuss the AdS/BCFT duality \cite{Takayanagi:2011zk} which is utilized to construct the gravity dual of CFTs in the presence of moving mirrors \cite{Akal:2020twv,Akal:2021foz}. In this context, consider a BCFT on a $d$ dimensional manifold $\Sigma$ with a boundary $\partial \Sigma$. The gravity dual of the $d$ dimensional BCFT is constructed by extending the manifold $ \Sigma$ to a $d+1$ dimensional bulk manifold whose boundary is the surface $Q\cup  \Sigma$, such that $Q$ is homologous to $\Sigma$. The surface $Q$ is called the end of the world brane (EOW) which obeys the following Neumann boundary condition to preserve the conformal invariance in the $BCFT$,
	\begin{equation}
	K_{ab}-h_{ab}K=-{\cal T}h_{ab},
	\end{equation}
	where $h_{ab}$ is the induced metric, $K_{ab}$ is the extrinsic curvature and $\mathcal{T}$ is the tension of the brane $Q$ which depends on the boundary conditions
	on $\partial \Sigma$. The  entanglement entropy for a single interval $A$ in a holographic $BCFT_{1+1}$ can be expressed  as follows \cite{Takayanagi:2011zk,Fujita:2011fp}
	\begin{equation}
	S_A=\mathrm{Min}[S_A^{\mathrm{con}},S_A^{\mathrm{dis}}],
	\end{equation}
	where $S_A^{\mathrm{con}}$ and $S_A^{\mathrm{dis}}$ are given by
	\begin{equation}
	S_A^{\mathrm{con}}=\frac{L(\Gamma_A^{\mathrm{con}})}{4G_N}, \quad S_A^{\mathrm{dis}}=\frac{L(\Gamma_A^{\mathrm{dis}})}{4G_N}.
	\end{equation}
	Here $L(\Gamma_A^{\mathrm{con}})$ and $L(\Gamma_A^{\mathrm{dis}} )$ are the lengths of the connected and disconnected geodesics homologous to $A$.  Having described the AdS/BCFT construction, we now review the general procedure to construct the gravity dual of the moving mirror setup by utilizing the Banados map which we discuss below.
	
	\subsubsection*{The Banados Map}
	Consider the Poincare $AdS_3$ metric which is given as
	\begin{align}
	ds^2=\frac{d\eta^2-dU\,dV}{\eta^2}\label{PAdS3}
	\end{align}
	The conformal transformations described in \cref{c-map} are dual to the following coordinate map in the bulk, known as the Banados map  \cite{Banados:1998gg,Roberts:2012aq,Shimaji:2018czt,Akal:2021foz}
	\begin{align}
	&U=p(u)\quad,\quad V=v+\frac{p''(u)}{2p'(u)}z^2\notag\\
	&\eta=z\sqrt{p'(u)}\,.\label{Banados-map}
	\end{align}
	Expressed in the above $(u,v,z)$ coordinates the bulk dual of the $BCFT_2$ describing the moving mirror set up is given by the following plane wave geometries in asymptotically $AdS_3$ spacetimes \cite{Akal:2020twv,Akal:2021foz}:
	\begin{align}
	ds^2=\frac{dz^2}{z^2}+T_+(u)du^2-\frac{1}{z^2}du\,dv\,.\label{plane-wave-MM-dual}\\
	\textrm{where} \quad T_{+}(u)=\frac{3\left(p^{\prime \prime}\right)^{2}-2 p^{\prime} p^{\prime \prime \prime}}{4 p^{\prime 2}} .
	\end{align}
	As described in \cite{Takayanagi:2017knl}, the profile for the EOW brane $Q$ is given by
	\begin{align}
	v=p(u)-\frac{p''(u)}{2p'(u)}z^2-2\lambda z\sqrt{p'(u)}\,,
	\end{align}
	where the constant $\lambda$ is related to the tension $\mathcal{T}$ of the brane as
	\begin{align}
	\lambda=\frac{\mathcal{T}}{\sqrt{1-\mathcal{T}^2}}.
	\end{align}
	Under, the Banados map \cref{Banados-map}, the brane has an AdS$_2$ geometry with the straight line profile given by
	\begin{align}
	V-U+2\lambda \eta=0\,.
	\end{align}
	The authors in \cite{Akal:2020twv,Akal:2021foz} evaluated the entanglement entropy by computing the lengths of the geodesics homologous to the subsystem in the above bulk dual geometry of the moving mirror setup, and demonstrated that the results match exactly with those obtained through the twist correlators in the corresponding dual $BCFT_{1+1}$ given in \cref{SAMM}.

	\subsection{Reflected entropy}\label{sec:S_R-holo}
	As described in the introduction, entanglement entropy is a unique entanglement measure for pure states only and is neither a correlation nor an entanglement measure for mixed states. In this context, here we review a  mixed state correlation measure termed reflected entropy. Consider a bipartite quantum system $A\cup B$ in a mixed state described by the density matrix $\rho_{AB}$. It is possible to canonically purify a given mixed state $\rho_{AB}$ into $\ket{\sqrt{\rho_{AB}}}$ in a doubled Hilbert space $\mathcal{H}_A\otimes\mathcal{H}_B\otimes\mathcal{H}_{A^\star}
	\otimes\mathcal{H}_{B^\star}$ where $A^\star$ and $B^\star$ are the mirror copies of $A$ and $B$ respectively\footnote{See \cite{Dutta:2019gen,Jeong:2019xdr} for more details about the construction of $\ket{\sqrt{\rho_{AB}}}$.}. 
	The reflected entropy $S_R(A: B)$ is then defined as \cite{Dutta:2019gen}
	\begin{align}
	S_R(A: B) = S(AA^*)_{\sqrt{\rho_{AB}}}.
	\end{align}
	Note that the reflected entropy is a measure of both classical and quantum correlations between the subsystems $A$ and $B$.
	
	Consider the configuration of two disjoint intervals $A \equiv [z_1,z_2]$ and $B \equiv [z_3,z_4]$ in a $CFT_{1+1}$. The reflected entropy may then be obtained from the R\'enyi reflected entropy  through a replica technique described in \cite{Dutta:2019gen}, in terms  of a four point twist field correlator given as follows 
	\begin{equation}
	\begin{aligned}
	S_R(A:B) &=\lim_{ n \to 1 }\lim_{m \to 1 }\,S_n\left(AA^*\right)_{\psi_m}\\ 
	&= \lim_{ n \to 1 }\lim_{m \to 1 } \frac{1}{1-n}\log\frac{\left<\sigma_{g^{}_A}(z_1)\sigma_{g_A^{-1}}(z_2)\sigma_{g^{}_B}(z_3)\sigma_{g_B^{-1}}(z_4)\right>_{\mathrm{CFT}^{\bigotimes mn}}}{\left(\left<\sigma_{g^{}_m}(z_1)\sigma_{g_m^{-1}}(z_2)\sigma_{g^{}_m}(z_3)\sigma_{g_m^{-1}}(z_4)\right>_{\mathrm{CFT}^{\bigotimes m}}\right)^n},\label{Renyi-reflected}
	\end{aligned}
	\end{equation}
	where the twist operators $\sigma_{g_A}$ and $\sigma_{g_B}$ are inserted at the endpoints of the intervals $A$ and $B$ and $m,n$ are the replica indices. The conformal dimensions for these twist operators are given by
	\begin{equation}\label{reflected-twist-field}
	h_{g_A^{-1}}=h_{g^{}_B}=\frac{n \,c}{24}\left(m-\frac{1}{m}\right),
	\quad h_{{g^{}_B} g_A^{-1}}=\frac{2 \,c}{24}\left(n-\frac{1}{n}\right), \quad h_{g_m}=\frac{c}{24}\left(m-\frac{1}{m}\right).
	\end{equation}
	
	Following this in \cite{Dutta:2019gen}, the authors also proposed a holographic construction for the reflected entropy. They demonstrated that the holographic reflected entropy is dual to twice the entanglement wedge cross section (EWCS) using gravitational path integral techniques, as follows
	\begin{align}
	S_{R}(A: B)=2 E_{W}(A: B)\,.
	\end{align}
	The authors also showed that the holographic computation exactly matches with the field theoretic replica technique results for the $AdS_3/CFT_2$ scenario.
	
	\subsection{Entanglement Negativity}
	
	Another quantum information theoretic measure which will be discussed in this article is the entanglement negativity introduced by Vidal and Werner in \cite{Vidal:2002zz}. This non-convex entanglement monotone \cite{Plenio:2005cwa} provides an upper bound to the distillable entanglement of a given mixed state. In this subsection we recapitulate the definition and corresponding replica technique to compute the entanglement negativity in $CFT_{1+1}$s. Furthermore, we discuss  the holographic construction for the entanglement negativity in the context of AdS/CFT correspondence. For a bipartite quantum system $A\cup B$ in a  mixed state $\rho_{AB}$, the entanglement negativity $\mathcal{E}(A:B)$ is defined as the trace norm of the partially transposed density matrix $\rho_{A\cup B}^{T_B}$ expressed as follows
	\begin{align}
	{\cal E}(A:B)= \log ||&\rho_{A\cup B}^{T_B}||\\
	\langle q_i^1q_j^2 \mid \rho_{A \cup B}^{T_{B}}\mid q_k^1q_l^2\rangle &= \langle q_i^1q_l^2\mid \rho_{A \cup B}\mid q_k^1q_j^2\rangle\nonumber.
	\end{align}
	Note that in the above equation the second line describes the partial transpose operation with respect to the subsystem $B$. The trace norm in the first line denotes the absolute sum of the eigenvalues of  $\rho_{A\cup B}^{T_B}$. Following this, in \cite{Calabrese:2012ew,Calabrese:2012nk,Calabrese:2014yza} a replica technique was developed to compute the entanglement negativity, which is expressed as follows
	\begin{align}
	\mathcal{E}(A:B)=\lim _{n_{e} \rightarrow 1} \ln \operatorname{Tr}\left(\rho_{A\cup B}^{T_B}\right)^{n_{e}}\,,
	\end{align}
	where $n_e$ indicates that the limit has to be considered as an analytic continuation of the even sequences of $n_e$ to $n_e=1$. This technique was then utilized to evaluate the entanglement negativity for various mixed states in $CFT_{1+1}$.  Specifically, the entanglement negativity for the mixed state configuration of disjoint intervals was described through the replica technique by a four point correlator of twist operators $\tau,\, \overline{\tau}$ as follows
	\begin{align}
	\operatorname{Tr}(\rho_{AB}^{T_{B}})^{n_e}=\langle \tau_{n_e}(u_{1}) \overline{\tau}_{n_e}(v_{1}) \overline{\tau}_{n_e}(u_{2}) \tau_{n_e}(v_{2})\rangle
	\end{align}
	
	Following these developments, various holographic proposals for computing the entanglement negativity of the mixed state configurations involving adjacent and the disjoint intervals in $AdS_3/CFT_2$ scenario was developed in \cite{Jain:2017aqk,Malvimat:2018txq}. These proposals involved specific algebraic sums of the holographic Renyi entropies of order half described by the lengths of backreacting cosmic branes homologous to the subsystems  \cite{KumarBasak:2021rrx,KumarBasak:2020ams}. For the mixed state of disjoint intervals in a $CFT_2$, the holographic entanglement negativity is expressed as follows
	\begin{align}
	{\cal E}=\frac{1}{2}\left[S^{(1/2)}(A\cup C)+S^{(1/2)}(B\cup C )-S^{(1/2)}(A \cup C \cup B)-S^{(1/2)}(C)\right]\label{HENDJ}.
	\end{align}
	In the context of $AdS_3/CFT_2$ and for spherical entangling surfaces in higher dimensions, the effect of the backreaction of the bulk cosmic brane reduces to a numerical proportionality factor ($\chi_d$) such that\footnote{Note that, for spherical entangling surfaces this backreaction pre-factor $\mathcal{X}_d$ is dependent on the spacetime dimensions, and may be expressed as follows \cite{Hung:2011nu,Rangamani:2014ywa,Dong:2016fnf}
		\begin{equation}
		\mathcal{X}_d = \left(\frac{1}{2}x_d^{d-2}\left(1 + x_d^2\right) -
		1\right), \,\,\,\,\, x_d = \frac{2}{d}\left(1 + \sqrt{1 - \frac{d}{2} +
			\frac{d^2}{4}}\right).
		\end{equation}
	} $S^{(1/2)}(X)=\chi_2 S(X)=\frac{3}{2}S(X)$\footnote{Note that in the context of $AdS/BCFT$ the relation between the length of the back reacted cosmic brane  and the length of the geodesic ${\cal L}^{1/2}=\frac{3}{2}{\cal L}$ is only for the dynamical part. This is because  the contribution from the boundary degrees of freedom contained in $S_{bdy}$ to the Renyi entropy is independent of the replica index $n$ as evident from \cref{SAMM}. }. Hence, the above expression reduces to an algebraic sum of the lengths of the geodesics given below
	\begin{align}\label{hendjcomb}
	{\cal E}=	\frac{3}{16 G_N}\left[{\cal L}_{A\cup C }+{\cal L}_{B\cup C }-{\cal L}_{A\cup C \cup B}-{\cal L}_{C}\right].
	\end{align}
	where ${\cal L}_{X}$ corresponds to the length of the geodesic homologous to the subsystem $X$. $C$ corresponds to the interval sandwiched between the disjoint intervals $A$ and $B$. Note that in the limit $C\to \emptyset$ the two interval $A$ and $B$ become adjacent and the proposal reduces to 
	\begin{align}
	{\cal E}&=\frac{1}{2}\left[S^{(1/2)}(A)+S^{(1/2)}(B )-S^{(1/2)}(A \cup B)\right]\label{HENADJ}.\\
	&=\frac{3}{16 G_N}\left[{\cal L}_{A }+{\cal L}_{B }-{\cal L}_{A \cup B}\right]\label{henadjH}.
	\end{align}

	This completes our review of the results which we will utilize in the present article. We now proceed to compute the reflected entropy in the moving mirror setup in the following section.
	\section{ Reflected Entropy for Moving Mirrors} \label{sec:Reflected-MM}
	
	In this section, we begin by computing the reflected entropy of  mixed states involving two adjacent and disjoint intervals in the radiation flux of  a moving mirror by employing the replica technique in a $BCFT_{1+1}$ in the large central charge limit. Following this, we determine the holographic reflected entropy of the above configurations  through the entanglement wedge cross section in the corresponding dual bulk $AdS_{3}$ geometry utilizing the Banados map.
	
	\subsection{Reflected entropy in a $BCFT_{1+1}$}
	In this subsection, we describe the field theoretic computations for the reflected entropy for various bipartite mixed state configurations in a moving mirror setup. As described in \cref{sec:S_R-holo}, the replica technique for computing the reflected entropy of two disjoint intervals $A$ and $B$ involves certain correlation functions of twist operators $\sigma_{g^{}_A}$ and $\sigma_{g^{}_B}$ inserted at the endpoints of the two intervals, whose conformal dimensions are given in \cref{reflected-twist-field}. In a $BCFT_{1+1}$  such correlation functions may be evaluated utilizing the \textit{doubling trick}  \cite{Cardy:2004hm,recknagel_schomerus_2013}. In the large central charge limit, the $BCFT_{1+1}$ correlators are expected to have specific factorization in the respective channels \cite{Sully:2020pza}. In the following, after describing the general method, we will systematically evaluate the twist correlators corresponding to different phases depending on subsystem sizes.

	\subsubsection{Two disjoint intervals} \label{sec:Disjoint}
\begin{figure}[h!]
		\centering
		\includegraphics[scale=0.5]{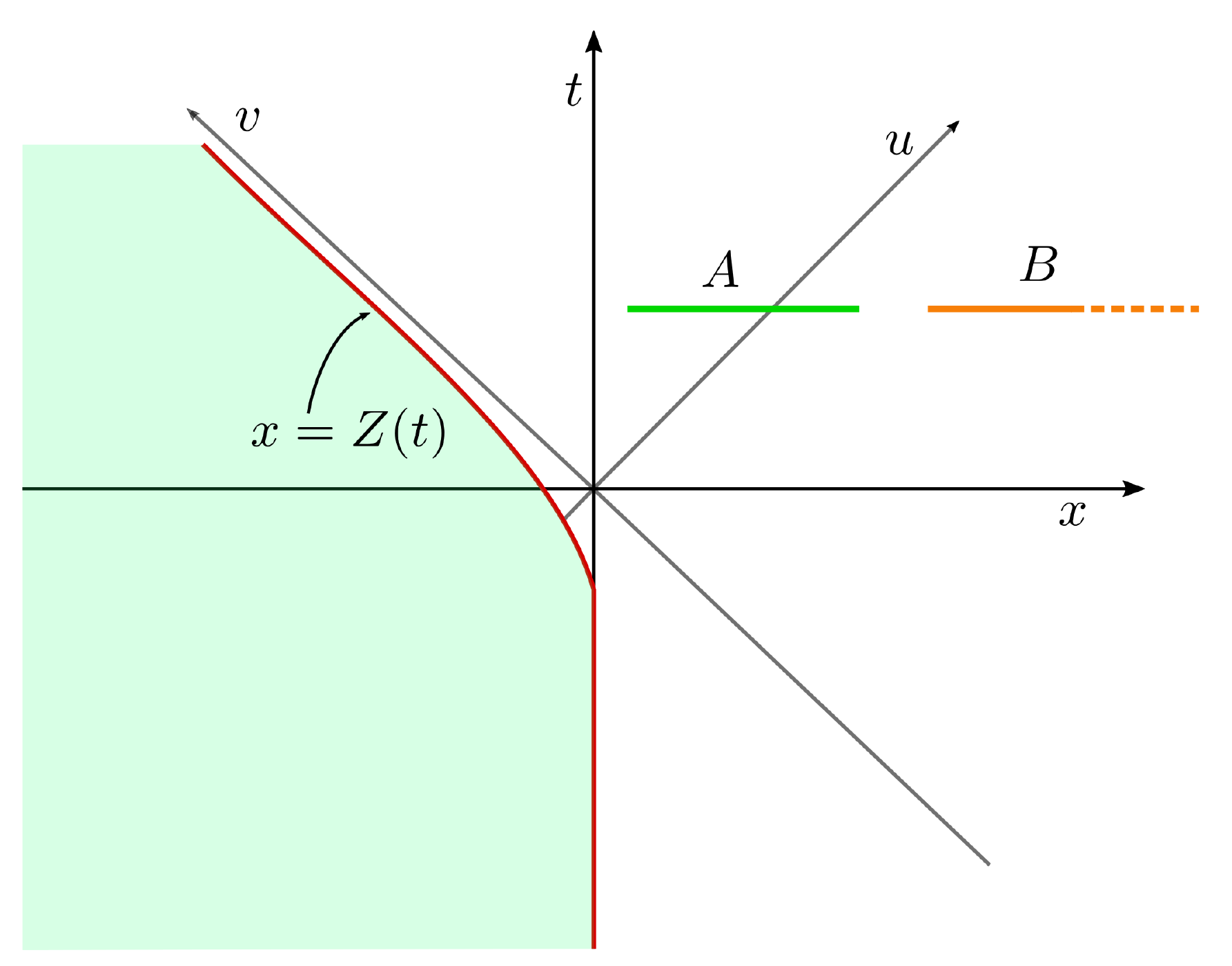}
		\caption{Schematics of two disjoint intervals $A=[(t,x_1),(t,x_2)]$ and $B=[(t,x_3),(t,\infty)]$ in the moving mirror BCFT. Figure modified from \cite{Akal:2021foz}.}
		\label{fig:disjoint}
	\end{figure}

	In this section we begin with the computation of the reflected entropy between two disjoint intervals $A=[(t,x_1),(t,x_2)]$ and $B=[(t,x_3),(t,\infty)]$ in the $(1+1)$-dimensional boundary conformal field theory (BCFT) with a boundary described by the moving mirror $x=Z(t)$. The schematics of the setup is sketched in \cref{fig:disjoint}.  
	
\begin{figure}[h!]
		\centering
		\includegraphics[scale=0.4]{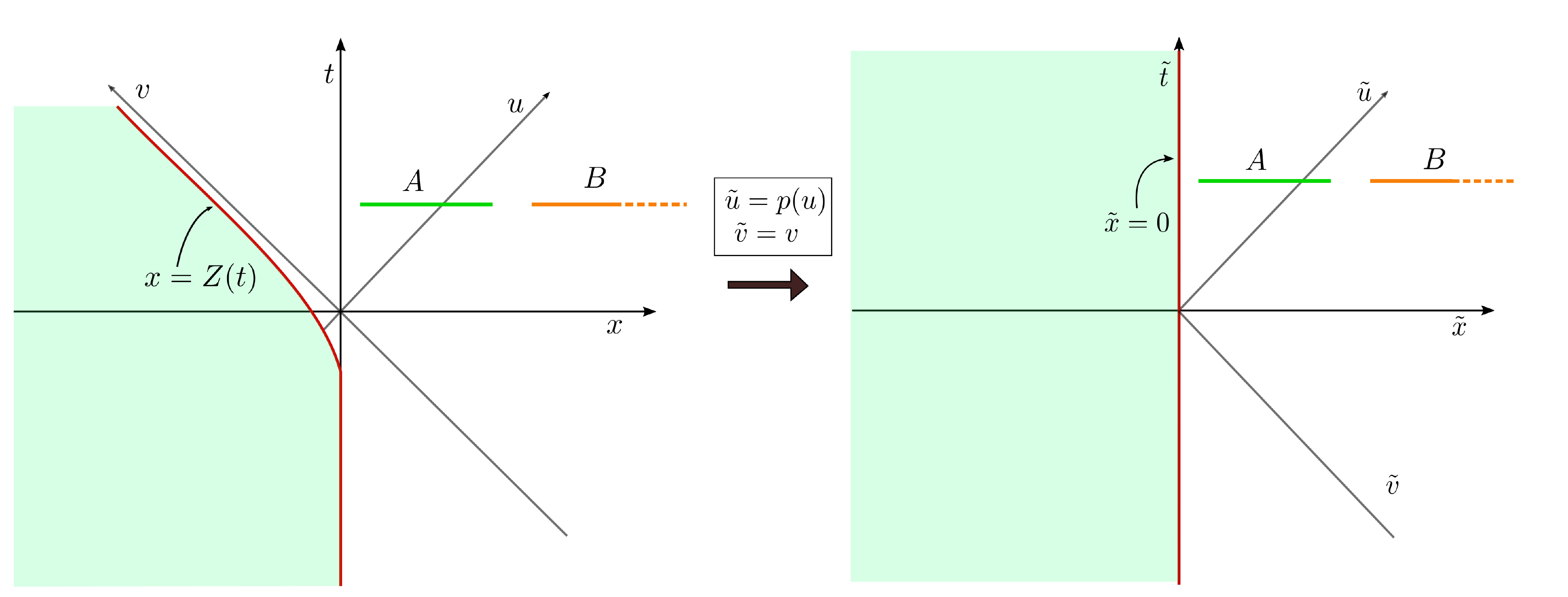}
		\caption{Conformal transformation to the static mirror setup. Schematics of two disjoint intervals $A=[(\tilde{t}_1,\tilde{x}_1),(\tilde{t}_2,\tilde{x}_2)]$ and $B=[(\tilde{t}_3,\tilde{x}_3),(\tilde{t}_4,\infty)]$ in the static mirror BCFT. Figure modified from \cite{Akal:2021foz}.}
		\label{fig:disj-static-map}
	\end{figure}

	As discussed in \cite{Akal:2020twv,Akal:2021foz}, it is convenient to employ the light cone coordinates $(u,v)$. We conformally map the static mirror setup described by the new set of coordinates $(\tilde{u},\tilde{v})$ as
	\begin{align}
	\tilde{u}=p(u)\quad,\quad \tilde{v}=v\,,\label{static-mirror-map}
	\end{align}
	which renders the mirror profile in the new coordinates as $\tilde{v}=\tilde{u}$. The conformal transformation in \cref{static-mirror-map} maps the moving mirror to a static one described by a $BCFT_2$ on the right half plane (RHP) and the intervals are now given by
	\begin{align*}
	A=[(\tilde{t}_1,\tilde{x}_1),(\tilde{t}_2,\tilde{x}_2)]~~,~~B=[(\tilde{t}_3,\tilde{x}_3),(\tilde{t}_4,\infty)]\,.
	\end{align*}
	Alternatively, in terms of the light-cone coordinates the intervals may be expressed as follows
	\begin{align*}
	A=[(\tilde{u}_1,\tilde{v}_1),(\tilde{u}_2,\tilde{v}_2)]~~,~~B=[(\tilde{u}_3,\tilde{v}_3),(\infty,\infty)]\,,
	\end{align*}
	as shown schematically in \cref{fig:disj-static-map}. We may now employ the standard complex coordinates $\tilde{z}=(\tilde{t},\tilde{x})$ to describe the $BCFT_2$ on the RHP and denote the endpoints of the two intervals as $A=[\tilde{z}_1,\tilde{z}_2]$ and $B=[\tilde{z}_3,\infty]$. Furthermore, for simplicity we will put the intervals on an equal time slice utilizing another conformal map:
	\begin{align*}
	A=[b_1,b_2]~~~,~~~B=[b_3,\infty]\,.
	\end{align*}
	At the end of the computations, the final result for the reflected entropy is obtained in the original coordinates $(x,t)$ by inverting the above conformal transformations.
	There are three possible phases of the reflected entropy for the present configuration which are motivated from the structure of the corresponding bulk entanglement wedge. In the following, we will systematically investigate these phases through appropriate replica techniques \cite{Dutta:2019gen} in the field theory.
	\subsubsection*{Phase-I}
	\label{sec:disjoint-phase-1}
\begin{figure}[h!]
		\centering
		\includegraphics[scale=0.8]{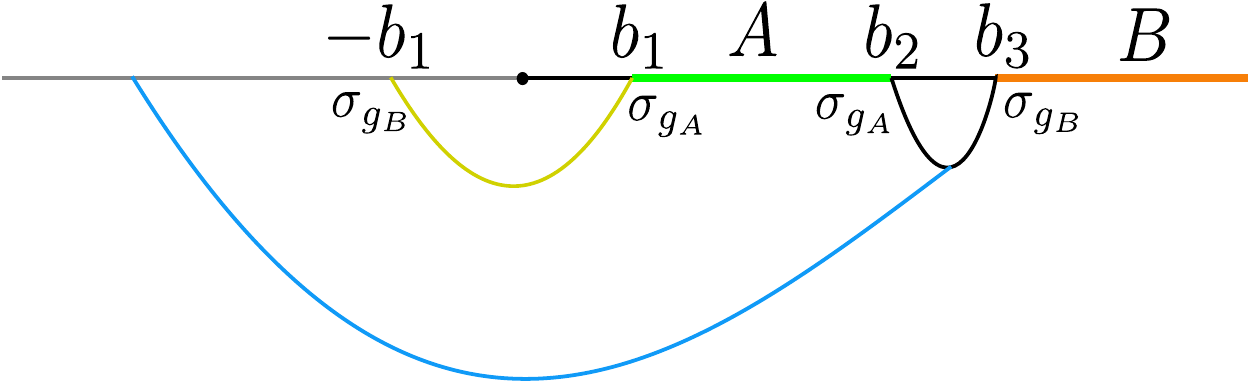}
		\caption{OPE channel corresponding to the present configuration.}
		\label{fig:disj-phase1-S_R}
	\end{figure}

	In this phase, we consider the subsystem $A$ to be very close to the boundary. Utilizing \cref{Renyi-reflected} the reflected entropy for this configuration may be obtained as \cite{Li:2021dmf}
	\begin{align}
	S_R(A:B)=\lim_{m,n\to 1}\frac{1}{1-n}\log\frac{\left<\sigma_{g^{}_A}(b_1)\sigma_{g_A^{-1}}(b_2)\sigma_{g^{}_B}(b_3)\right>_{\mathrm{BCFT}^{\bigotimes mn}}}{\left(\left<\sigma_{g^{}_m}(b_1)\sigma_{g_m^{-1}}(b_2)\sigma_{g^{}_m}(b_3)\right>_{\mathrm{BCFT}^{\bigotimes m}}\right)^n}\,,\label{disj-phase1-SR-calc}
	\end{align} 
	For this channel, the OPE configuration is described in \cref{fig:disj-phase1-S_R}. Utilizing the doubling trick described in \cite{Sully:2020pza}\footnote{For a more complete description of the doubling trick, see \cite{Cardy:2004hm,recknagel_schomerus_2013}.}, the three-point correlator of twist fields in the numerator of \cref{disj-phase1-SR-calc} may be recast into a six-point function in a chiral CFT$_2$ defined on the full complex plane as
	\begin{align}
	&\left<\sigma_{g^{}_A}(b_1)\sigma_{g_A^{-1}}(b_2)\sigma_{g^{}_B}(b_3)\right>_{\mathrm{BCFT}^{\bigotimes mn}}\notag\\&=\left<\sigma_{g^{}_A}(b_1)\sigma_{g^{-1}_B}(-b_1)\sigma_{g_A^{-1}}(b_2)\sigma_{g^{}_B}(-b_2)\sigma_{g^{}_B}(b_3)\sigma_{g^{-1}_B}(-b_3)\right>_{\mathrm{CFT}^{\bigotimes mn}}\label{3pt-disj}
	\end{align}
	
	Now, using the OPE channel sketched in \cref{fig:disj-phase1-S_R}, the above six-point twist correlator may be seen to factorize in the following way:
	\begin{align}
	\left<\sigma_{g^{}_A}(b_1)\sigma_{g^{-1}_B}(-b_1)\right>_{\mathrm{CFT}^{\bigotimes mn}}\left<\sigma_{g_A^{-1}}(b_2)\sigma_{g^{}_B}(-b_2)\sigma_{g^{}_B}(b_3)\sigma_{g^{-1}_B}(-b_3)\right>_{\mathrm{CFT}^{\bigotimes mn}}\,,
	\end{align}
	where the corresponding OPE coefficient includes the boundary degrees of freedom in terms of the boundary entropy $S_{\text{bdy}}$\footnote{Note that the constant $S_{\text{bdy}}$ encapsulates the boundary degrees of freedom and appears in the BOE expansion coefficient \cite{Cardy:2004hm,Sully:2020pza}.}.
	The twist correlator in the denominator of \cref{disj-phase1-SR-calc} admits similar doubling and factorization in the specific channel under consideration and hence we have the following expression for the reflected entropy between the two disjoint intervals as
	\begin{align}
	S_R(A:B)=\lim_{m,n\to 1}\frac{1}{1-n}\log\frac{\left<\sigma_{g_A^{-1}}(b_2)\sigma_{g^{}_B}(-b_2)\sigma_{g^{}_B}(b_3)\sigma_{g^{-1}_B}(-b_3)\right>_{\mathrm{CFT}^{\bigotimes mn}}}{\left(\left<\sigma_{g_m^{-1}}(b_2)\sigma_{g^{}_m}(-b_2)\sigma_{g^{}_m}(b_3)\sigma_{g^{-1}_m}(-b_3)\right>_{\mathrm{CFT}^{\bigotimes m}}\right)^n}\,.
	\end{align}
	The conformal block which provides the dominant contribution to the four-point function in the numerator is given by \cite{Dutta:2019gen,Jeong:2019xdr}
	\begin{align}
	\log\mathcal{F}(mnc,h,h_p,\eta)=-2h\log(\eta)+2h_p\log\left(\frac{1+\sqrt{1-\eta}}{2\sqrt{\eta}}\right)\,,\label{Conformal-block}
	\end{align}
	where $h$ is the conformal dimension of the external twist operators $\sigma_{g^{}_A}\,,\, \sigma_{g^{}_B}$, $h_p$ is the conformal dimension of the intermediate operator providing dominant contribution to the four-point function, and $\eta$ is the cross ratio corresponding to the disjoint intervals. These conformal dimensions are given as \cite{Dutta:2019gen}
	\begin{align}
	h=\frac{nc}{24}\left(m-\frac{1}{m}\right)~~,~~h_p=\frac{2c}{24}\left(n-\frac{1}{n}\right)\label{S_R-twist-dimensions}
	\end{align}
	Note that an overall factor of two is absent in the expression for the conformal block in \cref{Conformal-block}, as the correlator on the full complex plane is chiral. Furthermore, to compute the four-point twist correlator, we also need the OPE coefficient for the dominant channel, which involves contributions from the boundary entropy as well as the usual OPE coefficient determined in \cite{Dutta:2019gen}:
	\begin{align}
	C_{n,m}=e^{2(1-n)S_{\text{bdy}}}\,(2m)^{-4h}\,.
	\end{align}
	Utilizing \cref{S_R-twist-dimensions,Conformal-block} and the above OPE coefficients, the final expression for the reflected entropy between the two disjoint intervals may be written as
	\begin{align}
	S_R=\frac{c}{3}\log\left(\frac{b_2+b_3+2\sqrt{b_2b_3}}{b_3-b_2}\right)+2S_{\text{bdy}}\label{disj-phase1-SR-expr}
	\end{align}
	where we have used the following expression for the cross ratio
	\begin{align}
	\eta=\frac{(b_3+b_2)^2}{(b_3-b_2)^2}\,.
	\end{align}
	
	Now, for generic complex intervals $A=[\tilde{z}_1,\tilde{z}_2]$ and $B=[\tilde{z}_3,\infty]$ in the $BCFT_2$ defined on the RHP, we may write down the reflected entropy simply by modifying the cross ratio $\eta$ as
	\begin{align}
	S_R(A:B)=\frac{c}{3}\log\left(\frac{1+\sqrt{1-\tilde{x}}}{\sqrt{\tilde{x}}}\right)+2S_{\text{bdy}}\,,
	\end{align}
	where the modified cross ratio $\tilde{x}$ is given as
	\begin{align}
	\tilde{x}=\frac{(\tilde{z}_3-\tilde{z}_2^*)(\tilde{z}_3^*-\tilde{z}_2)}{(\tilde{z}_3-\tilde{z}_2)(\tilde{z}_3^*-\tilde{z}_2^*)}\equiv\frac{(\tilde{u}_3-\tilde{v}_2)(\tilde{v}_3-\tilde{u}_2)}{(\tilde{u}_3-\tilde{u}_2)(\tilde{v}_3-\tilde{v}_2)}\,,\label{cross ratio1}
	\end{align}
	where $\tilde{z}^*$ denotes the complex conjugate of $\tilde{z}$.
	Finally, inverting the static mirror map, the reflected entropy between the intervals $A=[(t,x_1),(t,x_2)]$ and $B=[(t,x_3),(t,\infty)]$ in the moving mirror BCFT may be obtained as
	\begin{align}
	S_R(A:B)&=\lim_{m,n\to 1}\frac{1}{1-n}\log\frac{\left<\sigma_{g_A^{-1}}(t,x_2)\sigma_{g^{}_B}(t,-x_2)\sigma_{g^{}_B}(t,x_3)\sigma_{g^{-1}_B}(t,-x_3)\right>_{\mathrm{CFT}^{\bigotimes mn}}}{\left(\left<\sigma_{g_m^{-1}}(t,x_2)\sigma_{g^{}_m}(t,-x_2)\sigma_{g^{}_m}(t,x_3)\sigma_{g^{-1}_m}(t,-x_3)\right>_{\mathrm{CFT}^{\bigotimes m}}\right)^n}\notag\\
	&=\frac{c}{3}\log\left(\frac{1+\sqrt{1-\zeta_1}}{2\sqrt{\zeta_1}}\right)+2S_{\text{bdy}}\,,\label{disj1-SR-final1}
	\end{align}
	where the modified cross ratio involves the mirror profile as
	\begin{align}
	\zeta_1=\frac{\left(p(t-x_3)-t-x_3\right)\left(t+x_3-p(t-x_2)\right)}{\left(x_3-x_2\right)\left(p(t-x_3)-p(t-x_2)\right)}\,.
	\end{align}

	%\begin{figure}[h!]
	%	\centering
	%	\includegraphics[scale=0.15]{EW-disj-phase1-schematic.png}
	%	\caption{Schematics of the bulk entanglement wedge corresponding to the mixed state configuration of the two disjoint intervals in the dual moving mirror BCFT.}
	%	\label{fig:EW-disj-phase1-schematic}
	%\end{figure}
	%

	\subsubsection*{Phase-II}
\begin{figure}[h!]
		\centering
		\includegraphics[scale=0.8]{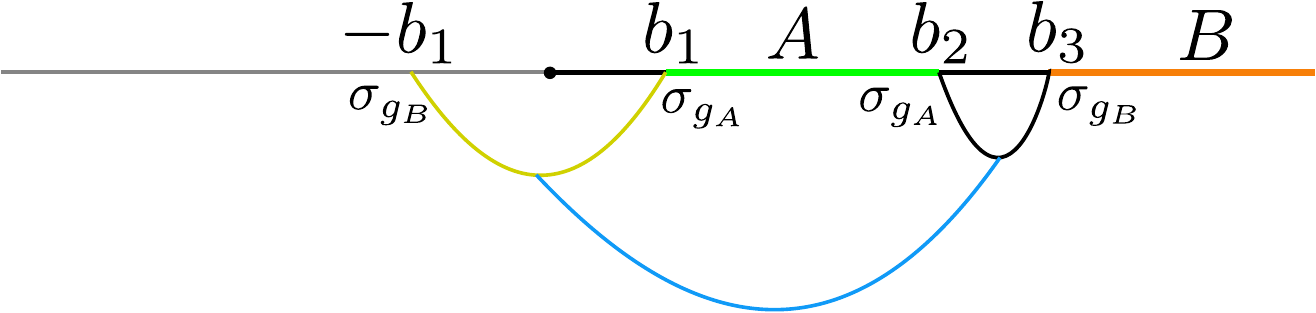}
		\caption{OPE channel corresponding to phase-II of the reflected entropy for two disjoint intervals.}
		\label{fig:SR-disj2-ope}
	\end{figure}

	In this subsection we are concerned with the case where the first interval $A$ is very small and close to the boundary of the moving mirror BCFT. For this phase the corresponding OPE channels are shown in \cref{fig:SR-disj2-ope}. In this channel upon utilizing the doubling trick, the three-point twist correlator relevant to the calculation of the reflected entropy reduces to the following chiral twist correlator in a CFT$_2$ defined on the full complex plane
	\begin{align}
	\left<\sigma_{g^{}_A}(b_1)\sigma_{g_A^{-1}}(b_2)\sigma_{g^{}_B}(b_3)\right>_{\mathrm{BCFT}^{\bigotimes mn}}=\left<\sigma_{g^{}_A}(b_1)\sigma_{g^{-1}_B}(-b_1)\sigma_{g_A^{-1}}(b_2)\sigma_{g^{}_B}(b_3)\right>_{\mathrm{CFT}^{\bigotimes mn}}\label{disj2-doubling0}
	\end{align}
	
	The dominant contribution to the four point function in \cref{disj2-doubling0} is obtained from the Virasoro conformal block given in \cref{Conformal-block}, with the cross ratio now given by
	\begin{align}
	\eta=\frac{(b_2-b_1)(b_1+b_3)}{(b_3-b_1)(b_1+b_2)}\,.\label{cross ratio2}
	\end{align}
	Therefore, the reflected entropy between the static intervals $A$ and $B$ for this phase is obtained as
	\begin{align}
	S_R=\frac{c}{3}\log\left(\frac{1+\sqrt{1-\eta}}{\sqrt{\eta}}\right)\,,
	\end{align}
	with the cross ratio $\eta$ given in \cref{cross ratio2}. As discussed earlier, the reflected entropy for the generic intervals $A=[\tilde{z}_1,\tilde{z}_2]$ and $B=[\tilde{z}_3,\infty]$ may be obtained similarly with the modified cross ratio given by
	\begin{align}
	\tilde{x}=\frac{(\tilde{z}_2-\tilde{z}_1)(\tilde{z}_3-\tilde{z}_1^*)}{(\tilde{z}_3-\tilde{z}_1)(\tilde{z}_2-\tilde{z}_1^*)}\equiv \frac{(\tilde{u}_2-\tilde{u}_1)(\tilde{u}_3-\tilde{v}_1)}{(\tilde{u}_3-\tilde{u}_1)(\tilde{u}_2-\tilde{v}_1)}\,.
	\end{align}
	Finally inverting the static mirror map in \cref{static-mirror-map}, the reflected entropy in phase-II for the two disjoint intervals $A$ and $B$ in the moving mirror BCFT is obtained to be
	\begin{align}
	S_R(A:B)&=\lim_{m,n\to 1}\frac{1}{1-n}\log\frac{\left<\sigma_{g_A^{-1}}(t,x_1)\sigma_{g^{}_A}(t,-x_1)\sigma_{g^{}_B}(t,x_2)\sigma_{g^{-1}_B}(t,x_3)\right>_{\mathrm{CFT}^{\bigotimes mn}}}{\left(\left<\sigma_{g_m^{-1}}(t,x_1)\sigma_{g^{}_m}(t,-x_1)\sigma_{g^{}_m}(t,x_2)\sigma_{g^{-1}_m}(t,x_3)\right>_{\mathrm{CFT}^{\bigotimes mn}}\right)^n}\notag\\
	&=\frac{c}{3}\log\left(\frac{1+\sqrt{1-\zeta_2}}{\sqrt{\zeta_2}}\right)\,,\label{disj1-SR-final2}
	\end{align}
	where the modified cross ratio involves the mirror profile $p(u)$ as
	\begin{align}
	\zeta_2=\frac{\left(p(t-x_2)-p(t-x_1)\right)\left(p(t-x_3)-t-x_1\right)}{\left(p(t-x_3)-p(t-x_1)\right)\left(p(t-x_2)-t-x_1\right)}\,.
	\end{align}

	\subsubsection*{Phase-III}
\begin{figure}[h!]
		\centering
		\includegraphics[scale=0.8]{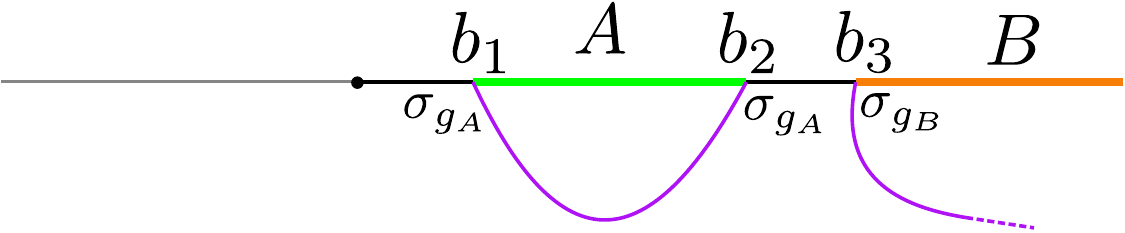}
		\caption{OPE channel corresponding to phase-III of the reflected entropy for two disjoint intervals.}
		\label{fig:SR-disj-phase3}
	\end{figure}

	This phase corresponds to the situation where the two intervals are in the bulk of the moving mirror BCFT and are sufficiently separated. For this phase, the corresponding OPE channels of the three-point correlator in \cref{3pt-disj} is sketched in \cref{fig:SR-disj-phase3}.
	
	In this phase the twist correlator in the chiral CFT on the full complex plane factorizes as follows
	\begin{align}
	&\left<\sigma_{g^{}_A}(b_1)\sigma_{g_A^{-1}}(b_2)\sigma_{g^{}_B}(b_3)\sigma_{g^{-1}_B}(\infty)\right>_{\mathrm{CFT}^{\bigotimes mn}}\notag\\&=\left<\sigma_{g^{}_A}(b_1)\sigma_{g_A^{-1}}(b_2)\right>_{\mathrm{CFT}^{\bigotimes mn}}\left<\sigma_{g^{}_B}(b_3)\sigma_{g^{-1}_B}(\infty)\right>_{\mathrm{CFT}^{\bigotimes mn}}\label{disj2-doubling}
	\end{align}
	Hence the reflected entropy for this phase vanishes as follows
	\begin{align}
	S_R=\lim_{m,n\to 1}\frac{1}{1-n}\log\frac{\left<\sigma_{g^{}_A}(b_1)\sigma_{g_A^{-1}}(b_2)\right>_{mn}\left<\sigma_{g^{}_B}(b_3)\sigma_{g^{-1}_B}(\infty)\right>_{mn}}{\left(\left<\sigma_{g^{}_m}(b_1)\sigma_{g_m^{-1}}(b_2)\right>_{m}\left<\sigma_{g^{}_m}(b_3)\sigma_{g^{-1}_m}(\infty)\right>_{m}\right)^n}= 0\,.
	\end{align}
	\subsubsection{Two adjacent intervals}
	\label{sec:Adjacent}
	We next focus on the case with two adjacent intervals  $A=[(t,0),(t,x_1)]$ and $B=[(t,x_1),(t,x_2)]$ in the moving mirror BCFT with the mirror profile given by $x=Z(t)$. As described before, we conformally map to the static mirror setup defined on the RHP utilizing the transformation in \cref{static-mirror-map}. The adjacent intervals under consideration are then given in the transformed coordinates as
	\begin{align*}
	A\equiv[0,\tilde{z}_1]= [(\tilde{t}_0,0),(\tilde{t}_1,\tilde{x}_1)]~~,~~B\equiv[\tilde{z}_1,\tilde{z}_2]=[(\tilde{t}_1,\tilde{x}_1),(\tilde{t}_2,\tilde{x}_2)]
	\end{align*}
	\begin{figure}[h!]
		\centering
		\includegraphics[scale=0.5]{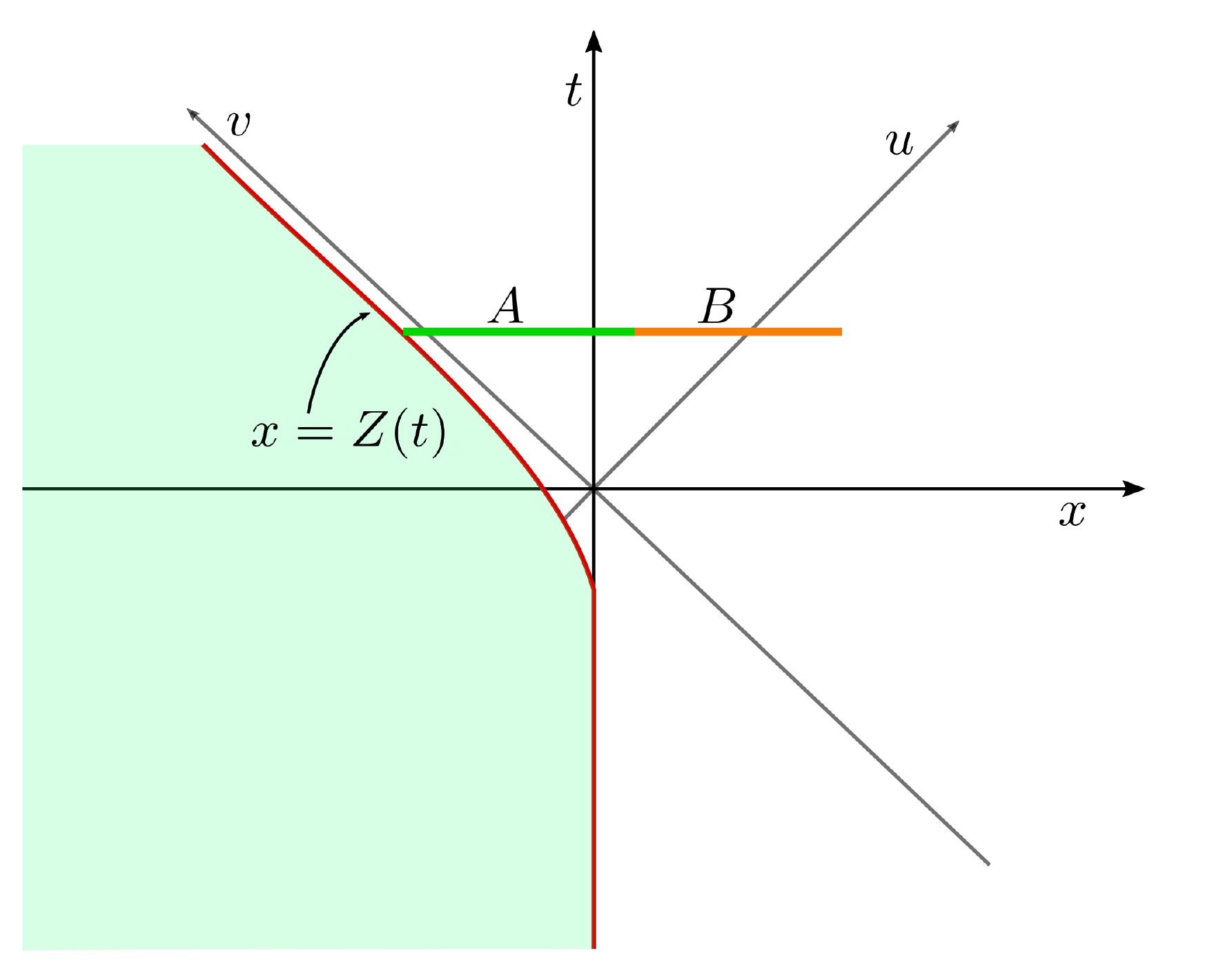}
		\caption{Schematics of two adjacent intervals $A=[(t,0),(t,x_1)]$ and $B=[(t,x_1),(t,x_2)]$ in the moving mirror BCFT. Figure modified from \cite{Akal:2021foz}.}
		\label{fig:adjacent}
	\end{figure}
	The reflected entropy for the configuration of two adjacent intervals $A=[0,\tilde{z}_1]$ and $B=[\tilde{z}_1,\tilde{z}_2]$ is computed in the replica technique through certain correlators of twist operators in the static $BCFT_2$ as
	\begin{align}
	S_R(A:B)=\lim_{m,n\to 1}\frac{1}{1-n}\log\frac{\left<\sigma_{g_A^{-1}g_B}(\tilde{z}_1)\sigma_{g^{}_B}(\tilde{z}_2)\right>_{\mathrm{BCFT}^{\bigotimes mn}}}{\left<\sigma_{g^{}_m}(\tilde{z}_2)\right>_{\mathrm{BCFT}^{\bigotimes mn}}}\,,\label{S_R-adj1}
	\end{align}
	where the conformal dimensions of the twist fields $\sigma_{g_B}$ and $\sigma_{g_A^{-1}g_B}$ are given by
	\begin{align}
	h_B=\frac{nc}{24}\left(m-\frac{1}{m}\right)~~,~~h_{AB}=\frac{2c}{24}\left(n-\frac{1}{n}\right)\,.
	\end{align}
	As earlier, the reflected entropy for the two adjacent intervals admit different phases which we investigate below.
	
	\subsubsection*{Phase-I}
	This phase corresponds to the scenario where the second interval $B$ is smaller such that it is close to the boundary of the moving mirror BCFT.
	For this phase, the two-point twist correlator in the numerator of \cref{S_R-adj1} is dominated by the boundary channel\footnote{Note that the conformal block for the $BCFT$ two-point function admits two different types of channels, namely the boundary operator expansion (BOE) and the bulk operator product expansion (OPE) \cite{Sully:2020pza}. The boundary channel refers to the BOE expansion for the $BCFT$ two-point function.} and hence one obtains
	\begin{align}
	\left<\sigma_{g_A^{-1}g_B}(\tilde{z}_1)\sigma_{g^{}_B}(\tilde{z}_2)\right>_{\mathrm{BCFT}^{\bigotimes mn}}&=\left<\sigma_{g_A^{-1}g_B}(\tilde{z}_1)\right>_{\mathrm{BCFT}^{\bigotimes mn}}\left<\sigma_{g^{}_B}(\tilde{z}_2)\right>_{\mathrm{BCFT}^{\bigotimes mn}}\notag\\
	&=\frac{e^{2(1-n)S_{\text{bdy}}}\epsilon^{2(h_A+h_{AB})}}{(2\,\text{Im}\,\tilde{z}_1)^{2h_{AB}}(2 \textrm{Im} \,\tilde{z}_2)^{h_B}}\,,\label{adj-factorization1}
	\end{align}
	where $S_{\text{bdy}}$ quantifies the boundary degrees of freedom  and $\epsilon$ is the UV cut-off in the field theory. Hence, utilizing \cref{adj-factorization1} the reflected entropy for this phase may be obtained from \cref{S_R-adj1} as
	\begin{align}
	S_R&=\frac{c}{3}\log\left(\frac{2\,\text{Im}\,\tilde{z}_1}{\epsilon}\right)+2S_{\text{bdy}}\notag\\
	&=\frac{c}{3}\log\left(\frac{\tilde{v}_1-\tilde{u}_1}{\epsilon}\right)+2S_{\text{bdy}}
	\end{align}
	Inverting the static mirror map, the reflected entropy between the the two intervals $A=[(t,0),(t,x_1)]$ and $B=[(t,x_1),(t,x_2)]$ in the moving mirror BCFT is obtained as follows
	\begin{align}
	S_R(A:B)=\frac{c}{3}\log\left[\frac{t+x_1-p(t-x_1)}{\epsilon\sqrt{p'(t-x_1)}}\right]+2S_{\text{bdy}}\,.\label{Adj-SR-phase1}
	\end{align}
	\subsubsection*{Phase-II} 
	In this phase, the far endpoint of the second interval $B$ is away from the boundary and therefore the OPE channel is favored. Hence, the corresponding two-point function in \cref{S_R-adj1} may be factorized as follows  \cite{Li:2021dmf}
	\begin{align}
	\left<\sigma_{g_A^{-1}g_B}(\tilde{z}_1)\sigma_{g^{}_B}(\tilde{z}_2)\right>_{\mathrm{BCFT}^{\bigotimes mn}}&=\left<\sigma_{g_A^{-1}g_B}(\tilde{z}_1)\sigma_{g^{}_B}(\tilde{z}_2)\sigma_{g^{-1}_B}(\tilde{z}_2^*)\right>_{\mathrm{CFT}^{\bigotimes mn}}\notag\\
	&=(2m)^{-4h_A}(\tilde{z}_1-\tilde{z}_2)^{-4h_A}(\tilde{z}_1-\tilde{z}_2^*)^{-4h_A}(2 \,\text{Im}\,\tilde{z}_2)^{4h_A-4h_{AB}}
	\end{align}
	Subsequently, the reflected entropy between $A$ and $B$ is obtained as
	\begin{align}
	S_R=\frac{c}{3}\log\left[\frac{2(\tilde{z}_1-\tilde{z}_2)(\tilde{z}_1-\tilde{z}_2^*)}{\epsilon(2\,\text{Im}\,\tilde{z}_2)}\right]
	\end{align}
	Finally, reversing the static mirror map, we obtain the reflected entropy in the moving mirror BCFT as
	\begin{align}
	S_R(A:B)=\frac{c}{3}\log\left[\frac{2\left(p(t-x_1)-p(t-x_2)\right)\left(p(t-x_1)-t-x_2\right)}{\epsilon\sqrt{p'(t-x_1)}\left(p(t-x_1)-t-x_1\right)}\right]\,.\label{SR-adj1-final}
	\end{align}

	\subsection{Entanglement wedge cross-section in holographic moving mirrors}
	In this section, we investigate the phase structure of the holographic reflected entropy for bipartite mixed states involving two disjoint and two adjacent intervals, through the bulk entanglement wedge cross section in the dual plane wave geometries. As described earlier, the holographic dual of the $BCFT_2$ with a boundary described by the moving mirror profile $v=p(u)$ is given by the metric in \cref{plane-wave-MM-dual}. As described in \cite{Takayanagi:2017knl,Nguyen:2017yqw}, the entanglement wedge is a co-dimension one bulk region of spacetime dual to the reduced density matrix of the bipartite state under consideration. It was further established in \cite{Dutta:2019gen}, that twice the area of the minimal cross-section of the entanglement wedge (EWCS) is holographically dual to the reflected entropy for the bipartite state. In the following, we proceed to compute the EWCS corresponding to various bipartite state configurations in the moving mirror $BCFT_2$. 
	
	\begin{figure}[h!]
		\centering
		\includegraphics[scale=0.8]{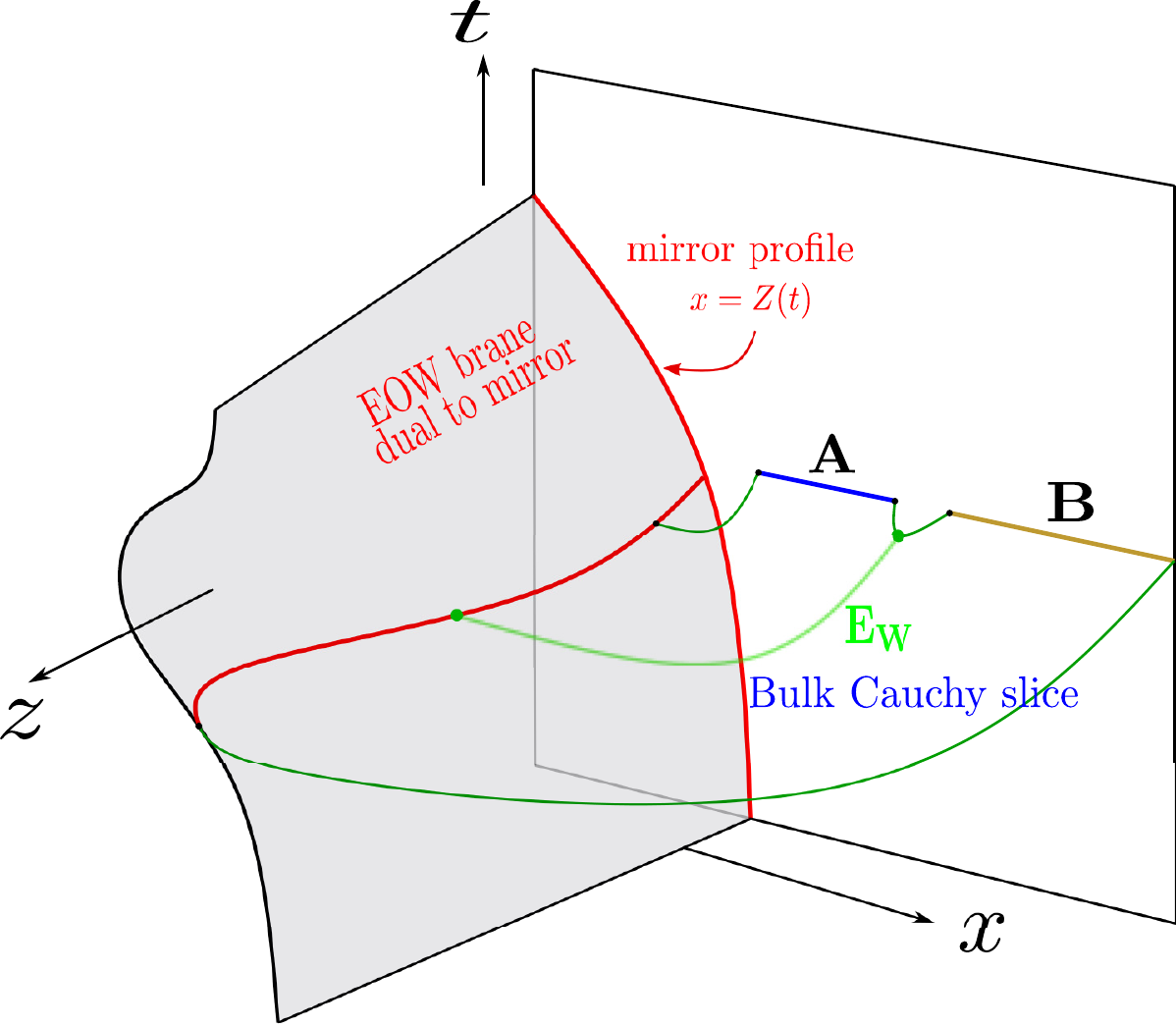}
		\caption{Schematics of the bulk entanglement wedge for the mixed state configuration of the two disjoint intervals in the dual moving mirror BCFT. The entanglement wedge is the region on the bulk Cauchy slice at coordinate time $t$ which is enclosed by the subsystems $A$ and $B$ and the RT surfaces of $A\cup B$. The EWCS for this configuration ends on the EOW brane which is holographically dual to the moving mirror boundary. Figure modified from \cite{Akal:2021foz}.}
		\label{fig:disj-EW-MM}
	\end{figure}
	
	\subsubsection{Two Disjoint Intervals}
\begin{figure}[h!]
		\centering
		\includegraphics[scale=0.8]{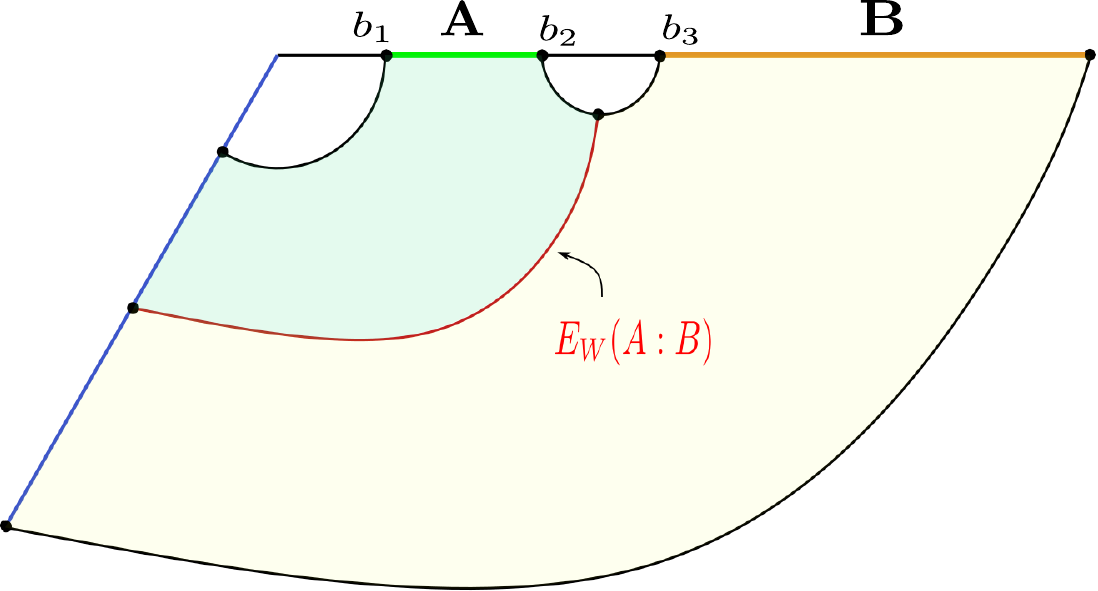}
		\caption{Schematics of the bulk entanglement wedge corresponding to the mixed state configuration of the two disjoint intervals in the dual moving mirror BCFT. Figure modified from \cite{Li:2021dmf}.}
		\label{fig:EW-disj-phase1-schematic}
	\end{figure}

	We begin with the configuration of two disjoint intervals $A=[(t,x_1),(t,x_2)]$ and $B=[(t,x_3),(t,\infty)]$.
	The schematics of the EWCS for $A$ and $B$ is depicted in \cref{fig:disj-EW-MM} where the EWCS ends on the EOW brane $Q$. In order to compute the bulk EWCS, we first utilize the Banados map in \cref{Banados-map} to transform the geometry to that of the standard $AdS_3$/$BCFT_2$ setup as shown in \cref{fig:EW-disj-phase1-schematic}. Note that, for simplicity we consider the intervals on an equal time slice. 
	
	As described earlier in subsection \ref{sec:Disjoint}, there are three possible phases of the bulk EWCS for the two disjoint intervals under consideration. In the following, we systematically investigate the  EWCS for these three phases.
	
	\subsubsection*{Phase-I}
\begin{figure}[ht]
		\centering
		\includegraphics[scale=1.15]{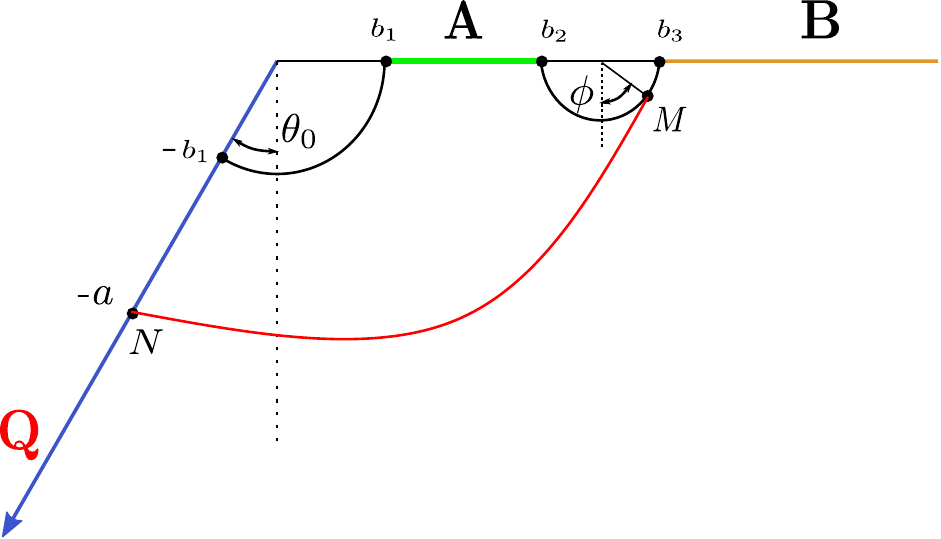}
		\caption{Calculation of the bulk entanglement wedge corresponding to the mixed state configuration of the two disjoint intervals $A=[b_1,b_2]$ and $B=[b_3,\infty]$ in the usual $AdS_3$/$BCFT_2$ setup. Figure modified from \cite{Li:2021dmf}.}
		\label{fig:EW-disj-phase1-calculation}
	\end{figure}

	When the interval $A$ is very close to the boundary, we have a connected entanglement wedge configuration. Furthermore, if $A$ is large enough, the EWCS ends on the EOW brane $Q$ as shown in \cref{fig:disj-EW-MM}. In the following, we compute the EWCS for the two disjoint intervals $A=[b_1,b_2]$ and $B=[b_3,\infty]$ in the dual $BCFT_2$ for which the bulk dual is described by Poincar\'e $AdS_3$ geometry in \cref{PAdS3} as depicted in \cref{fig:EW-disj-phase1-calculation}. 
	
	The EWCS is given by the length of a geodesic connecting the point $M$ on the RT surface and a point $N$ on the EOW brane making an angle $\theta_0$ with the vertical, as shown in \cref{fig:EW-disj-phase1-calculation}. The coordinates of the point $M$ is given by
	\begin{align*}
	M: (x,t,z)\equiv(R+r\sin\phi,0,r\cos\phi)\,,
	\end{align*}
	where $r=\frac{1}{2}(b_3-b_2)$ and $R=\frac{1}{2}(b_3+b_2)$. Similarly, the coordinates of the point $N$ on the brane are given by
	\begin{align*}
	N: (x,t,z)\equiv(-a\sin\theta_0,0,a\cos\theta_0)\,,
	\end{align*}
	where $a$ denotes the distance of the point $N$ on the EOW brane, from the origin. The length of the geodesic connecting $M$ and $N$ is hence obtained as \cite{Ryu:2006bv}
	\begin{align}
	L_{MN}=\cosh^{-1}\left[\frac{\left(\frac{b_2+b_3}{2}+\frac{b_3-b_2}{2}\sin\phi+a\sin\theta_0\right)^2+\left(\frac{b_3-b_2}{2}\cos\phi\right)^2+a^2\cos^2\theta_0}{2\left(\frac{b_3-b_2}{2}\cos\phi\right)a\cos\theta_0}\right]\label{L_PQ}
	\end{align}
	
	The EWCS is obtained by extremizing the above expression with respect to the positions of the endpoints $M$ and $N$. The extremization leads to the following values of the unknown parameters $a$ and $\phi$:
	\begin{align}
	a=\sqrt{b_2b_3}~~,~~\phi=\sin^{-1}\left(\frac{b_3-b_2}{b_3+b_2}\right)\,.
	\end{align}
	Substituting these into \cref{L_PQ}, the EWCS corresponding to the disjoint intervals may be obtained as
	\begin{align}
	L^{\text{min}}_{MN}&=\cosh^{-1}\left[\frac{\sec\theta_0\left(b_2+b_3+2\sqrt{b_2b_3}\sin\theta_0\right)}{b_3-b_2}\right]\notag\\
	&=\cosh^{-1}\left(\frac{1}{\cos\theta_0}\right)+\log\left(\frac{b_2+b_3+2\sqrt{b_2b_3}}{b_3-b_2}\right)\,.
	\end{align}
	The first term in the above expression is related to the boundary entropy as \cite{Takayanagi:2011zk,Fujita:2011fp}
	\begin{align}
	S_{\text{bdy}}=\frac{1}{4G_N}\tanh^{-1}(\sin\theta_0)\,.\label{S-bdy}
	\end{align}
	Hence the EWCS for the configuration of two disjoint intervals $A=[b_1,b_2]$ and $B=[b_3,\infty]$ is given by
	\begin{align}
	E_W(A:B)=\frac{1}{4G_N}L^{\text{min}}_{MN}=\frac{1}{4G_N}\log\left(\frac{b_2+b_3+2\sqrt{b_2b_3}}{b_3-b_2}\right)+S_{\text{bdy}}\,.
	\end{align}
	The above expression for the EWCS matches exactly with half the reflected entropy computed in \cref{disj-phase1-SR-expr} upon utilizing the standard Brown-Henneaux relation in $AdS_3$/CFT$_2$ \cite{Brown:1986nw} which serves as a consistency check. As in the field theoretic calculations, the above expression may be conveniently expressed in terms of the cross ratio involved. Therefore  utilizing the conformal symmetry of the dual $BCFT_2$, we may obtain the EWCS corresponding to the generic complex intervals $A=[\tilde{z}_1,\tilde{z}_2]$ and $B=[\tilde{z}_3,\infty]$ in the field theory as follows
	\begin{align}
	E_W(A:B)=\frac{1}{4G_N}\log\left(\frac{1+\sqrt{1-\tilde{x}}}{\sqrt{\tilde{x}}}\right)+S_{\text{bdy}}\,,
	\end{align}
	where the modified cross ratio $\tilde{x}$ is given in \cref{cross ratio1}. 
	
	Finally, we may utilize the Banados map given in \cref{Banados-map}, to transform back to the original plane wave geometry dual to the moving mirror BCFT. From \cref{Banados-map,cross ratio1} the cross ratio corresponding to the setup of two disjoint intervals in question may be obtained as
	\begin{align}
	\zeta&=\frac{(U_3-V_2)(V_3-U_2)}{(U_3-U_2)(V_3-V_2)}\notag\\
	&=\frac{\left(p(t-x_3)-t-x_3\right)\left(t+x_3-p(t-x_2)\right)}{\left(x_3-x_2\right)\left(p(t-x_3)-p(t-x_2)\right)}\,,
	\end{align}
	and hence, the EWCS is given by
	\begin{align}
	E_W(A:B)=\frac{1}{4G_N}\log\left(\frac{1+\sqrt{1-\zeta}}{2\sqrt{\zeta}}\right)+S_{\text{bdy}}\,.
	\end{align}
	Once again, the holographic reflected entropy matches exactly with the field theoretic replica technique calculations in the large central charge limit. This serves as a strong consistency check for our holographic construction for the bulk entanglement wedge. 
	\subsubsection*{Phase-II}
\begin{figure}[h!]
		\centering
		\includegraphics[scale=0.8]{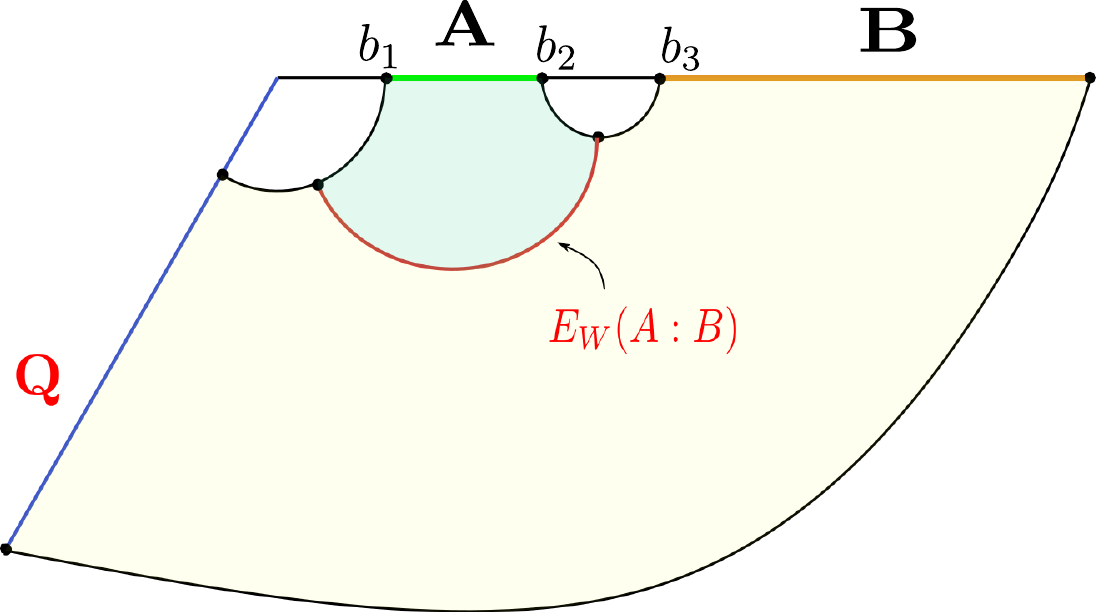}
		\caption{Schematics of the entanglement wedge cross section corresponding to phase-II conformally mapped the static mirror setup. Figure modified from \cite{Li:2021dmf}.}
		\label{fig:EW-disj-phase2}
	\end{figure}

	Now we consider the case where the first interval $A$ is very small. In this phase the bulk entanglement wedge is connected, but due to the small size of the subsystem $A$, the EWCS ends on the RT surface of $A$. The configuration is similar to the one shown in \cref{fig:disj-EW-MM}.
	Utilizing the Banados map in \cref{Banados-map}, we may again transform the bulk geometry to that of the standard Poincar\'e $AdS_3$ with the metric given in \cref{PAdS3}. The schematics of the static geometry corresponding to the intervals $A=[b_1,b_2]$ and $B=[b_3,\infty]$ on an equal time slice is depicted in \cref{fig:EW-disj-phase2}.
	
	In this phase the EWCS resides entirely in the bulk Poincar\'e $AdS_3$ geometry and hence does not involve any contribution from the EOW brane $Q$. Therefore, using standard $AdS_3$/CFT$_2$ results  we obtain the EWCS as follows \cite{Takayanagi:2017knl,Nguyen:2017yqw}
	\begin{align}
	E_W=\frac{1}{4G_N}\log\left(1+\xi+2\sqrt{\xi(\xi+1)}\right)\,,
	\end{align}
	where the cross ratio $\xi$ is given by
	\begin{align}
	\xi=\frac{2b_1(b_3-b_2)}{(b_2-b_1))(b_1+b_3)}\,.
	\end{align}
	Note that, the above expression for the EWCS matches exactly with the reflected entropy for this phase computed through the field theoretic replica technique. Now for generic complex intervals $A=[\tilde{z}_1,\tilde{z}_2]$ and $B=[\tilde{z}_3,\infty]$, the EWCS may be obtained utilizing the conformal symmetry of the dual $BCFT$ with the modification of the cross ratio $\tilde{\xi}$ as
	\begin{align}
	\tilde{\xi}=\frac{(\tilde{z}_1-\tilde{z}_1^*)(\tilde{z}_3-\tilde{z}_2)}{(\tilde{z}_2-\tilde{z}_1)(\tilde{z}_3-\tilde{z}_1^*)}\equiv \frac{(\tilde{u}_1-\tilde{v}_1)(\tilde{u}_3-\tilde{u}_2)}{(\tilde{u}_2-\tilde{u}_1)(\tilde{u}_3-\tilde{v}_1)}\,.
	\end{align}
	Under the Bandos map given in \cref{Banados-map}, the EWCS for the disjoint intervals $A$ and $B$ in phase-II for the moving mirror BCFT is therefore given by
	\begin{align}
	E_W(A:B)=\frac{1}{4G_N}\log\left(1+\xi_2+2\sqrt{\xi_2(\xi_2+1)}\right)\,,
	\end{align}
	with
	\begin{align}
	\xi_2&=\frac{(U_1-V_1)(U_3-U_2)}{(U_2-U_1)(U_3-V_1)}\notag\\
	&=\frac{\left(p(t-x_1)-t-x_1\right)\left(p(t-x_3)-p(t-x_2)\right)}{\left(p(t-x_2)-p(t-x_1)\right)\left(p(t-x_3)-t-x_1\right)}\,,
	\end{align}
	Once again this exactly matches with half of the reflected entropy given in \cref{disj1-SR-final2} which serves as yet another consistency check of the holographic construction.

	\subsubsection*{Phase-III}
\begin{figure}[h!]
		\centering
		\includegraphics[scale=0.8]{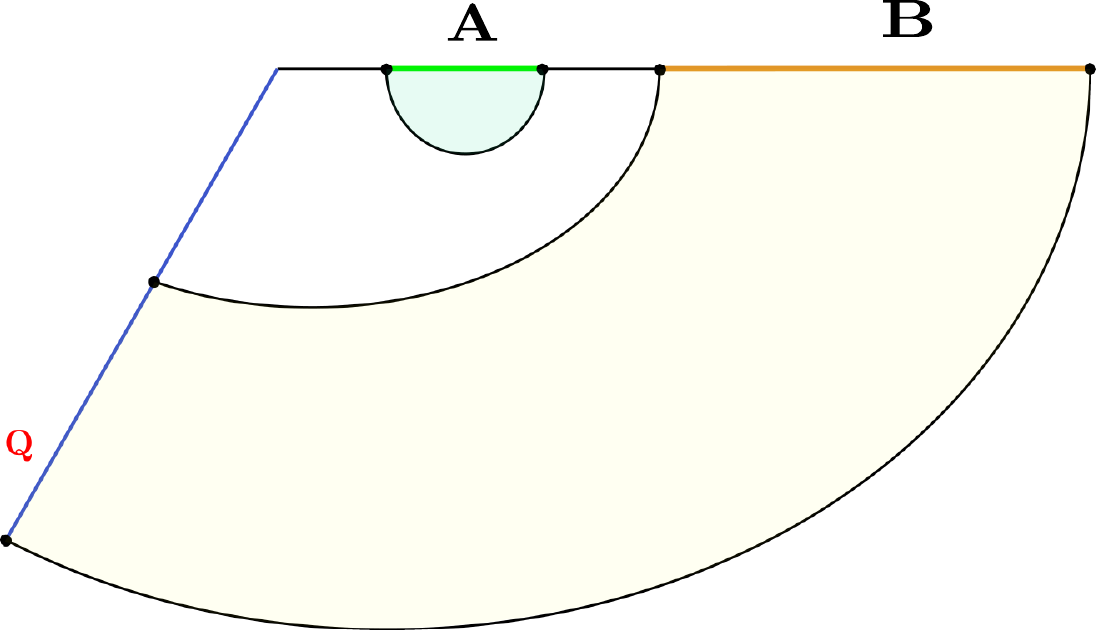}
		\caption{Schematics of the entanglement wedge cross section corresponding to phase-III conformally mapped the static mirror setup. Figure modified from \cite{Li:2021dmf}.}
		\label{fig:EW-disj-phase3}
	\end{figure}

	There exists another possible phase for the reflected entropy of the two disjoint intervals $A$ and $B$. This trivial phase corresponds to the disconnected entanglement wedge in the dual bulk geometry as shown in \cref{fig:EW-disj-phase3}. Consequently, the cross section vanishes, $E_W=0$, for this configuration.
	
	\subsubsection{Two Adjacent Intervals}
	We now proceed to compute the bulk EWCS for the mixed state configuration of two adjacent intervals $A=[(t,0),(t,x_1)]$ and $B=[(t,x_1),(t,x_2)]$ described in subsection \ref{sec:Adjacent}.
	
	\subsubsection*{Phase-I}
\begin{figure}[h!]
		\centering
		\includegraphics[scale=1.2]{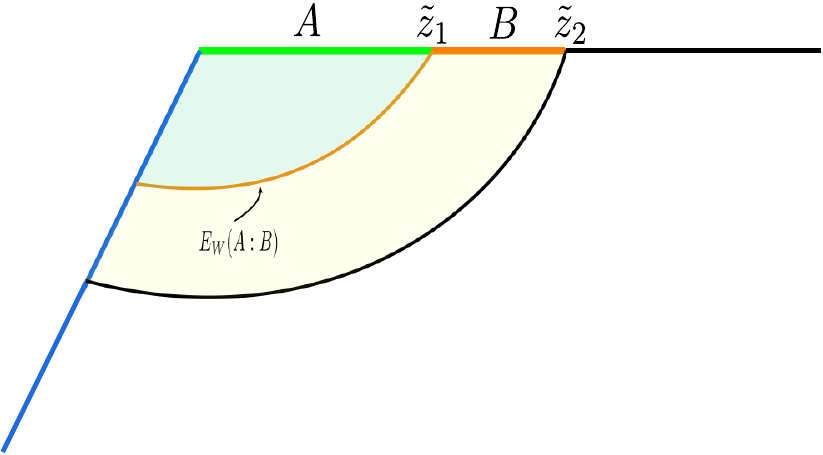}
		\caption{Schematics of the entanglement wedge cross section for the two adjacent intervals in phase-I conformally mapped the static mirror setup.}
		\label{fig:Adj-phase1}
	\end{figure}

	As shown in \cref{fig:Adj-phase1}, the minimal cross section of the entanglement wedge between $A$ and $B$ is given by the RT surface starting from the point $\tilde{z}_1$ and ending on the EOW brane .Hence, the EWCS is obtained through the length of this RT surface described in \cite{Akal:2020twv,Akal:2021foz} as follows
	\begin{align}
	E_W(A:B)=\frac{1}{4G_N}\log\left[\frac{t+x_1-p(t-x_1)}{\epsilon\sqrt{p'(t-x_1)}}\right]+S_{\text{bdy}}\,.
	\end{align}
	Note that the holographic reflected entropy obtained from the above expression matches exactly with the field theoretic replica technique computations in \cref{Adj-SR-phase1}.
	
	\subsubsection*{Phase-II}
\begin{figure}[h!]
		\centering
		\includegraphics[scale=1.2]{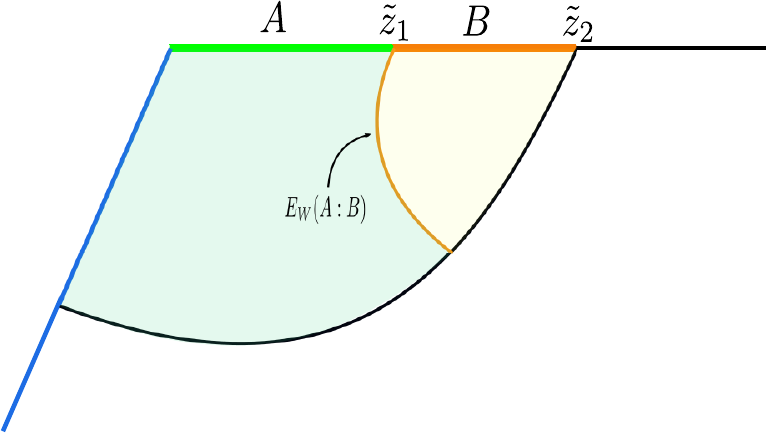}
		\caption{Schematics of the entanglement wedge cross section for the two adjacent intervals in phase-II conformally mapped the static mirror setup.}
		\label{fig:Adj-phase2}
	\end{figure}

	In this phase, the EWCS ends on the RT surface corresponding to $A\cup B$ as depicted in \cref{fig:Adj-phase2}. As the EWCS resides entirely inside the bulk Poincar\'e $AdS_3$ geometry away from the EOW brane, one may apply the EWCS computed in the context of $AdS_3$/CFT$_2$ \cite{Takayanagi:2017knl,Nguyen:2017yqw} to obtain
	\begin{align}
	E_W(A:B)=\frac{1}{4G_N}\log\left[\frac{(\tilde{u}_1-\tilde{u}_2)(\tilde{u}_1-\tilde{v}_2)}{\epsilon\left(\tilde{u}_1-\tilde{v}_1\right)}\right]\,.
	\end{align}
	Finally, utilizing the Banados map in \cref{Banados-map}, we obtain the EWCS for the two adjacent intervals $A$ and $B$ in the moving mirror BCFT as
	\begin{align}
	E_W(A:B)=\frac{1}{4G_N}\log\left[\frac{2\left(p(t-x_1)-p(t-x_2)\right)\left(p(t-x_1)-t-x_2\right)}{\epsilon\sqrt{p'(t-x_1)}\left(p(t-x_1)-t-x_1\right)}\right]
	\end{align}
	Once again, the above expression matches exactly with the field theoretic computation in \cref{SR-adj1-final} upon using the Brown-Henneaux formula \cite{Brown:1986nw}.

	\subsection{Page curves for reflected entropy}	
	In this subsection, we provide the analogues of the Page curves for the reflected entropy for various bipartite states involving two disjoint and two adjacent intervals in a holographic moving mirror $BCFT_2$, computed in the previous subsections.

	\subsubsection{Escaping mirror}
	As described earlier, the radiation emitted by the escaping mirror mimics that of an eternal black hole. We may now obtain the holographic reflected entropy or the bulk EWCS for the mixed state configurations involving two disjoint and two adjacent intervals by substituting the mirror profile given in \cref{escprof} into the expressions for the reflected entropies for the various phases obtained in the earlier subsections. The behavior of the reflected entropy with time for various scenarios are plotted in \cref{fig:SR-esc-disj-Page,fig:Adj-esc}.
	\subsubsection*{Disjoint intervals}
\begin{figure}[H]
		\centering
		\begin{subfigure}[b]{0.49\textwidth}
			\centering
			\includegraphics[width=\textwidth]{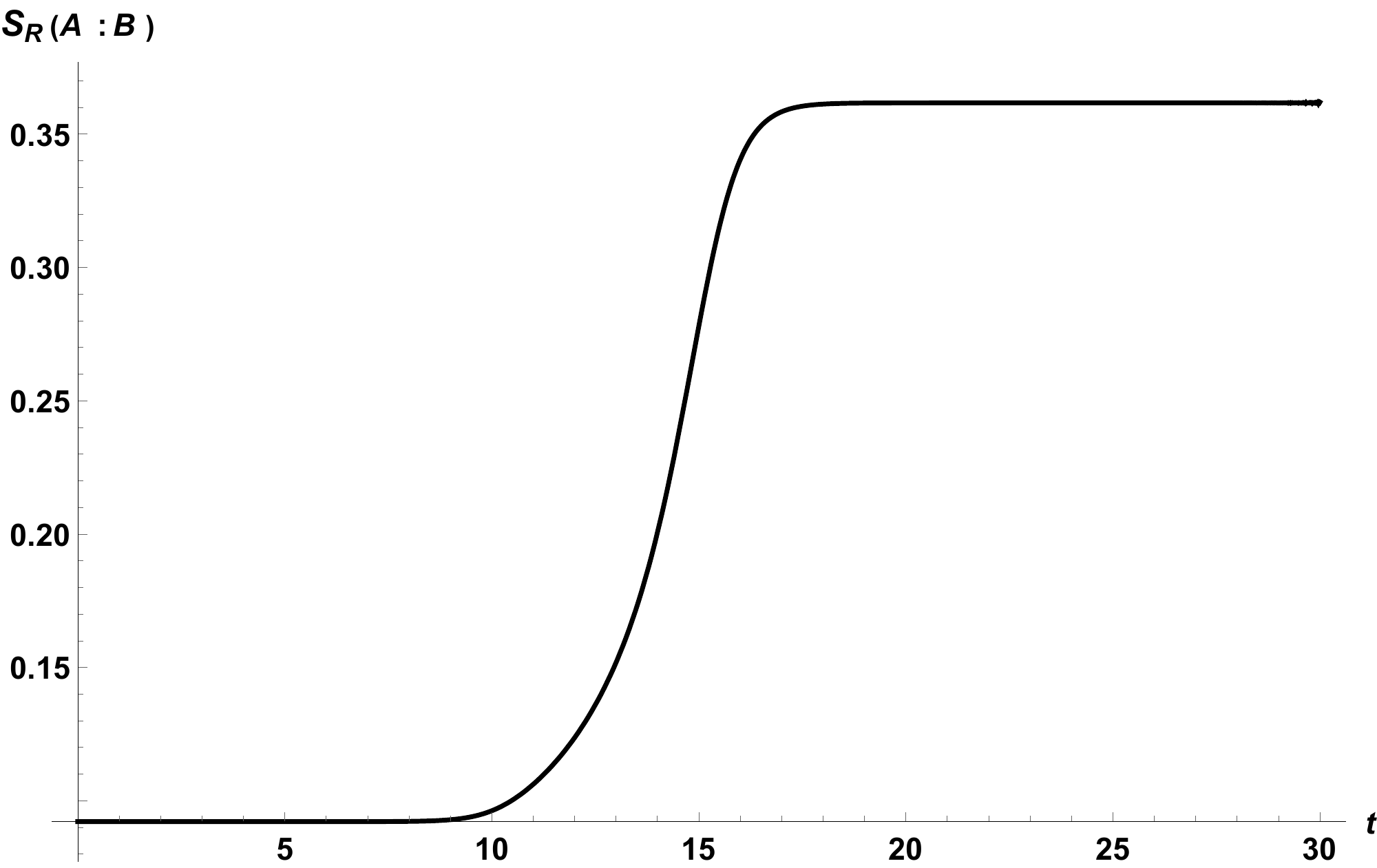}
			\caption{Fixed Intervals: \\$A=[10,15]$ and $B=[15.5,\infty]$;\\ $\beta=0.5$ , $S_{\text{bdy}}=3$ , $c=1$}
			\label{SR-esc-disj1-1}
		\end{subfigure}
		\hfill
		\begin{subfigure}[b]{0.49\textwidth}
			\centering
			\includegraphics[width=\textwidth]{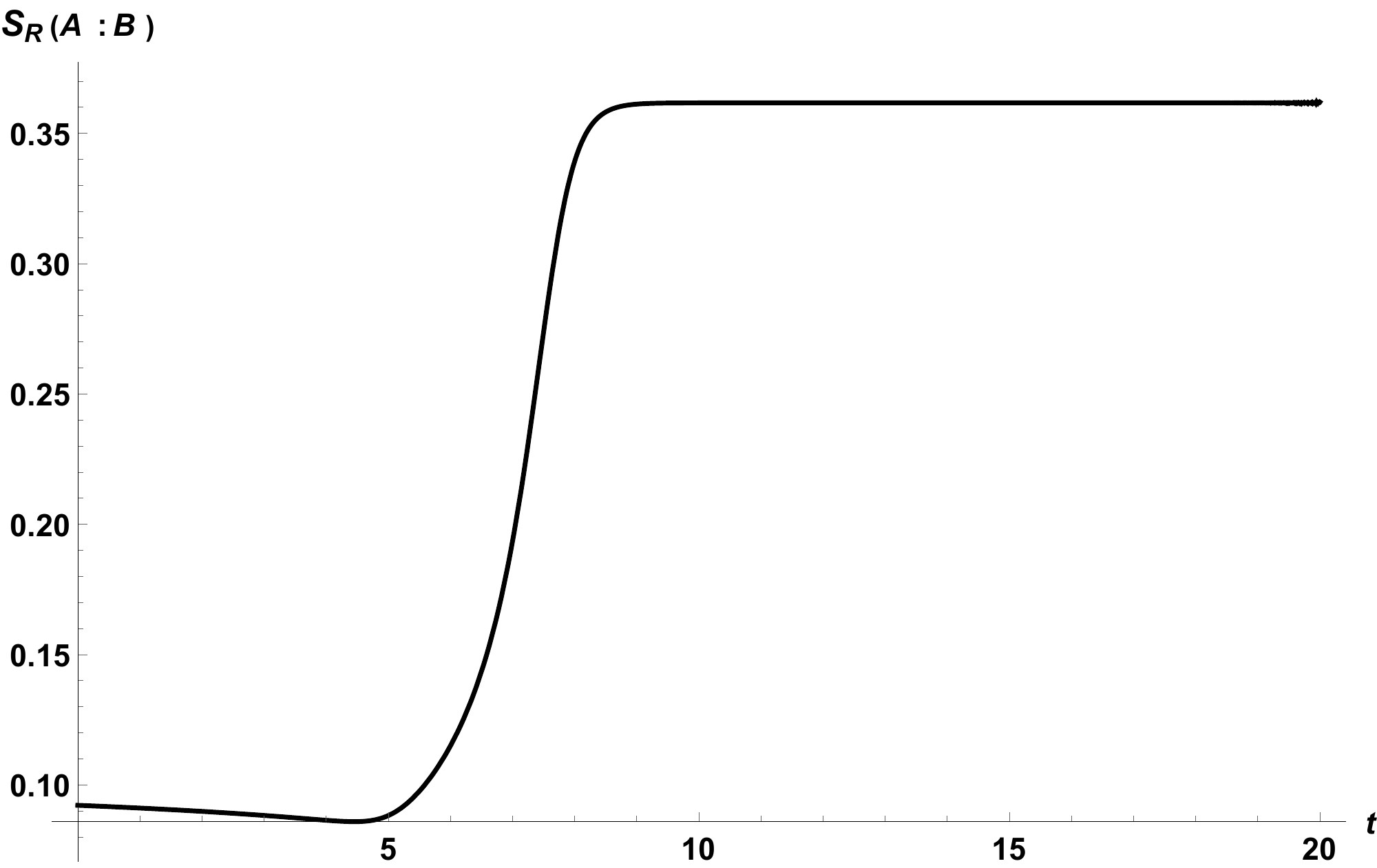}
			\caption{ Co-moving intervals: $A=[Z(t)+10,Z(t)+15]$ and $B=[Z(t)+15.5,\infty]$; \\$\beta=0.5$ , $S_{\text{bdy}}=3$ , $c=1$}
			\label{SR-esc-disj1-2}
		\end{subfigure}
		\caption{Page curves for the reflected entropy between two disjoint intervals in a escaping mirror $BCFT_2$.}
		\label{fig:SR-esc-disj-Page}
	\end{figure}

	We begin with the configuration of two disjoint intervals $A=[(t,x_1),(t,x_2)]$ and $B=[(t,x_3),(t,\infty)]$ as described in  subsection \ref{sec:Disjoint}. The case of two intervals $A=[10,15]$ and $B=[15.5,\infty]$ fixed with respect to the mirror is depicted in \cref{SR-esc-disj1-1}, while in \cref{SR-esc-disj1-2} we sketch the Page curve for two co-moving intervals $A=[Z(t)+10,Z(t)+15]$ and $B=[Z(t)+15.5,\infty]$, with the mirror profile given as $x=Z(t)$.
	Note that the reflected entropy for both the case with fixed and co-moving intervals follow very similar behavior, although the phase transitions occur at different times.

	\subsubsection*{Adjacent intervals}
\begin{figure}[h!]
		\centering
		\includegraphics[scale=0.4]{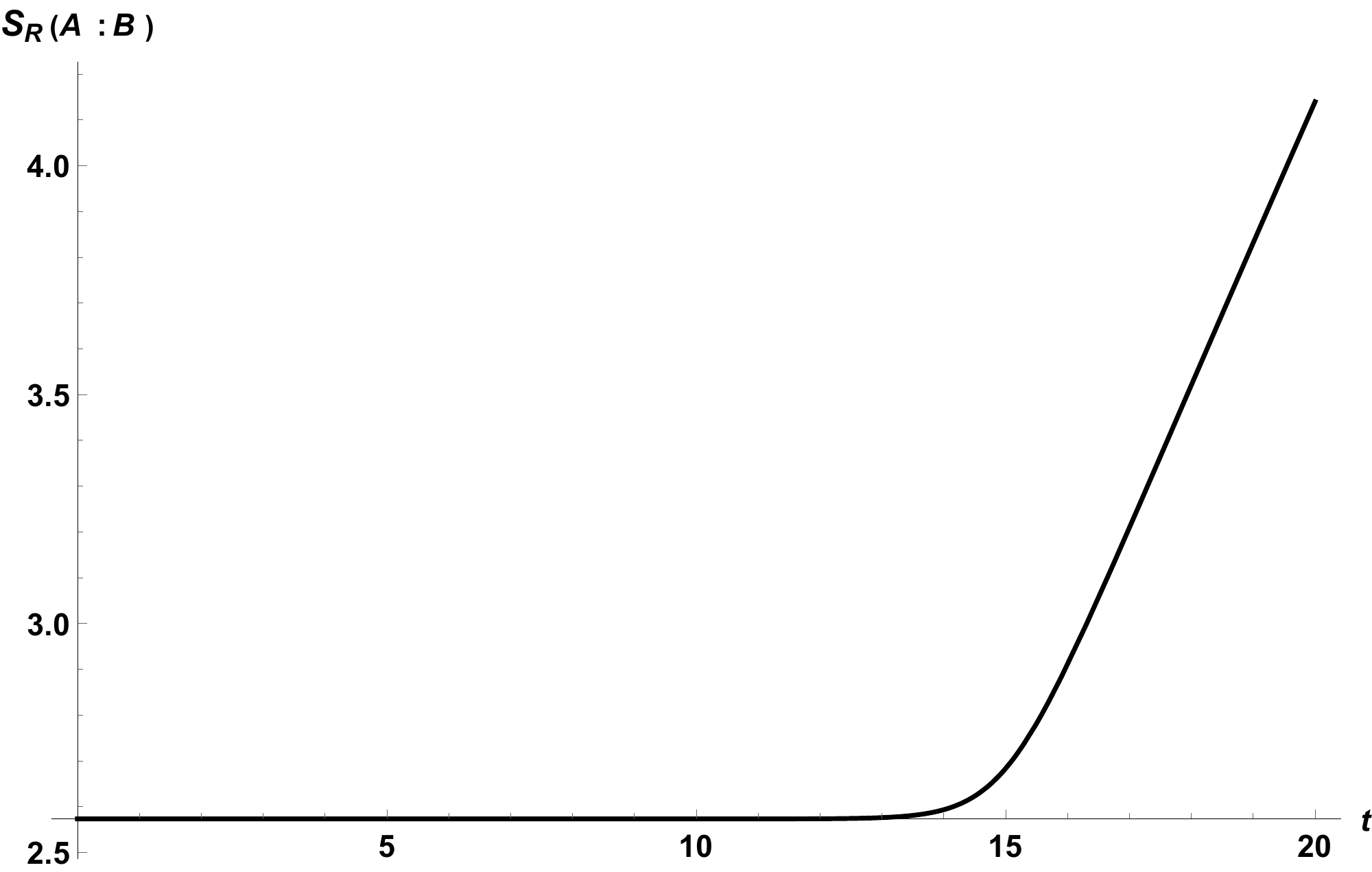}
		\caption{Page curve for the reflected entropy between two fixed adjacent intervals $A=[0,15]$ and $B=[15,30]$ in a escaping mirror $BCFT_2$ ; $\beta=0.5$ , $S_{\text{bdy}}=0.2$ , $c=1$.}
		\label{fig:Adj-esc}
	\end{figure}

	Next we consider the case of two adjacent intervals $A=[(t,0),(t,x_1)]$ and $B=[(t,x_1),(t,x_2)]$, described in \cref{sec:Adjacent}, in the escaping mirror $BCFT_2$ with the mirror profile given in \cref{escprof}. The analogue of the Page curve for the reflected entropy for this configuration is depicted in \cref{fig:Adj-esc}.
	
	\subsubsection{Kink mirror}
	The kink mirror with the profile \cref{kinkprof} provides a toy model for an evaporating black hole. Substituting \cref{kinkprof} into the various expressions for the holographic reflected entropies obtained in the earlier subsections, we may obtain the corresponding Page like curves for the evaporating black hole scenario as shown in \cref{fig:SR-kink-disj_Page,fig:Adj-kink}.
	
	\subsubsection*{Disjoint intervals}
\begin{figure}[H]
		\centering
		\begin{subfigure}[b]{0.49\textwidth}
			\centering
			\includegraphics[width=\textwidth]{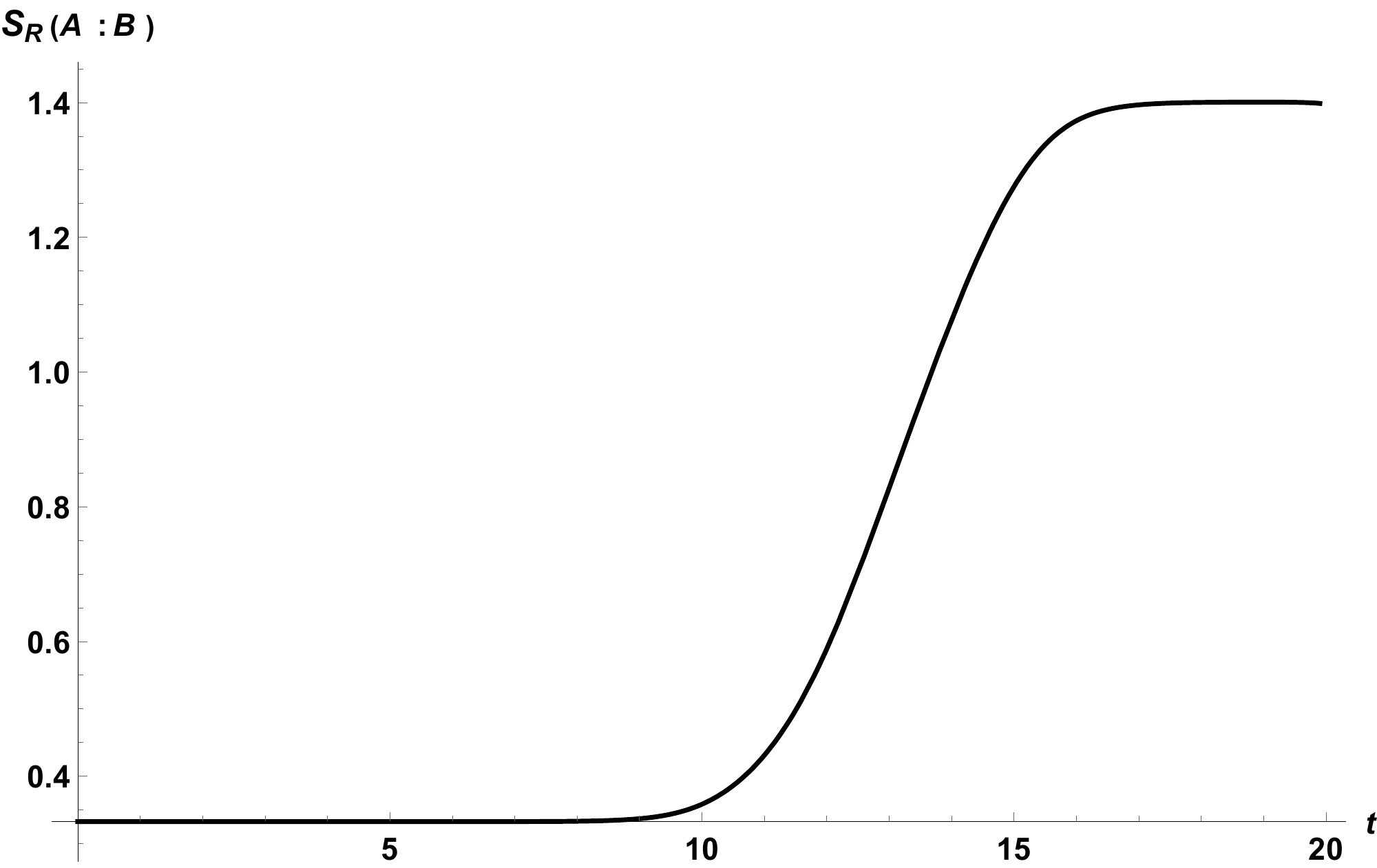}
			\caption{Fixed Intervals: \\$A=[10,12]$ and $B=[15.5,\infty]$;\\ $\beta=0.5$ , $u_0=20$, $S_{\text{bdy}}=4$ , $c=1$.}
			\label{SR-esc-disj2-1}
		\end{subfigure}
		\hfill
		\begin{subfigure}[b]{0.49\textwidth}
			\centering
			\includegraphics[width=\textwidth]{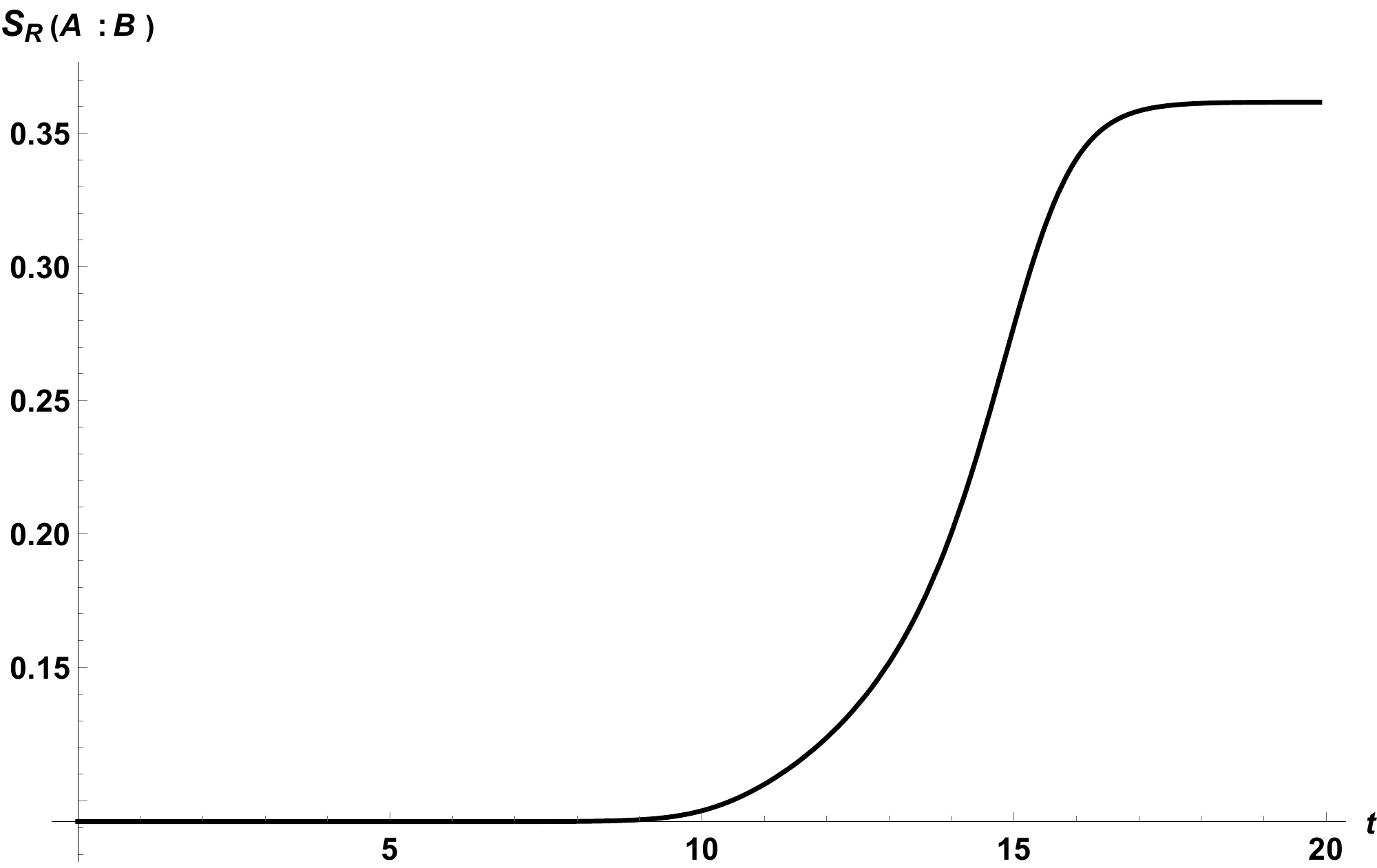}
			\caption{ Co-moving intervals: $A=[Z(t)+10,Z(t)+15]$ and $B=[Z(t)+15.5,\infty]$; \\
				$\beta=0.5$ , $u_0=20$, $S_{\text{bdy}}=2$ , $c=1$.}
			\label{SR-esc-disj2-2}
		\end{subfigure}
		\caption{Page curves for the reflected entropy between two disjoint intervals in a kink mirror BCFT. }
		\label{fig:SR-kink-disj_Page}
	\end{figure}

	For the case of two disjoint intervals $A=[(t,x_1),(t,x_2)]$ and $B=[(t,x_3),(t,\infty)]$ in the kink mirror $BCFT_2$, the analogue of the Page curves for the reflected entropy are depicted in \cref{fig:SR-kink-disj_Page}. Once again the behavior of the reflected entropy for both fixed and co-moving intervals follow very similar profiles.

	\subsubsection*{Adjacent intervals}
\begin{figure}[H]
		\centering
		\includegraphics[scale=0.45]{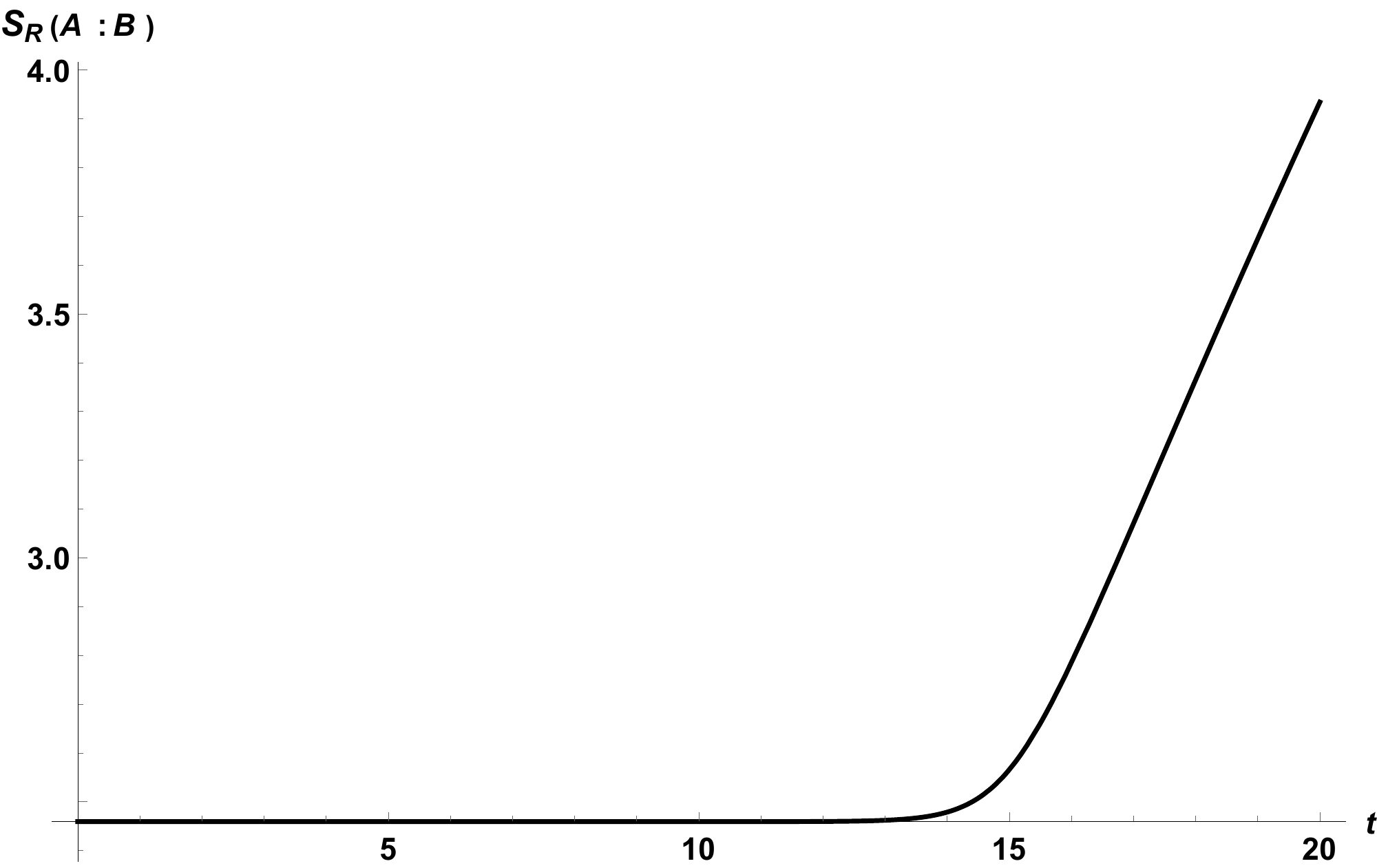}
		\caption{Page curve for the reflected entropy between two fixed adjacent intervals $A=[0,15]$ and $B=[15,25]$ in a escaping mirror $BCFT_2$ ; $\beta=0.5$ , $u_0=20$ , $S_{\text{bdy}}=0.4$ , $c=1$.}
		\label{fig:Adj-kink}
	\end{figure}

	Finally, we consider the case of two adjacent intervals $A=[(t,0),(t,x_1)]$ and $B=[(t,x_1),(t,x_2)]$ in a kink mirror BCFT, for which the analogue of the Page curve for the reflected entropy is depicted in \cref{fig:Adj-kink}.
	
	\subsubsection{Discussion }
	Remarkably, the analogues of the Page curves obtained above follow very closely the expected curves\footnote{Note that similar curves were also obtained in \cite{Akers:2021pvd} in the context of random tensor networks mimicking a holographic duality.} obtained in \cite{Akers:2022max} for another model of black hole evaporation involving JT gravity coupled to end-of-the-world branes \cite{Penington:2019kki}. As described in \cite{Akers:2022max}, a phase transition in the reflected entropy curve occurs due to a transition between the disconnected and connected sets of replica geometries. In the disconnected phase of the replica geometry, the radiation subsystems purify themselves, while in the connected phase the canonical purification occurs through a closed universe. The dominance of the connected replica geometries may be interpreted as the appearance of an island cross-section in the effective lower dimensional picture. 
	
	In a similar manner, the phase transition in the analogues of the Page curves obtained above may be interpreted in terms of the transitions between the different phases of the bulk EWCS. For example, for the case of two disjoint intervals in phase-I, the bulk EWCS ends on the EOW brane as shown in \cref{fig:EW-disj-phase1-schematic} which corresponds to the presence of a non-trivial island cross-section in the effective $2d$ setup. On the other hand, in phase-II sketched in \cref{fig:EW-disj-phase2}, the EWCS ends on the RT surface of $A$ which is reminiscent of the disconnected geometry in \cite{Akers:2022max}. It will be interesting to explore such interpretations in terms of the appearance of islands in the effective semi-classical picture, in the light of the CFT modes radiated and reflected through the moving mirror boundary, similar to the discussion in \cite{Akal:2021foz} for the case of entanglement entropy. We leave this exciting issue for future explorations.
	
	\section{Entanglement Negativity for Moving Mirrors}\label{secENMM}
	Having described the analogues of Page curves for reflected entropy, we now turn our attention to  entanglement negativity which is another significant mixed state entanglement measure  in quantum information theory.
	In this section we determine this quantity for various mixed state configurations in the radiation flux of a moving mirror. 
	We  begin with the computation of the entanglement negativity for mixed states involving two  adjacent and disjoint intervals by utilizing the replica technique  in  the large central charge limit of a $BCFT_{1+1}$. Following this, we apply the results obtained to compute the corresponding entanglement negativity in for the moving mirror configurations. As described earlier, here we first map the moving mirror scenario to the static mirror configuration where we compute the entanglement negativity and finally transform back to the moving mirror coordinates.
	Subsequently, we will utilize the holographic proposals described in \cite{Jain:2017aqk,Malvimat:2018txq} to compute the entanglement negativity of the above mixed state configurations in the corresponding dual bulk $AdS_{3}$ geometry utilizing the Banados map and demonstrate that the results evaluated exactly agree  with those obtained from the replica technique  in a $BCFT$.

	\subsection{Entanglement Negativity in a $BCFT_{1+1}$ }
	We  now describe our computation of the entanglement negativity for adjacent and disjoint intervals embedded in the radiation flux of the moving mirror by utilizing the replica technique in a $BCFT_{1+1}$ with large central charge.

	\subsubsection{Adjacent Intervals}
	\begin{figure}[H]
		\centering
		\includegraphics[scale=0.7]{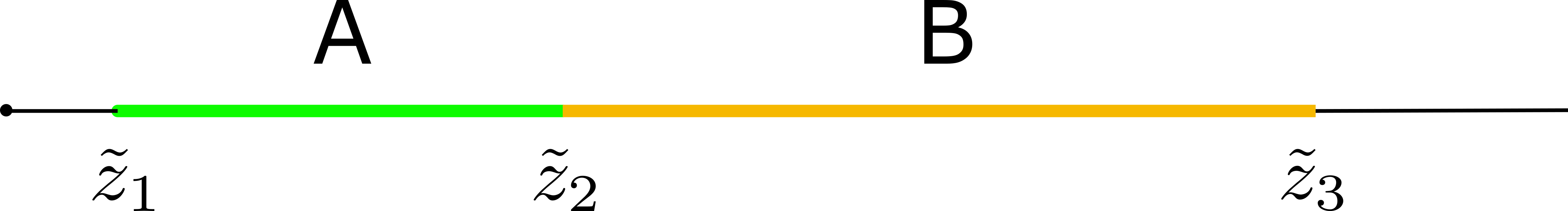}
		\caption{Configuration involving two adjacent intervals away from the boundary in a $BCFT_{1+1}$}
	\end{figure}
	
	To begin with we compute the entanglement negativity for two adjacent intervals $A=[\tilde{z}_1,\tilde{z}_2]$ and $B=[\tilde{z}_2,\tilde{z}_3]$ in a $BCFT_{1+1}$. The entanglement negativity of the configuration in question is described by the following three point twist field correlator
	\begin{align}\label{ENadj2}
	{\cal E}=\lim _{n_{e} \rightarrow 1} \log\left\langle\tau_{n_{e}}\left(\tilde{z}_{1}\right) \bar{\tau}_{n_{e}}^{2}\left(\tilde{z}_{2}\right) \tau_{n_{e}}\left(\tilde{z}_{3}\right)\right\rangle_{B C F T^{\otimes_{n_{e}}}}
	\end{align}
	The above three point twist correlator is given by a six point function in the chiral $CFT$ which can be obtained through the Cardy's doubling trick \cite{Cardy:2004hm}
	\begin{align}
	\left\langle\tau_{n_{e}}\left(\tilde{z}_{1}\right) \bar{\tau}_{n_{e}}^{2}\left(\tilde{z}_{2}\right) \tau_{n_{e}}\left(\tilde{z}_{3}\right)\right\rangle_{B C F T^{\otimes_{n_{e}}}}=\left\langle\bar{\tau}_{n_{e}}\left(\tilde{z}_{1}^{*}\right) \tau_{n_{e}}^{2}\left(\tilde{z}_{2}^{*}\right) \bar{\tau}_{n_{e}}\left(\tilde{z}_{3}^{*}\right) \tau_{n_{e}}\left(\tilde{z}_{1}\right) \overline{\tau}_{n_{e}}^{2}\left(\tilde{z}_{2}\right) \tau_{n_{e}}\left(\tilde{z}_{3}\right)\right\rangle_{C F T^{\otimes_{n_{e}}}}
	\end{align}
	Generically the above three point correlator (six point function in the chiral CFT) depends on the full operator content of the theory and is difficult to determine.  Furthermore, in the large central charge limit the above three point correlator is then expected to factorize  in two different ways depending on the position of $A$ and the size of $B$  as we explain below.

	\section*{ $A$ close to the boundary}
	
	When we take one of the end points of subsystem $A$ close to the boundary then the required three point function factorizes as follows in the large-$c$ limit
	\begin{align}\label{3ptF1}
	\left\langle\tau_{n_{e}}\left(\tilde{z}_{1}\right) \bar{\tau}_{n_{e}}^{2}\left(\tilde{z}_{2}\right) \tau_{n_{e}}\left(\tilde{z}_{3}\right)\right\rangle_{B C F T^{\otimes_{n_{e}}}}\simeq\left\langle\tau_{n_{e}}\left(\tilde{z}_{1}\right)\right\rangle_{B C F T^{\otimes_{n_{e}}}}\left\langle\bar{\tau}_{n_{e}}^{2}\left(\tilde{z}_{2}\right) \tau_{n_{e}}\left(\tilde{z}_{3}\right)\right\rangle_{B C F T^{\otimes_{n_{e}}}}
	\end{align}
	The one point function above ( two point twist correlator in the Chiral CFT ) is  completely fixed by the conformal symmetry to be as follows
	\begin{align}
	\left\langle\tau_{n_{e}}\left(\tilde{z}_{1}\right)\right\rangle_{B C F T^{\otimes_{n_{e}}}}&= \left\langle\tau_{n_{e}}\left(\tilde{z}_{1}\right)\overline{\tau}_{n_{e}}\left(\tilde{z}_{1}^*\right)\right\rangle_{ C F T}
	\end{align}
	The two point twist correlator in \cref{3ptF1} lead to two different phases for entanglement negativity resulting from the bulk and the boundary channels which we describe below.
	\subsubsection*{Phase-I:  Boundary channel ($A$ close to the boundary and  $B$ large)}
	We now consider the boundary channel which corresponds to subsystem $B$ being large and the two point function described in \cref{3ptF1} factorizes into the product of two one point functions. This is expressed as follows\footnote{Note that the OPE coefficient for the one point function of $\tau_{n_{e}}^{2}$ in \cref{P1BC} can be obtained   through an analysis similar to that described in \cite{Sully:2020pza}.}
	\begin{align}\label{P1BC}
	\left\langle\bar{\tau}_{n_{e}}^{2}\left(\tilde{z}_{2}\right) \tau_{n_{e}}\left(\tilde{z}_{3}\right)\right\rangle_{B C F T^{\otimes_{n_{e}}}}&=\left\langle\bar{\tau}_{n_{e}}^{2}\left(\tilde{z}_{2}\right)\right\rangle_{B C F T^{\otimes_{n_{e}}}} \left\langle\tau_{n_{e}}\left(\tilde{z}_{3}\right)\right\rangle_{B C F T^{\otimes_{n_{e}}}}\nonumber\\
	&= \frac{e^{2\left(1-n_{e} / 2\right)S_{\text{bdy}}}}{\left(\frac{2 \text{Im} \tilde{z}_{2}}{\epsilon_{2}}\right)^{2 \Delta_{n_{e}}^{(2)}}} \cdot \frac{e^{\left(1-n_{e}\right)S_{\text{bdy}}}}{\left(\frac{2 \text{Im} \tilde{z}_{3}}{\epsilon_{3}}\right)^{2 \Delta_{n}} } 
	\end{align}
	This leads to the following expression for the entanglement negativity
	\begin{align}\label{negAcloseBlarge}
	{\cal E}=\frac{c}{4} \log \left(\frac{2 \text{Im}(\tilde{z}_{2})}{\epsilon_{2}}\right)+S_{\text{bdy}},
	\end{align}
	where $\epsilon_{i}$ is the cut off \footnote{Note that in this case, we allow for position-dependent cut-offs in the $BCFT$ to incorporate the effects of the non-trivial conformal map to the static mirror configuration.} at the point $\tilde{z}_i$.
	Having obtained the result for the entanglement negativity in a $BCFT_{1+1}$ we may now  reverse the static mirror map in \cref{static-mirror-map} as we did in the computation of reflected entropy in section \ref{sec:Disjoint}. Utilizing the reverse static mirror map, the entanglement negativity may be expressed as,
	\begin{align}\label{negadjmm1}
	{\cal E}=\frac{c}{4}\log\left[\frac{t+x_2-p(t-x_2)}{\epsilon\sqrt{p'(t-x_2)}}\right]+S_{\text{bdy}}\,.
	\end{align}
	\subsubsection*{Phase-II: Bulk channel ($A$ close to the boundary and $B$ small)}
	In the bulk channel which corresponds to the interval $B$ being small, the two point function is given by the following expression
	\begin{align}
	\left\langle\bar{\tau}_{n_{e}}^{2}\left(\tilde{z}_{2}\right) \tau_{n_{e}}\left(\tilde{z}_{3}\right)\right\rangle_{B C F T^{\otimes_{n_{e}}}}&=
	\langle\bar{\tau}_{n_{e}}^{2}(\tilde{z}_{2}) \tau_{n_{e}}(\tilde{z}_{3})\tau_{n_{e}}^{2}(\tilde{z}_{2}^*) \bar{\tau}_{n_{e}}(\tilde{z}_{3}^*)\rangle_{C F T^{\otimes_{n_{e}}}}\\
	&=\left(\frac{ \operatorname{Im} (\tilde{z}_{2})\left(\tilde{z}_{3}-\tilde{z}_{2}\right)^{2}}{\epsilon_{2}^{2}\, \operatorname{Im} (\tilde{z}_{3})}\right)^{\Delta_{n_{e}}^{(2)}}
	\end{align}
	Note that in order to obtain the last expression we have utilized the large central charge limit of the above four point function which was determined in \cite{Malvimat:2017yaj}.
	Using the above expression in \cref{3ptF1} and substituting the result obtained in  \cref{ENadj2}, the entanglement negativity may then be computed as follows
	\begin{align}\label{negAcloseBsmall}
	{\cal E}=\frac{c}{8} \log \left[\frac{ \operatorname{Im} (\tilde{z}_{2})\left(\tilde{z}_{3}-\tilde{z}_{2}\right)^{2}}{\epsilon_{2}^{2} \operatorname{Im} (\tilde{z}_{3})}\right]
	\end{align}
	As earlier we once again inverse the static mirror map in \cref{static-mirror-map} to obtain the following expression for  the entanglement negativity
	\begin{align}\label{negadjmm2}
	{\cal E}=\frac{c}{8} \log\left[\frac{(t+x_2-p(t-x_2))(p(t-x_3)-p(t-x_2))^2}{\epsilon^2 (t+x_3-p(t-x_3)) \sqrt{p'(t-x_2)p'(t-x_3)}}\right].
	\end{align}
	
	\smallskip
	
	\section*{ $A$ away from the boundary }
	
	For the case involving the subsystems away from the boundary the three point function factorizes as follows
	\begin{align}
	\left\langle\tau_{n_{e}}\left(\tilde{z}_{1}\right) \bar{\tau}_{n_{e}}^{2}\left(\tilde{z}_{2}\right) \tau_{n_{e}}\left(\tilde{z}_{3}\right)\right\rangle_{B C F T^{\otimes_{n_{e}}}}\simeq\left\langle\tau_{n_{e}}\left(\tilde{z}_{1}\right) \bar{\tau}_{n_{n}}^{2}\left(\tilde{z}_{2}\right)\right\rangle_{B C F T^{\otimes_{n_{e}}}}\left\langle\tau_{n_{e}}\left(\tilde{z}_{3}\right)\right\rangle_{B C F T^{\otimes_{n_{e}}}}
	\end{align}
	Once again the two point function on the RHS of the above equation can be obtained in the bulk and boundary channels which results in two different phases that are described below.
	
	\subsubsection*{Phase-III: Boundary Channel ($A$ away from the boundary and $B$ large)}
	In the boundary channel, utilizing the factorizations as earlier we obtain the entanglement negativity to be as follows
	\begin{align}\label{AawayBlarge}
	{\cal E}=\frac{c}{8} \log \left[\frac{\left(\tilde{z}_{2}-\tilde{z}_{1}\right)^{2}\left(2 \operatorname{Im} \tilde{z}_{2}\right)}{\epsilon_{2}^{2}\left(2 \operatorname{Im} \tilde{z}_{1}\right)}\right]
	\end{align}
	We then reverse the static mirror map in \cref{static-mirror-map},  to obtain the following expression for the entanglement negativity 
	\begin{align}\label{negadjmm3}
	{\cal E}=\frac{c}{8} \log\left[\frac{(t+x_2-p(t-x_2))(p(t-x_2)-p(t-x_1))^2}{\epsilon^2 (t+x_1-p(t-x_1)) \sqrt{p'(t-x_1)p'(t-x_2)}}\right].
	\end{align}

	\subsubsection*{Phase-IV:  Bulk channel ($A$ away from the boundary and $B$ small)}
	In the bulk channel on the other hand, the entanglement negativity is given by the following expression
	\begin{align}\label{AsmallBsmall}
	{\cal E}=\frac{c}{4} \log \left[\frac{\left(\tilde{z}_{2}-\tilde{z}_{1}\right)\left(\tilde{z}_{3}-\tilde{z}_{2}\right)}{\epsilon_{2}^{2}\left(\tilde{z}_{3}-\tilde{z}_{1}\right)}\right] .
	\end{align}
	As earlier, we use inverse static mirror map  to obtain the entanglement negativity to be as follows
	\begin{align}\label{negadjmm4}
	{\cal E}=\frac{c}{4}\log\left[\frac{(p(t-x_2)-p(t-x_1))(p(t-x_3)-p(t-x_2))^2}{\epsilon^2 (p(t-x_3)-p(t-x_1)) p'(t-x_2)}\right].
	\end{align}
	
	\subsubsection{Disjoint Intervals}\label{sec:endj}
	\begin{figure}[H]
		\centering
		\includegraphics[scale=0.7]{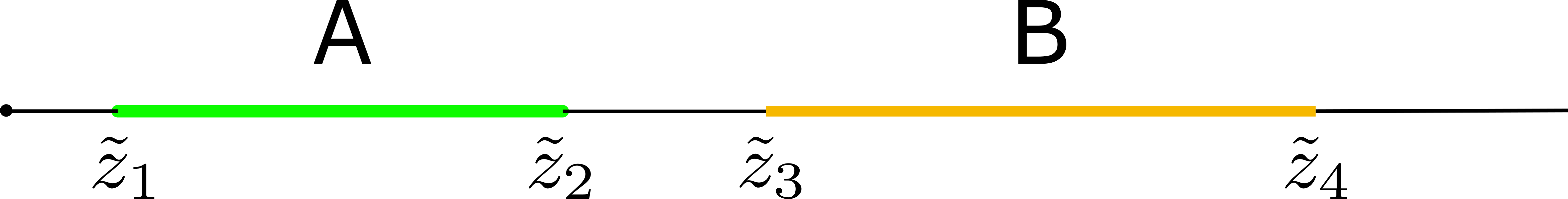}
		\caption{Configuration involving two disjoint intervals away from the boundary in a $BCFT_{1+1}$}
	\end{figure}
	
	Having obtained the entanglement negativity for two adjacent intervals in various channels, we now turn our attention to the entanglement negativity for two disjoint intervals described by $A=[\tilde{z}_1,\tilde{z}_2]$ and $B=[\tilde{z}_3,\tilde{z}_4]$ by utilizing the replica technique. The entanglement negativity for this mixed state configuration is described by the following four point twist correlator
	\begin{align}\label{ENdj2}
	{\cal E}=\lim _{n_{e} \rightarrow 1} \log[ \left\langle\tau_{n_{e}}\left(\tilde{z}_{1}\right) \bar{\tau}_{n_{e}}\left(\tilde{z}_{2}\right)\bar{\tau}_{n_{e}}\left(\tilde{z}_{3}\right) \tau_{n_{e}}\left(\tilde{z}_{4}\right)\right\rangle_{B C F T^ {\otimes_{n_{e}}}}]
	\end{align}
	
	\subsection*{Proximity limit: $A$, $B$ large separated by  $C$ small}
	
	The above four point function is described by a eight point twist correlator in the chiral CFT which is extremely difficult to determine even in the large central charge limit. However, here we consider the two intervals to be large compared to the length of the separation between them or in other words we take  the proximity limit $\tilde{z}_2\to \tilde{z}_3$. In such a scenario,  the required four point function factorizes as follows in the large-$c$ limit

	\begin{align}\label{disj_BCFT}
	\left\langle\tau_{n_{e}}\left(\tilde{z}_{1}\right) \bar{\tau}_{n_{e}}\left(\tilde{z}_{2}\right)\bar{\tau}_{n_{e}}\left(\tilde{z}_{3}\right) \tau_{n_{e}}\left(\tilde{z}_{4}\right)\right\rangle_{B C F T^{\otimes_{n_{e}}}}\simeq\left\langle\tau_{n_{e}}\left(\tilde{z}_{1}\right)\right\rangle_{B C F T^{\otimes_{n_{e}}}}&\left\langle\bar{\tau}_{n_{e}}\left(\tilde{z}_{2}\right) \bar{\tau}_{n_{n}}\left(\tilde{z}_{3}\right)\right\rangle_{B C F T^{\otimes_{n_{e}}}}\nonumber \\ &\left\langle\tau_{n_{e}}\left(\tilde{z}_{4}\right)\right\rangle_{B C F T^{\otimes_{n_{e}}}}
	\end{align}
	The two point twist correlators in \cref{disj_BCFT} is given by a four point twist correlator in the chiral $CFT$ which can be obtained through the doubling trick
	\begin{align}
	\left\langle\bar{\tau}_{n_{e}}\left(\tilde{z}_{2}\right)  \bar{\tau}_{n_{e}}\left(\tilde{z}_{3}\right)\right\rangle_{B C F T^{\otimes_{n_{e}}}}=\left\langle\tau_{n_{e}}\left(\tilde{z}_{2}^{*}\right) \bar{\tau}_{n_{e}}\left(\tilde{z}_{2}\right) \bar{\tau}_{n_{e}}\left(\tilde{z}_{3}\right)\tau_{n_{e}}\left(\tilde{z}_{3}^{*}\right) \right\rangle_{C F T^{\otimes_{n_{e}}}}
	\end{align}
	In the large central charge limit, this four point twist correlator can be computed in the $t-channel$ using the monodromy technique \cite{Malvimat:2018txq} to be as follows,
	\begin{align}
	\left\langle\bar{\tau}_{n_{e}}\left(\tilde{z}_{2}\right)  \bar{\tau}_{n_{e}}\left(\tilde{z}_{3}\right)\right\rangle_{B C F T^{\otimes_{n_{e}}}}&=\left[\frac{4 \operatorname{Im} (\tilde{z}_{2})\operatorname{Im} (\tilde{z}_{3})}{ \left(\tilde{z}_{3}-\tilde{z}_{2}\right)^{2}}\right]^{\Delta_{n_{e}}^{(2)}}
	\end{align}
	Using the above result  in \cref{ENadj2}, the entanglement negativity may be computed to be as follows
	\begin{align}\label{negdisj}
	{\cal E}=\frac{c}{8} \log \left[\frac{4 \operatorname{Im} (\tilde{z}_{2})\operatorname{Im} (\tilde{z}_{3})}{ \left(\tilde{z}_{3}-\tilde{z}_{2}\right)^{2}}\right]
	\end{align}
	In \cref{negdisj}, we  apply the inverse static mirror map to obtain the following expression for the entanglement negativity,
	\begin{align}\label{negdismm}
	{\cal E}=\frac{c}{8} \log\left[\frac{(t+x_2-p(t-x_2))(t+x_3-p(t-x_3))}{ (p(t+x_3)-p(t-x_2))^2 }\right].
	\end{align}

	\subsection{Holographic Entanglement Negativity  for Moving Mirror}
	
	Having determined the entanglement negativity through replica technique in a $BCFT_{1+1}$ we now proceed to  obtain the holographic entanglement negativity for mixed state configurations involving two  adjacent and disjoint intervals  by using the proposal described by  \cref{henadjH} and \cref{hendjcomb} in the dual bulk $AdS_3$ geometry. To this end we utilize the following expressions for the length of the back reacted cosmic branes ${\cal L}_{X}^{(\frac{1}{2})}$ corresponding to the Renyi entropy of order half for a subsystem $X=[z_i,z_j]$
	\begin{align}
	{\cal L}_{X}^{(\frac{1}{2})}=Min[{\cal L}_{X}^{(\frac{1}{2}), \,con},{\cal L}_{X}^{(\frac{1}{2}),\,dis}]
	\end{align}
	In the above equation con and dis denote the connected and disconnected lengths of the cosmic branes respectively. In the $AdS_3$ geometry dual to a $BCFT_{1+1}$ these lengths are given as follows
	\begin{align}
	{\cal L}_{X}^{(\frac{1}{2}), \,con}&= \frac{3}{2}\log[\frac{(z_i-z_j)^2}{\epsilon_{i}\epsilon_{j}}]\\
	{\cal L}_{X}^{(\frac{1}{2}),\,dis}&=\frac{3}{2}\log\left[ \frac{4 \operatorname{Im} \tilde{z}_{i}\operatorname{Im} \tilde{z}_{j}}{\epsilon_{i}\epsilon_{j}}\right]+2S_{\text{bdy}}
	\end{align}
	In order to arrive at the above expression  we have utilized the fact that in $AdS_3/CFT_2$ the effect of the backreaction reduced to a proportionality factor as described in section \ref{sec:review} which as follows
	\begin{align}
	{\cal L}_{X}^{(\frac{1}{2})}=\frac{3}{2}{\cal L}_{X}.
	\end{align}
	where ${\cal L}_{X}$ is the length of the geodesic corresponding to the subsystem-X.
	Note that for the disconnected the numerical factor is only for the dynamical part as the contribution from the $S_{bdy}$ to the length of the backreacting cosmic brane is independent of the replica index $n$ as described by \cref{SAMM}. As described in  section \ref{sec:review} we then utilize the Banados map to obtain the corresponding results for holographic moving mirrors.

	\subsubsection{Adjacent Intervals}
	Here, we compute the holographic entanglement negativity for two adjacent intervals from the dual bulk $AdS_3$ utilizing the proposal described earlier in \cref{HENADJ} and \cref{henadjH}
	
	\subsubsection*{Phase-I:  Boundary channel ($A$ close to the boundary and  $B$ large)}
	\begin{figure}[H]
		\centering
		\includegraphics[scale=0.7]{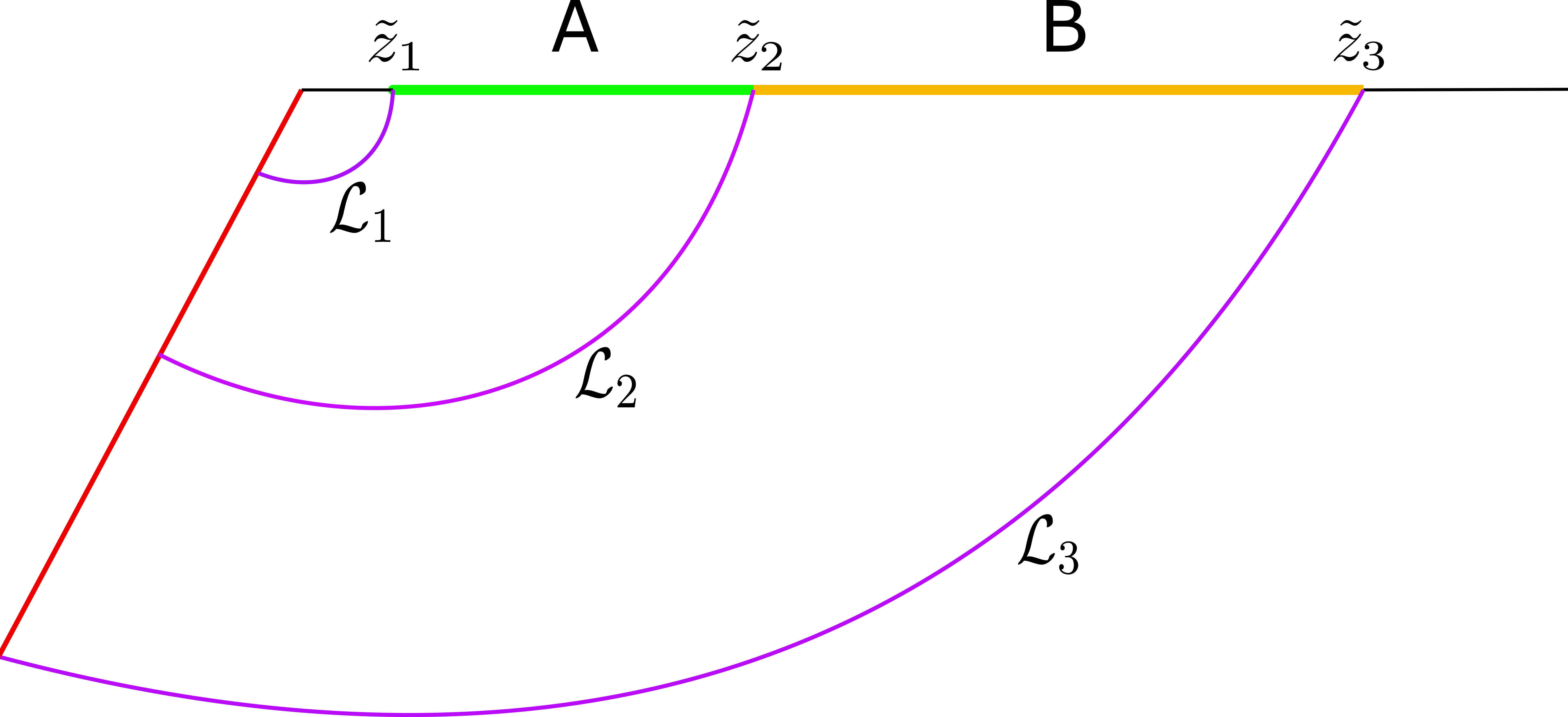}
		\caption{Geodesics contributing to the holographic entanglement negativity of adjacent intervals in Phase-I}\label{enadjp1}
	\end{figure}
	In phase-I,  the interval $A$ is close to the boundary and the interval $B$ is large. In such a scenario, the geodesics contributing to the holographic entanglement negativity are depicted in fig.\ref{enadjp1}. The length of the geodesics homologous to the various subsystems involved are as follows
	\begin{align}
	{\cal L}_A&=	{\cal L}_1+	{\cal L}_2\nonumber\\
	{\cal L}_B&=	{\cal L}_2+	{\cal L}_3\nonumber\\
	{\cal L}_{AB}&=	{\cal L}_1+	{\cal L}_3
	\end{align}
	Substituting the above expressions in \cref{henadjH}, we obtain the holographic entanglement negativity to be as follows
	\begin{align}
	{\cal E}&=\frac{3}{8 G_N}{\cal L}_2\nonumber\\
	&=\frac{3}{8 G_{N}}\left[\log \frac{2 \operatorname{Im} \tilde{z}_{2}}{\epsilon_{2}}\right]+S_{\text{bdy}}
	\end{align}
	where 
	Note that upon using the Brown-Henneaux formula \cite{Brown:1986nw}, the above result exactly matches with \cref{negAcloseBlarge} which we obtained using the replica technique in the dual $BCFT_{1+1}$. Furthermore, using the Banados map given in \cref{Banados-map} we obtain the following expression for  entanglement negativity 
	\begin{align}
	{\cal E}=\frac{3}{8 G_N}\log\left[\frac{t+x_2-p(t-x_2)}{\epsilon\sqrt{p'(t-x_2)}}\right]+S_{\text{bdy}}\,.
	\end{align}
	which matches identically with \cref{negadjmm1}.
	\subsubsection*{Phase-II:  Bulk  channel ($A$ close to the boundary and  $B$ small)}
	\begin{figure}[H]
		\centering
		\includegraphics[scale=0.7]{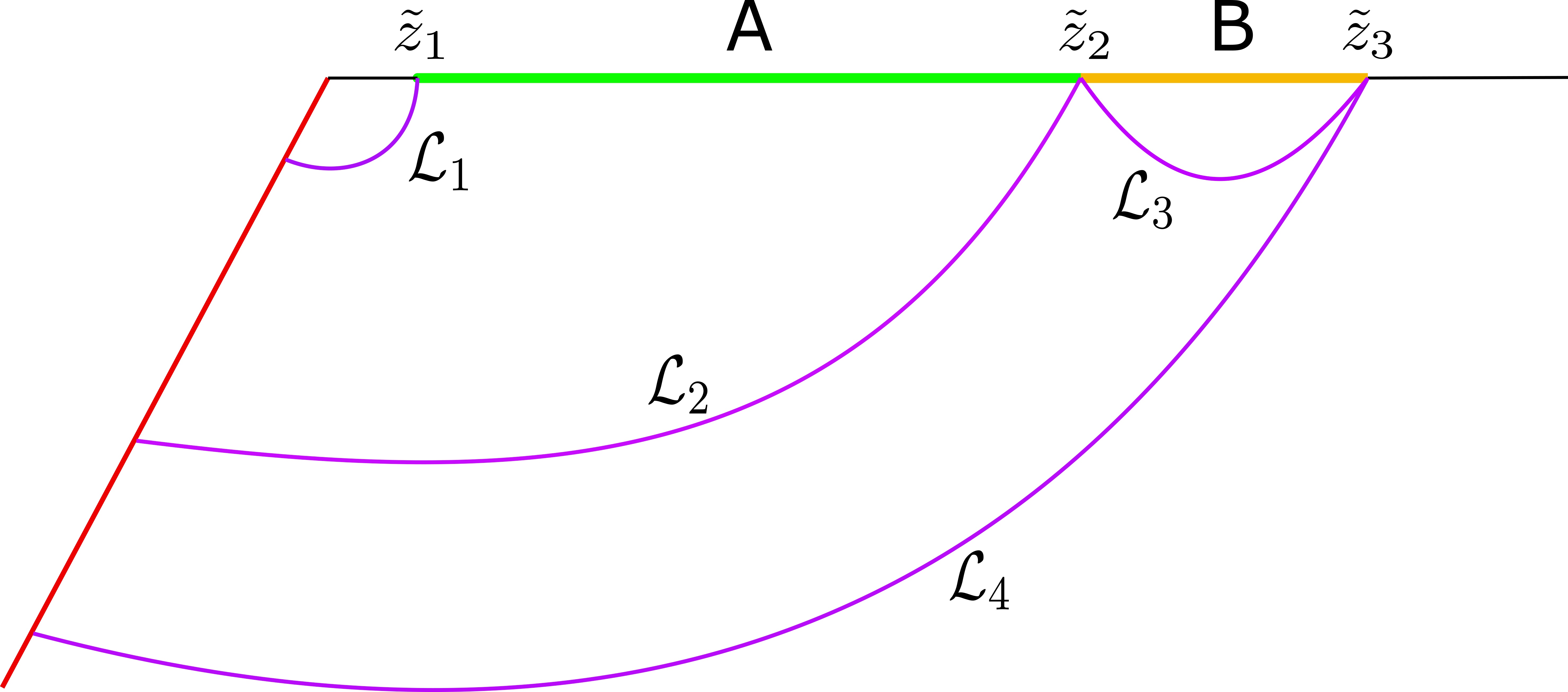}
		\caption{Geodesics contributing to the holographic entanglement negativity of adjacent intervals in Phase-I}\label{enadjp2}
	\end{figure}
	In this phase, we consider an  interval $A$ which is away from the boundary with a small interval $B$. For such a configuration, the geodesics leading to the holographic entanglement negativity are shown  in fig.\ref{enadjp2} and their lengths are as follows
	\begin{align}
	{\cal L}_A&=	{\cal L}_1+	{\cal L}_2\nonumber\\
	{\cal L}_B&=	{\cal L}_3\nonumber\\
	{\cal L}_{AB}&=	{\cal L}_1+	{\cal L}_4
	\end{align}
	Utilizing the above expressions  in \cref{henadjH} we determine the holographic entanglement negativity as 
	\begin{align}\label{henadjmm2}
	{\cal E}&=\frac{3}{16 G_N}\bigg[{\cal L}_2+{\cal L}_3-{\cal L}_4\bigg]\nonumber\\
	&=\frac{3}{16 G_{N}} \log \left[\frac{\operatorname{Im} \tilde{z}_{2}\left(\tilde{z}_{3}-\tilde{z}_{2}\right)^{2}}{\epsilon_{2}^{2} \operatorname{Im} \tilde{z}_{3}}\right]
	\end{align}
	The above expression once again matches precisely with the replica technique result obtained in \cref{negAcloseBsmall}. Subsequently, we utilize  the Banados map in \cref{henadjmm2} to obtain the following expression for the entanglement negativity 
	\begin{align}
	{\cal E}=\frac{3}{16G_N}\log\left[\frac{(t+x_2-p(t-x_2))(p(t-x_3)-p(t-x_2))^2}{\epsilon^2 (t+x_3-p(t-x_3)) \sqrt{p'(t-x_2)p'(t-x_3)}}\right].
	\end{align}
	Observe that once again the above expression obtained from holography agrees exactly with the corresponding replica technique result given in \cref{negadjmm2}.
	\subsubsection*{Phase-III:  Boundary channel ($A$ away from the boundary and  $B$ large)}
	\begin{figure}[H]
		\centering
		\includegraphics[scale=0.7]{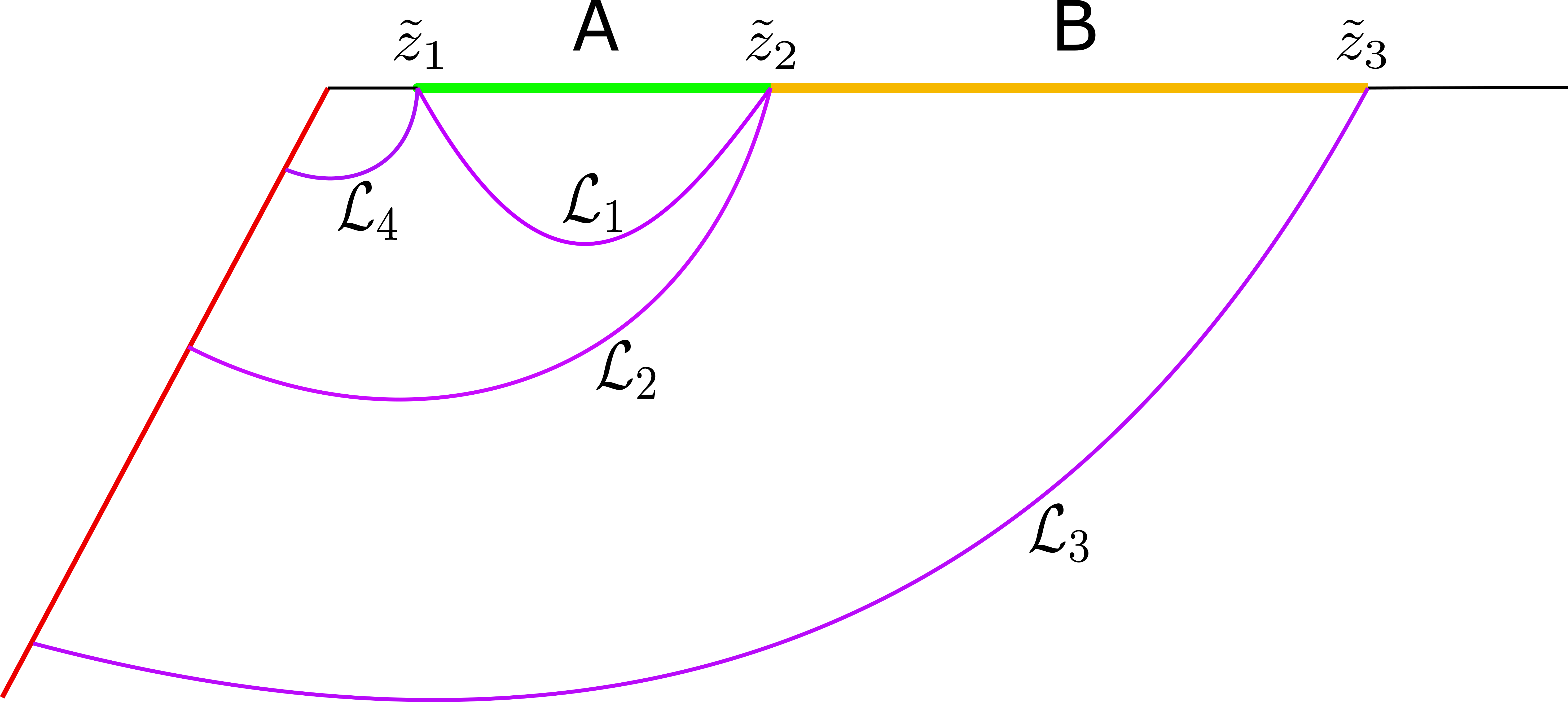}
		\caption{Geodesics contributing to the holographic entanglement negativity of adjacent intervals in Phase-I}\label{enadjp3}
	\end{figure}
	In phase-III,  the interval $A$ is away from the boundary and the interval $B$ is small. As illustrated in figure \ref{enadjp3}, the lengths of the geodesics contributing to the holographic entanglement negativity in this phase are as follows
	\begin{align}
	{\cal L}_A&=	{\cal L}_1\nonumber\\
	{\cal L}_B&=	{\cal L}_2+	{\cal L}_3\nonumber\\
	{\cal L}_{AB}&=	{\cal L}_3+	{\cal L}_4
	\end{align}
	Using the above geodesic lengths in \cref{henadjH} we determine the holographic entanglement negativity to be as
	\begin{align}
	{\cal E}&=\frac{3}{16 G_N}\bigg[{\cal L}_1+{\cal L}_2-{\cal L}_4\bigg]\nonumber\\
	&=\frac{3}{16 G_{N}} \log \left[\frac{\left(\tilde{z}_{2}-\tilde{z}_{1}\right)^{2} \operatorname{Im}  \tilde{z}_{2}}{\epsilon_{2}^{2} \operatorname{Im} \tilde{z}_{1}}\right]
	\end{align}
	Once again the above expression agrees with the corresponding result determined through the twist correlators in a $BCFT_{1+1}$ in \cref{AawayBlarge}. Furthermore, upon using Banados map we obtain the entanglement negativity 
	\begin{align}
	{\cal E}=\frac{3}{16G_N}\log\left[\frac{(t+x_2-p(t-x_2))(p(t-x_2)-p(t-x_1))^2}{\epsilon^2 (t+x_1-p(t-x_1)) \sqrt{p'(t-x_1)p'(t-x_2)}}\right].
	\end{align}
	The above expression precisely agrees with the corresponding replica technique result  given in \cref{negadjmm3} upon using the Brown-Henneaux relation.
	\subsubsection*{Phase-IV: Bulk channel ($A$ away from the boundary and  $B$ large)}
	\begin{figure}[H]
		\centering
		\includegraphics[scale=0.7]{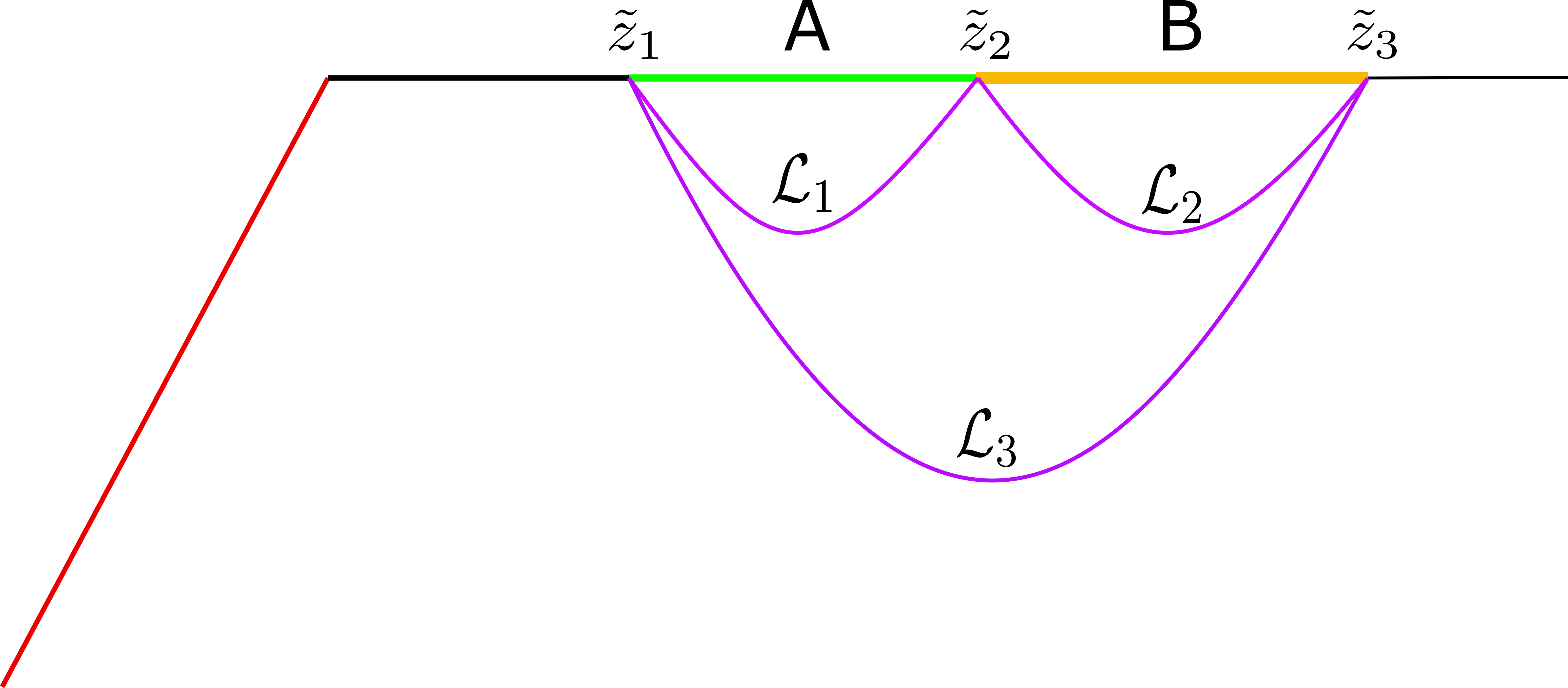}
		\caption{Geodesics contributing to the holographic entanglement negativity of adjacent intervals in Phase-I}\label{enadjp30}
	\end{figure}
	In phase-III,  the interval $A$ is away from the boundary and the interval $B$ is large. Note that here none of the geodesics  contributing to the holographic entanglement negativity  intersect the EOW brane as illustrated in fig.\ref{enadjp30}. The length of the required geodesics are as follows
	\begin{align}
	{\cal L}_A&=	{\cal L}_1\nonumber\\
	{\cal L}_B&=	{\cal L}_2\nonumber\\
	{\cal L}_{AB}&=	{\cal L}_3
	\end{align}
	Utilizing the above lengths in \cref{henadjH} we obtain the holographic negativity to be 
	\begin{align}\label{henadjmmhol4}
	{\cal E}&=\frac{3}{8 G_N}\bigg[{\cal L}_1+{\cal L}_2-{\cal L}_3\bigg]\nonumber\\
	&=\frac{3}{8 G_{N}} \log \left[\frac{\left(\tilde{z}_{2}-\tilde{z}_{1}\right)\left(\tilde{z}_{3}-\tilde{z}_{2}\right)}{\epsilon_{2}^{2}\left(\tilde{z}_{3}-\tilde{z}_{1}\right)}\right]
	\end{align}	
	The above expression once again agrees exactly  with the entanglement negativity computed in $BCFT$ given in \cref{AsmallBsmall}. By using the Banados map given in \cref{Banados-map} we obtain the following result for the holographic
	entanglement negativity 
	\begin{align}
	{\cal E}=\frac{3}{8G_N}\log\left[\frac{(p(t-x_2)-p(t-x_1))(p(t-x_3)-p(t-x_2))^2}{\epsilon^2 (p(t-x_3)-p(t-x_1)) p'(t-x_2)}\right].
	\end{align}
	which once again exactly matches with the corresponding replica technique result in \cref{negadjmm4}.
	\subsubsection{Disjoint Intervals}
	
	We now determine the entanglement negativity for two disjoint intervals in the proximity limit  ($A$,$B$ large separated by a small $C$)  by utilizing the holographic proposal given by \cref{HENDJ} and \cref{hendjcomb} described in \cite{Malvimat:2018txq}.

	\begin{figure}[H]
		\centering
		\includegraphics[scale=0.7]{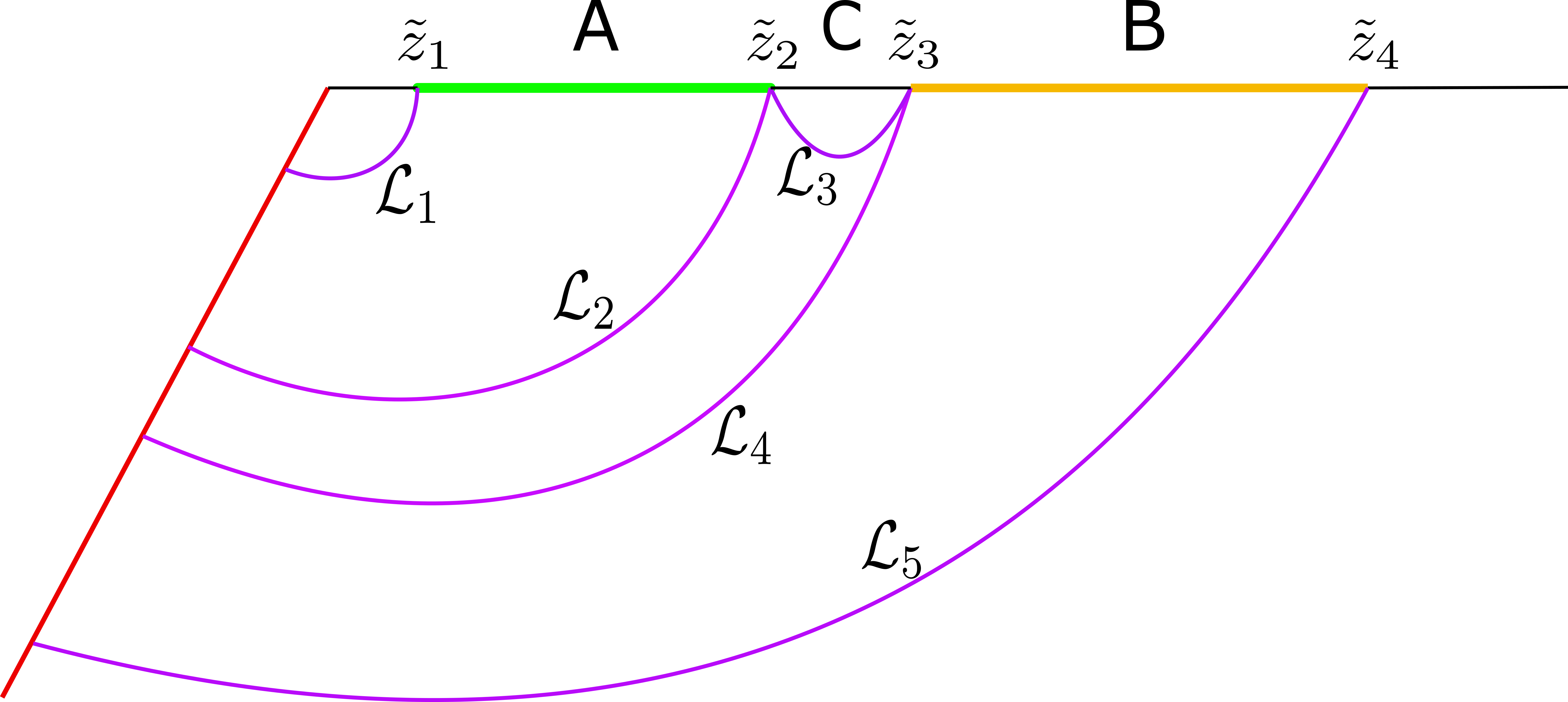}
		\caption{Geodesics contributing to the holographic entanglement negativity of disjoint intervals in proximity}\label{dj1}
	\end{figure}
	As depicted in the fig.\ref{dj1} the length of various geodesics homologous to the subsystems involved are as follows
	\begin{align}\label{lengthdisj}
	{\cal L}_{AC}&=	{\cal L}_1+{\cal L}_4\nonumber\\
	{\cal L}_{BC}&=	{\cal L}_2+{\cal L}_5\nonumber\\
	{\cal L}_C&=	{\cal L}_3\nonumber\\
	{\cal L}_{ABC}&=	{\cal L}_1+{\cal L}_5
	\end{align}
	Utilizing the above lengths in \cref{hendjcomb} we obtain the holographic entanglement negativity to be as follows
	\begin{align}\label{negdisholo}
	{\cal E}&=\frac{3}{16 G_N}\bigg[{\cal L}_{AC}+{\cal L}_{BC}-{\cal L}_C-{\cal L}_{ABC}\bigg]\nonumber\\
	&=\frac{3}{16 G_{N}} \log \left[\frac{4 \operatorname{Im} (\tilde{z}_{2})\operatorname{Im} (\tilde{z}_{3})}{ \left(\tilde{z}_{3}-\tilde{z}_{2}\right)^{2}}\right]
	\end{align}	
	The above expression once again precisely matches with that computed using the replica technique in a $BCFT_{1+1}$  given in \cref{negdisj}. Following this, utilising the Banados map in \cref{negdisholo} , the holographic entanglement negativity may be computed as 
	\begin{align}
	{\cal E}=\frac{3}{16G_N}\log\left[\frac{(t+x_2-p(t-x_2))(t+x_3-p(t-x_3))}{ (p(t+x_3)-p(t-x_2))^2 }\right].
	\end{align}
	Note that the above result once again precisely agrees with that obtained through replica technique in \cref{negdismm}.
	\subsection{Page Curves for Entanglement Negativity in Moving Mirrors}
	In previous subsections we demonstrated that the entanglement negativity obtained through the replica technique in a $BCFT_{1+1}$ match precisely with the corresponding holographic results for the mixed state configurations of two adjacent and  disjoint intervals in various phases. We also showed that upon using the required conformal transformation in the $BCFT$ and the Banados map in the dual bulk $AdS_3$ we can get the corresponding results in the moving mirror setup.
	In this subsection we obtain the analogues of the Page curves for the entanglement negativity through the holographic proposals described in \cref{HENADJ} and \cref{HENDJ} for the mixed state configurations of  adjacent and disjoint intervals in the context of  moving mirrors utilizing the results for the holographic entanglement entropy obtained in  \cite{Akal:2020twv,Akal:2021foz} and given by \cref{SAMM}.  Note that in previous subsection we determined the entanglement negativity for various phases in different regimes of the lengths of the subsystems involved. However, here we directly obtain the time evolution of the holographic entanglement negativity for all the phases of the configurations in question by directly substituting the expression for the length of a geodesic given by \cref{SAMM} for various subsystems in \cref{HENADJ} and \cref{HENDJ}

	\subsubsection{ Kink Mirror}
	
\begin{figure}[H]
		\centering
		\begin{subfigure}[b]{0.49\textwidth}
			\centering
			\includegraphics[scale=0.35]{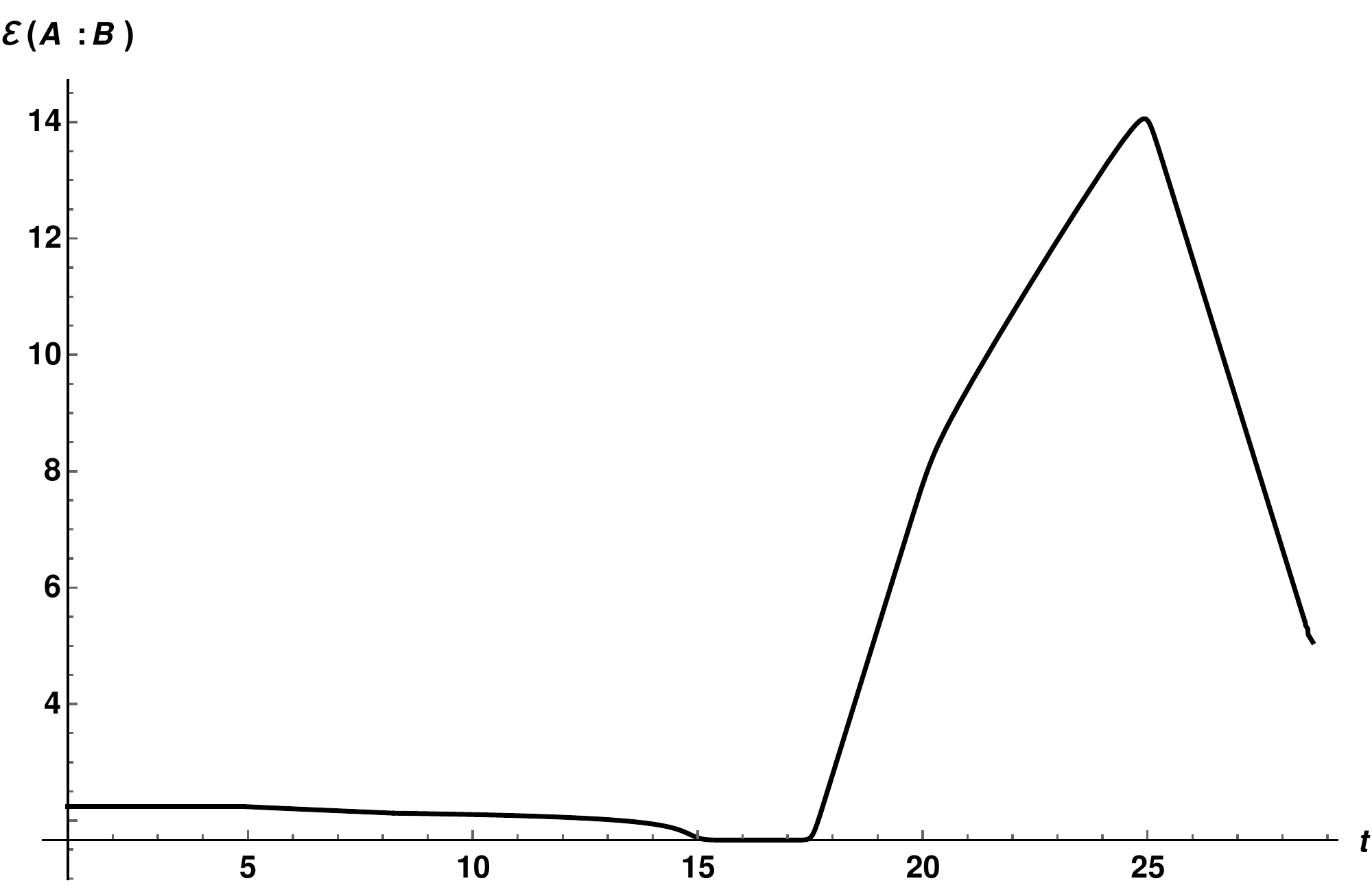}
			\caption{Fixed intervals: $A=[0.1,15]$, $B=[15,25]$, \\ $\beta=0.1$, $\epsilon=0.001$, $u_0=20$, $S_{\text{bdy}}=0$, \\ $c=1$}
			\label{neg_kink_adj_fixed}
		\end{subfigure}
		\hfill
		\begin{subfigure}[b]{0.49\textwidth}
			\centering
			\includegraphics[scale=0.35]{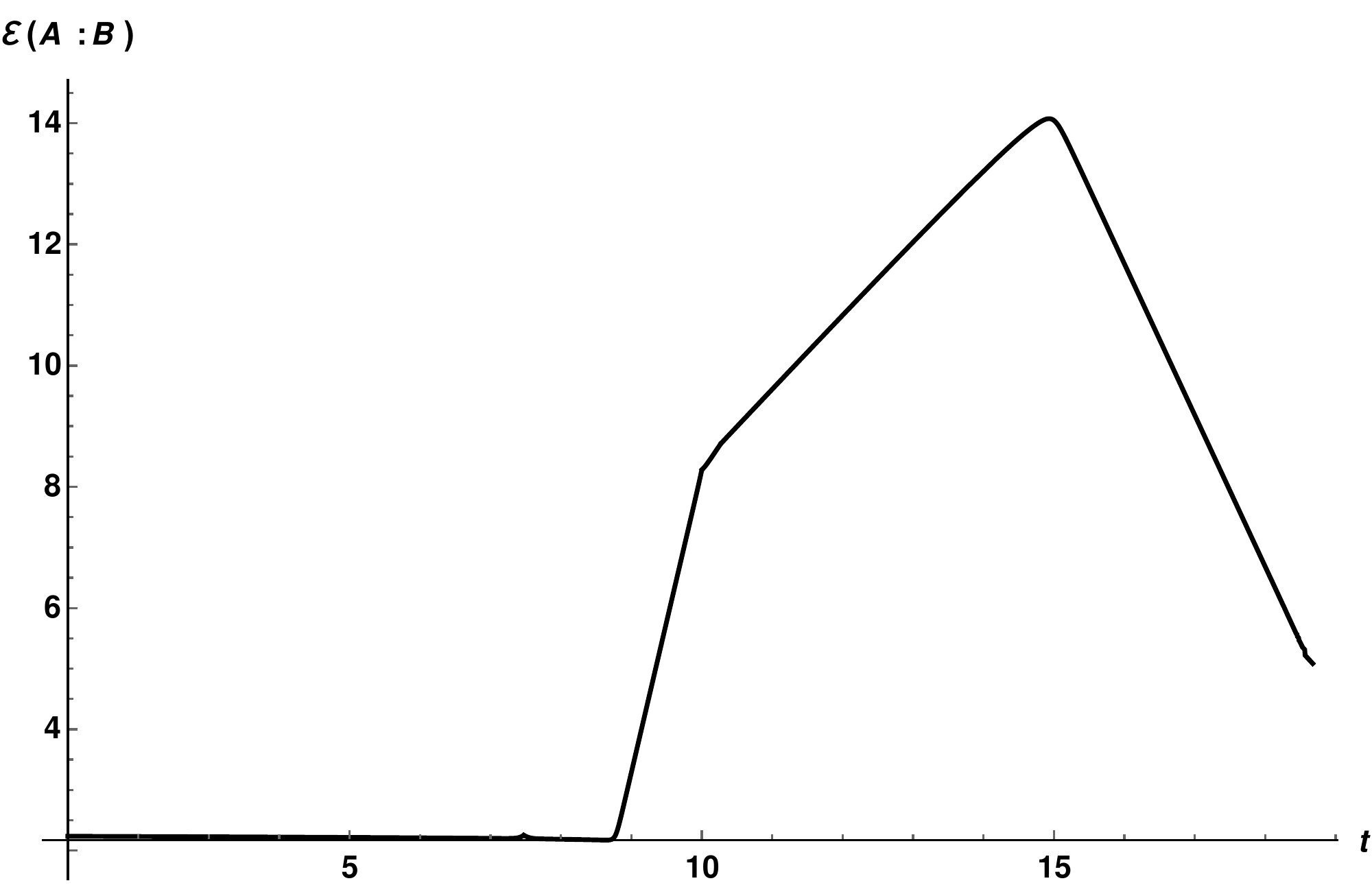}
			\caption{Co-moving intervals: $A=[Z(t)+0.1,Z(t)+15]$, $B=[Z(t)+15,Z(t)+25]$, $\beta=0.1$, $\epsilon=0.001$, $u_0=20$, $S_{\text{bdy}}=2$, $c=1$}
			\label{neg_kink_adj_comov}
		\end{subfigure}
		\caption{Page curve for the entanglement negativity between two adjacent intervals in a kink mirror BCFT.}
		\label{EN-kink-adj1}
	\end{figure}

\begin{figure}[H]
		\centering
		\begin{subfigure}[b]{0.49\textwidth}
			\centering
			\includegraphics[scale=0.35]{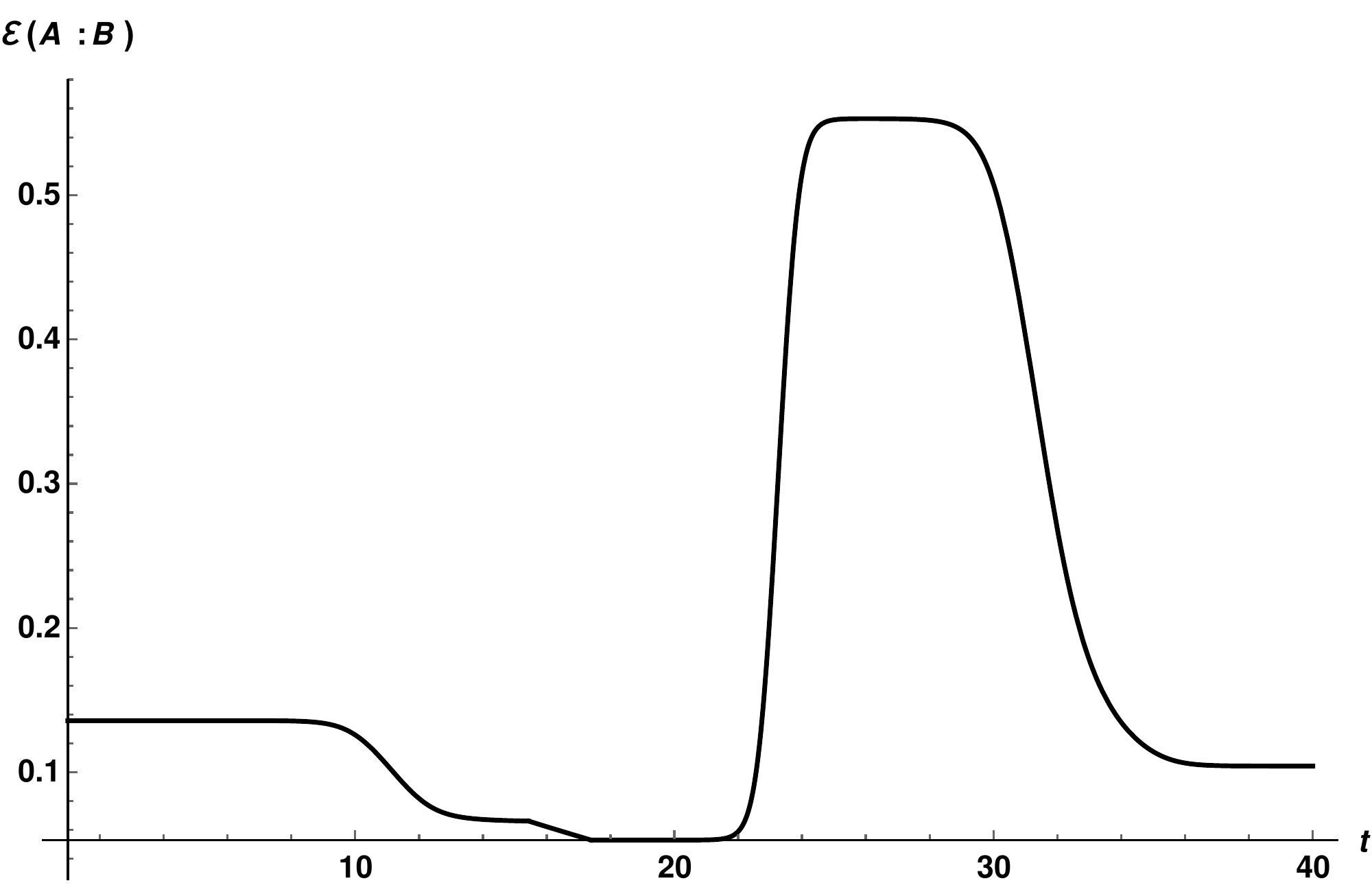}
			\caption{Fixed intervals: $A=[10,12]$, $B=[15.5,80]$, \\ $\beta=0.5$, $\epsilon=0.001$, $u_0=20$, $S_{\text{bdy}}=0$, \\ $c=1$}
			\label{neg_kink_disj_fixed}
		\end{subfigure}
		\hfill
		\begin{subfigure}[b]{0.49\textwidth}
			\centering
			\includegraphics[scale=0.35]{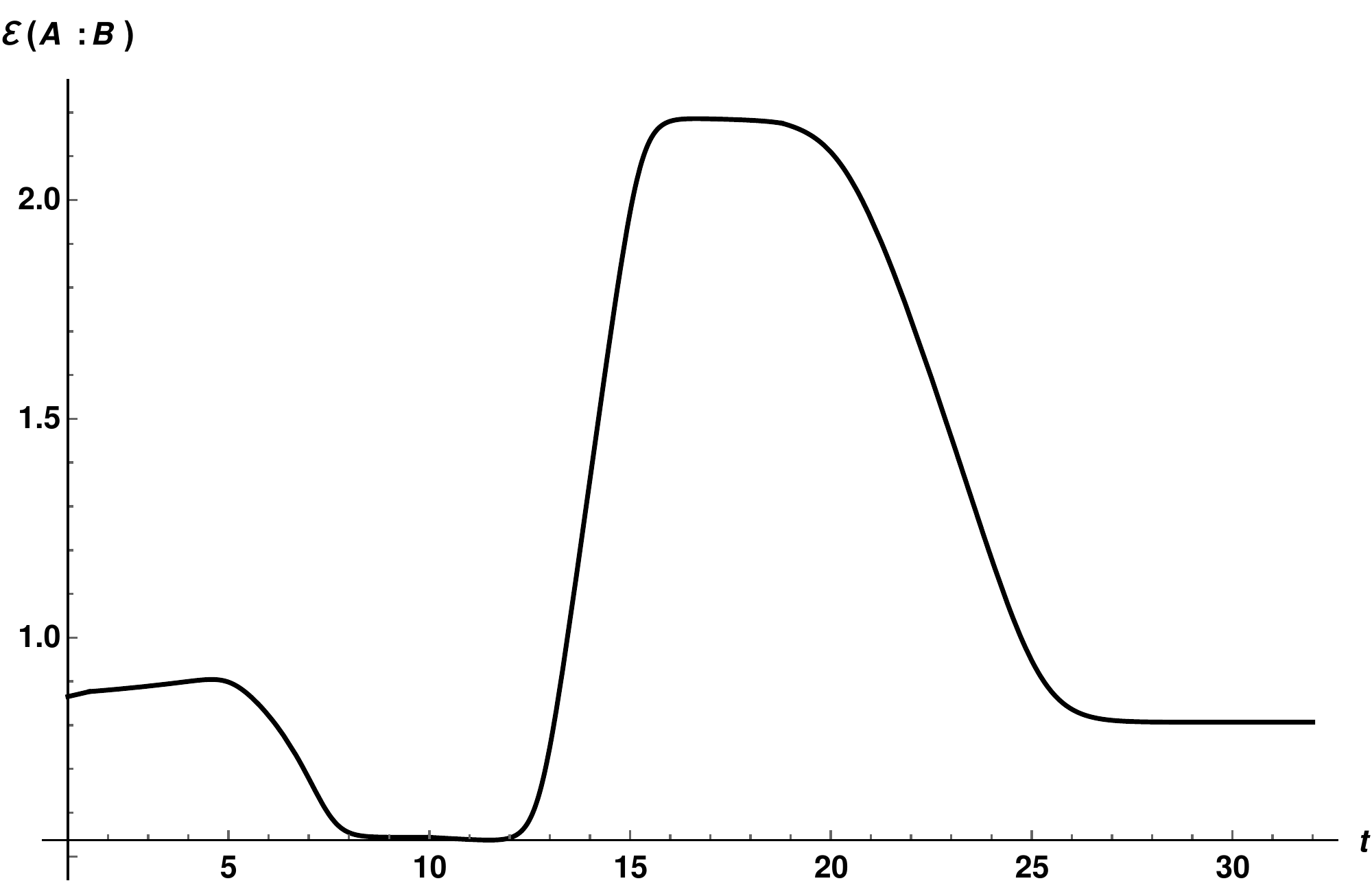}
			\caption{Co-moving intervals: $A=[Z(t)+10,Z(t)+15]$, $B=[Z(t)+15.5,Z(t)+80]$, $\beta=0.5$, $\epsilon=0.001$, $u_0=20$, $S_{\text{bdy}}=2$, $c=1$}
			\label{neg_kink_disj_comov}
		\end{subfigure}
		\caption{Page curve for the entanglement negativity between two disjoint intervals in a kink mirror BCFT.}
		\label{EN-kink-disj1}
	\end{figure}

	As described in section \ref{sec:review}, the radiation from a kink mirror mimics the Hawking radiation emitted by  an evaporating single sided black hole. The Page curves for the entanglement entropy of a single interval in the radiation flux of the kink mirror was obtained in \cite{Akal:2020twv,Akal:2021foz}. Here, we  determine the behaviour of  entanglement negativity for  adjacent and disjoint intervals in the radiation flux of a moving kink mirror whose profile is given by \cref{kinkprof}. Substituting the kink mirror profile in \cref{SAMM} we obtain the holographic entanglement entropies  for various subsystems involved in the combination occurring in the holographic entanglement negativity proposal in \cref{henadjH} and \cref{hendjcomb}. The behavior of  the entanglement negativity for two adjacent and disjoint intervals is depicted in the plots given in figures \ref{EN-kink-adj1} and \ref{EN-kink-disj1}. Quite interestingly, the increasing part of the plots for the time evolution  of the entanglement negativity for the adjacent intervals in figures \ref{neg_kink_adj_fixed} and \ref{neg_kink_adj_comov} closely resembles the expected curve from random matrix theory \cite{Shapourian:2020mkc}. This was also reproduced in the context of JT black hole coupled to a thermal bath described by a holographic $2d$ CFT in \cite{KumarBasak:2021rrx} where the behavior was also interpreted in terms of the island contributions. It would be very interesting to explore the island interpretation for this and the rest of the cases listed below which is an open issue for future investigations.
	
	\subsubsection*{ Adjacent Intervals}

	%\subsubsection*{Subsystem Size}
	%\subsubsection*{Comoving}
	%\subsubsection*{Time Evolution}
	%\subsubsection*{Subsystem Size}
	\subsubsection*{ Disjoint Intervals}
	%\subsubsection*{Fixed Intervals}
	%\subsubsection*{Time Evolution}
	
	%\subsubsection*{Subsystem Size}
	%\subsubsection*{Comoving}
	%\subsubsection*{Time Evolution}
	
	%\subsubsection*{Subsystem Size}
	\subsubsection{ Escaping mirror}
\begin{figure}[H]
		\centering
		\begin{subfigure}[b]{0.49\textwidth}
			\centering
			\includegraphics[scale=0.35]{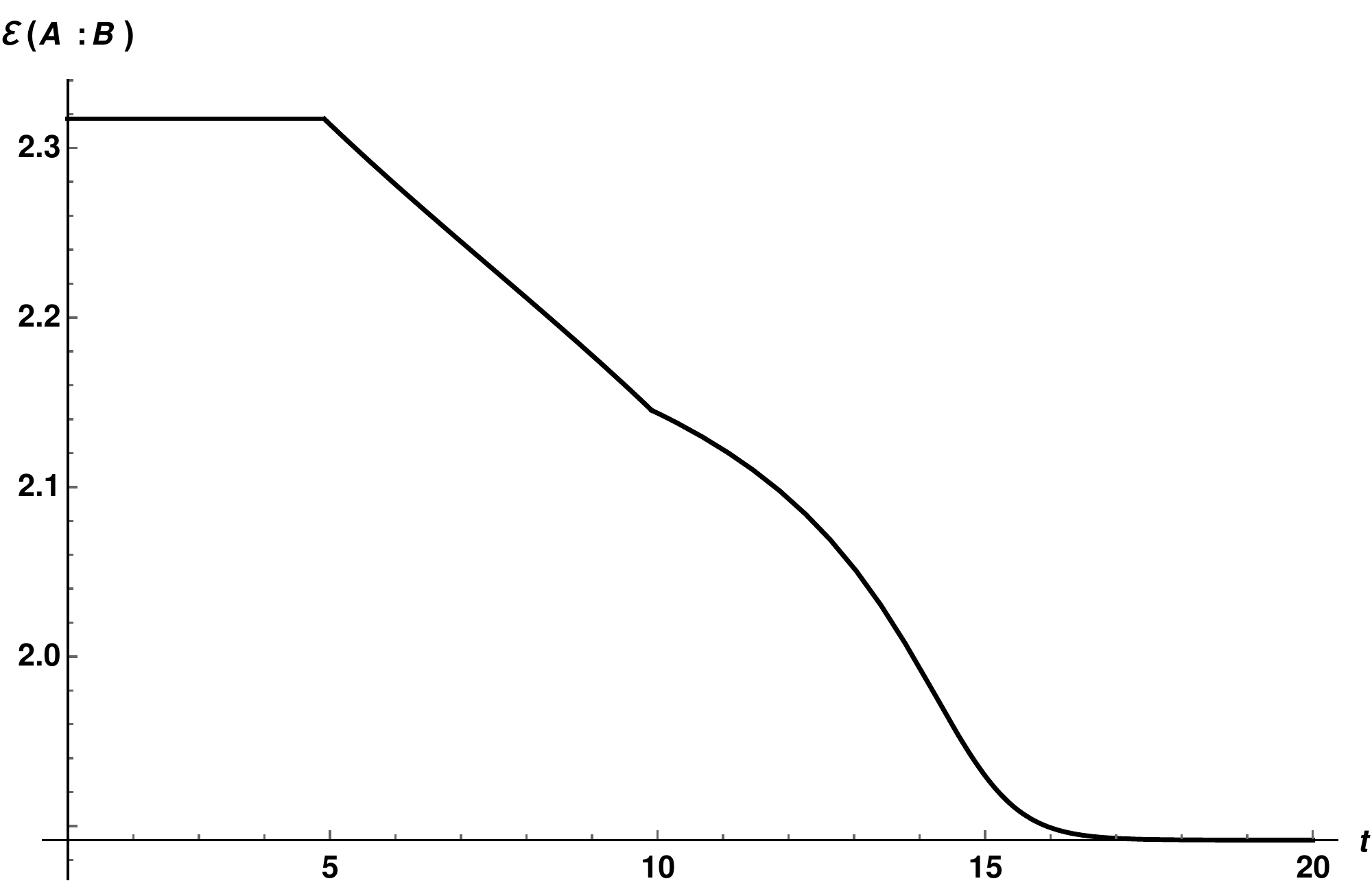}
			\caption{Fixed intervals: $A=[0.1,15]$, $B=[15,30]$, \\ $\beta=0.5$, $\epsilon=0.001$, $u_0=20$, $S_{\text{bdy}}=0.4$, \\ $c=1$}
			\label{neg_escaping_adj_fixed}
		\end{subfigure}
		\hfill
		\begin{subfigure}[b]{0.49\textwidth}
			\centering
			\includegraphics[scale=0.35]{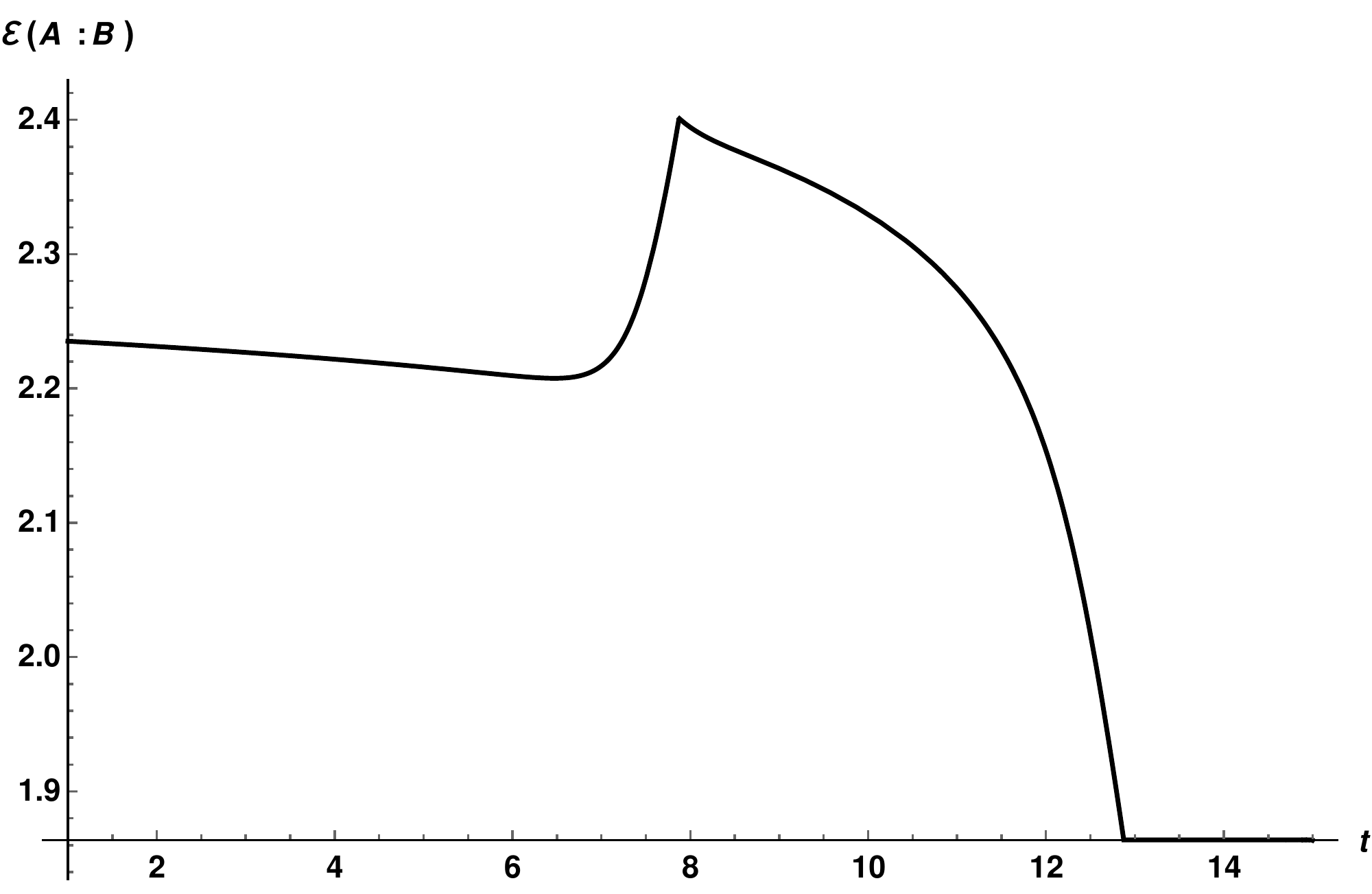}
			\caption{Co-moving intervals: $A=[Z(t)+0.1,Z(t)+15]$, $B=[Z(t)+15,Z(t)+25]$, $\beta=0.5$, $\epsilon=0.001$, $u_0=20$, $S_{\text{bdy}}=0$, $c=1$}
			\label{neg_escaping_adj_comov}
		\end{subfigure}
		\caption{Page curve for the entanglement negativity between two adjacent intervals in an escaping mirror BCFT.}
		\label{EN-esc-Adj1F}
	\end{figure}

\begin{figure}[H]
		\centering
		\begin{subfigure}[b]{0.49\textwidth}
			\centering
			\includegraphics[scale=0.35]{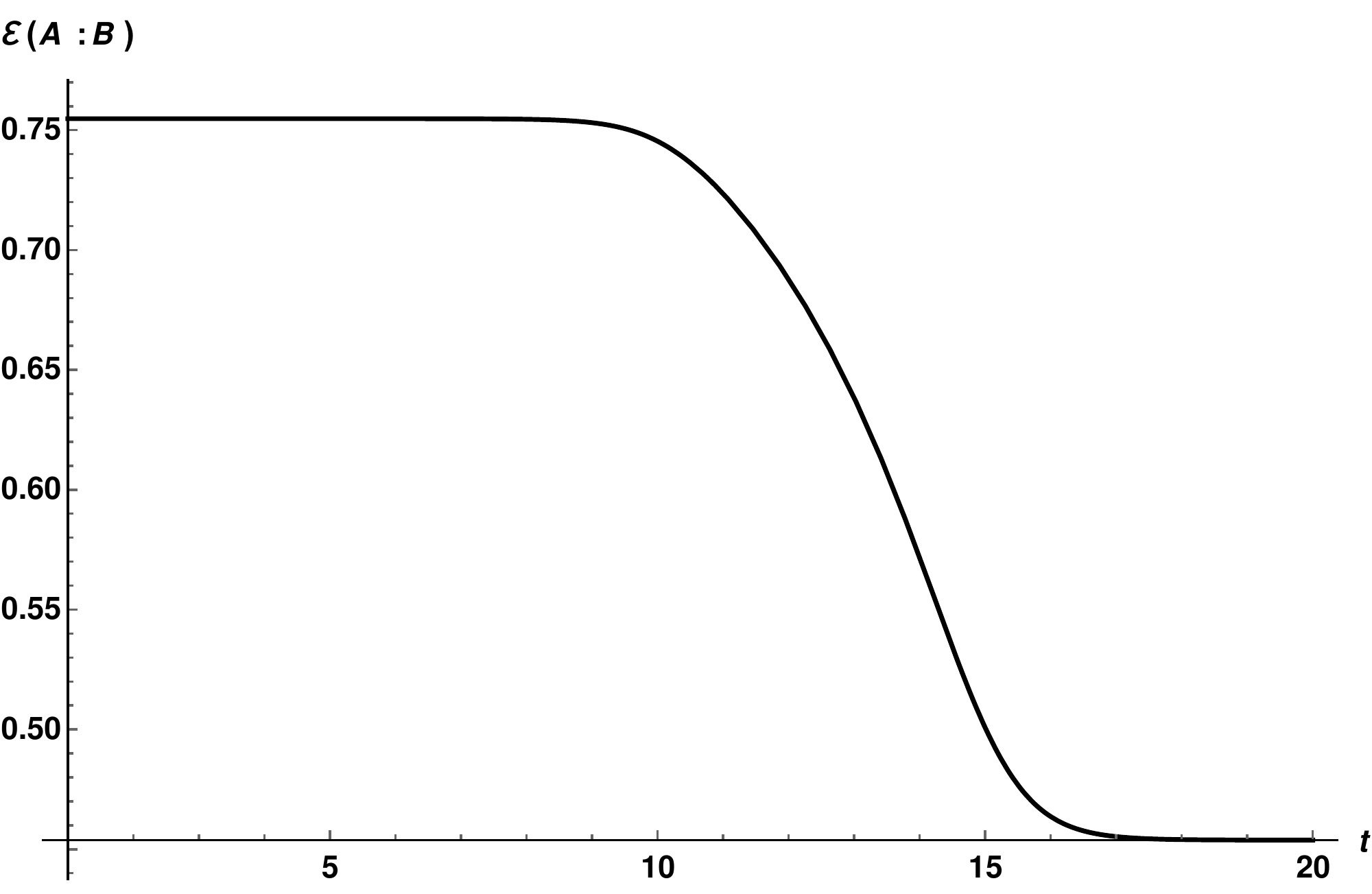}
			\caption{Fixed intervals: $A=[10,15]$, $B=[15.5,50]$, \\ $\beta=0.5$, $\epsilon=0.001$, $u_0=20$, $S_{\text{bdy}}=0$, \\ $c=1$}
			\label{neg_escaping_disj_fixed}
		\end{subfigure}
		\hfill
		\begin{subfigure}[b]{0.49\textwidth}
			\centering
			\includegraphics[scale=0.35]{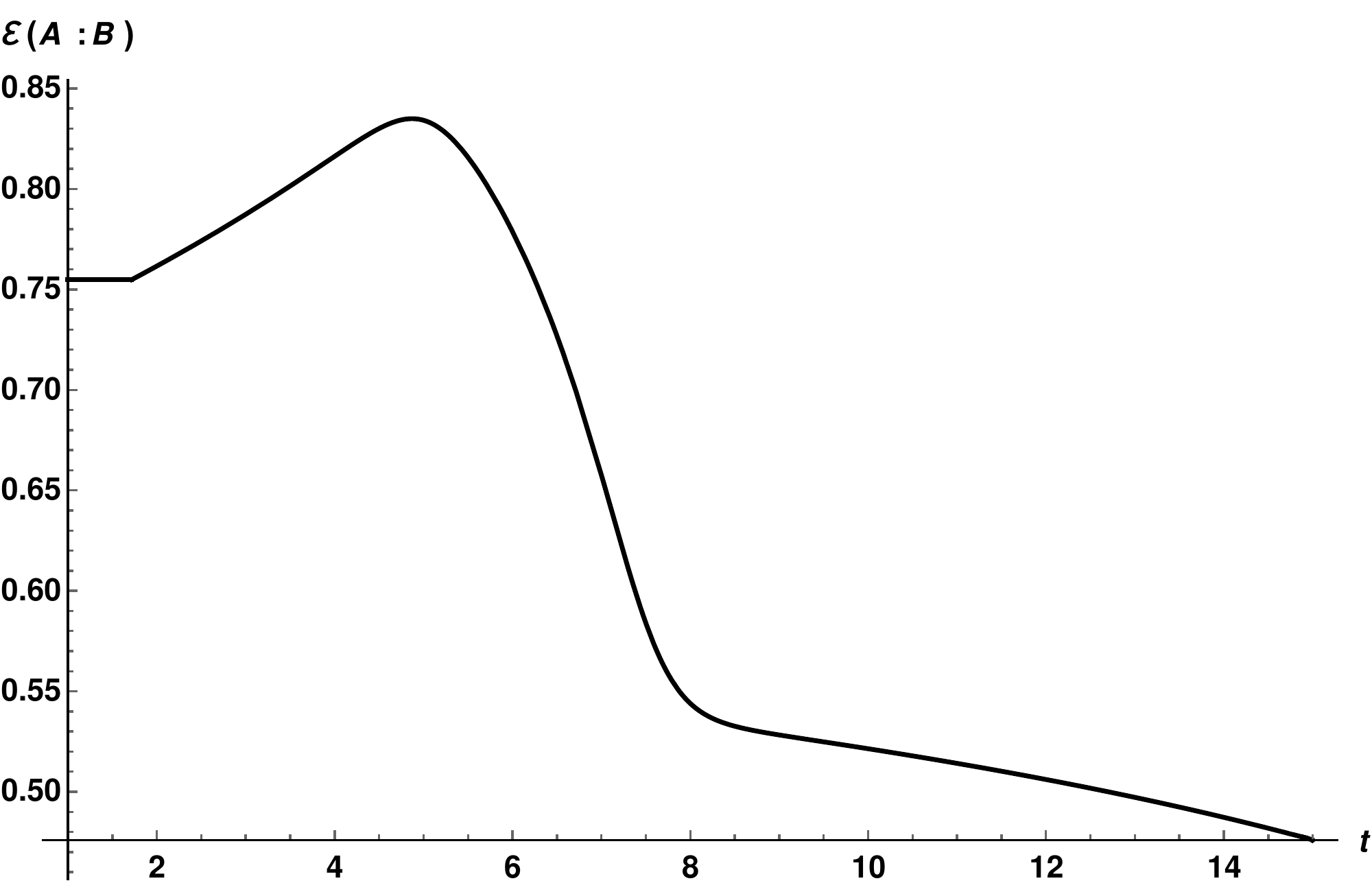}
			\caption{Co-moving intervals: $A=[Z(t)+10,Z(t)+15]$, $B=[Z(t)+15.5,Z(t)+80]$, $\beta=0.5$, $\epsilon=0.001$, $u_0=20$, $S_{\text{bdy}}=2$, $c=1$}
			\label{neg_escaping_disj_comov}
		\end{subfigure}
		\caption{Page curve for the entanglement negativity between two disjoint intervals in an escaping mirror BCFT.}
		\label{EN-esc-disj1C}
	\end{figure}

	As reviewed earlier in section \ref{sec:review}, the radiation from an escaping mirror  whose profile is given in \cref{escprof} mimics the Hawking radiation from an eternal black hole. Utilizing this mirror profile in \cref{SAMM} and substituting the length of the geodesics ( or the holographic entanglement entropies ) for various subsystems in  \cref{henadjH} and \cref{hendjcomb} we obtain the behavior for the holographic entanglement negativity for adjacent and disjoint intervals depicted in figures \ref{EN-esc-Adj1F} and \ref{EN-esc-disj1C}. We describe the analogues of Page curves for the entanglement negativity  for various scenarios below.

	\subsubsection*{ Adjacent Intervals}
	%\subsubsection*{ Fixed Intervals}
	%\subsubsection*{Time Evolution}
	
	\subsubsection*{ Disjoint Intervals}

	\section{Summary and discussions}
	\label{sec:sum}
	
	In this section we summarize and discuss the results of our computation. In this article we have obtained the behavior for the  reflected entropy and the entanglement   negativity  of mixed state configurations involving two adjacent and disjoint intervals in the radiation flux of  moving mirrors. These measures were computed for the configurations in the radiation from kink and escaping mirror which mimic the Hawking radiation from  evaporating and eternal black holes respectively.    In this context,  the replica technique was utilized to obtain the reflected entropy  for required mixed states of adjacent and disjoint  intervals  in the large central charge limit of a $BCFT_{1+1}$ describing the moving mirror setup. In order to ease our computation we considered one of the intervals to be semi-infinite. Following this we evaluated the holographic reflected entropy in the corresponding  bulk gravity dual  involving  the $AdS_3$ geometry by determining the entanglement wedge cross section corresponding to the mixed states under consideration. It was demonstrated that the results obtained from the replica technique in the large-$c$ limit agree exactly with the holographic computation of twice the EWCS for various phases of the required mixed states. Following this,  we obtained the Page curves corresponding to  the reflected entropy of  fixed and  co-moving intervals for the corresponding mirror profiles. Quite interestingly, the behaviour we obtained closely resembles the expected Page curve for the reflected entropy determined from earlier investigations. 
	
	Subsequently, we have computed the entanglement negativity for mixed state configurations involving two adjacent and disjoint intervals in the radiation flux of a moving mirror through the corresponding replica technique in the large-$c$ limit of a $BCFT_{1+1}$. We then obtained the corresponding  entanglement negativity utilizing the holographic proposals involving an algebraic sum of the lengths of backreacting cosmic branes. We demonstrated that the holographic entanglement negativity matches precisely with the large-$c$ result obtained through twist field correlators in the $BCFT_{1+1}$ for various phases of the mixed states under consideration. Following this we obtained the Page curves for entanglement negativity of adjacent and disjoint intervals in the radiation flux of the kink and the escaping mirrors for fixed and co-moving intervals.

	The structure of mixed state entanglement in Hawking radiation seems to contain rich phase information which requires further investigations in the various black hole evaporation models \cite{Akers:2019nfi,Penington:2019kki,Balasubramanian:2020hfs,Verheijden:2021yrb,Dong:2021oad}. It would be quite interesting to investigate this significant issue to elucidate our understanding of the information recovery from the black hole interior. It would also be fascinating to explore the behavior of the reflected entropy and the entanglement negativity in the context of doubly holographic models and compare them with those obtained through the $AdS/BCFT$ duality.  We would like to return to these exciting issues in the near future.
	
	\section{Acknowledgement}
	
	The work of GS is partially supported by the Jag Mohan Chair Professor
	position at the Indian Institute of Technology, Kanpur.
	
	\bibliographystyle{utphys}
	\bibliography{SRENMM}
	
\end{document}